\documentclass[a4paper,12pt]{report}

\usepackage{a4wide}
\usepackage{graphicx}
\usepackage{grffile}
\usepackage{xcolor}
\usepackage[latin1]{inputenc}
\usepackage{tabularx}
\usepackage{hyperref}
\usepackage{latexsym}
\usepackage[strings]{underscore}
\usepackage{caption}
\usepackage{fancyhdr}
\newcommand{\abs}[1]{| #1 |}

\newcommand{\const}{{\mbox{const}}}

\newcommand {\dA} {$\bf dA\:$}

\makeatletter
\def\@makechapterhead#1{%
  {\parindent \z@ \raggedright \normalfont
    \ifnum \c@secnumdepth >\m@ne
        \Large\bfseries \space  \thechapter
    \fi
    \interlinepenalty\@M
    \Large \bfseries ~~ #1\par\nobreak
    \vskip 20\p@
  }}
\def\@makeschapterhead#1{%
  {\parindent \z@ \raggedright
    \normalfont
    \interlinepenalty\@M
    \Huge \bfseries  #1\par\nobreak
    \vskip 20\p@
  }}
\makeatother

\newcommand {\new}[1]	{\textcolor{violet}{#1}}

\begin{document}

\setlength{\parindent}{0mm}

\title{Design and Construction of a Device for Measuring Light-Scattering on Anisotropic Materials}
\author{Diplomathesis\\
	Peter Apian-Bennewitz \thanks{ info@pab-opto.de }\\
	April 1990 \\
	Department of Physics\\
	University of Freiburg im Breisgau\\
	and\\
	Fraunhofer Institute for Solar Energy Systems\\
	Freiburg im Breisgau
	}
\date{English translation : February 2021\\Peter Apian-Bennewitz\\~\\
	{\small last updated \today}\\~\\
	{\small Creative Common Licence \href{https://creativecommons.org/licenses/by-nd/4.0/}{\bf CC BY-ND 4.0}}
}
\maketitle

\pagenumbering{alph}
{
\renewcommand{\baselinestretch}{-1.5}\normalsize
\tableofcontents
}
\renewcommand{\baselinestretch}{1.0}\normalsize

\pagenumbering{arabic}
\setcounter{page}{3}

\section{Overview}

This text describes the construction of a device for the optical characterisation of scattering, inhomogeneous, anisotropic media. It measures
the transmission and reflexion dependent on two outgoing and two incoming angles, integral over a spectral range from 400 nm to 700 nm. The sample
size is 40 x 40 x 10 cm. Besides transmission and reflexion values, the scattering characteristic \new{(BSDF)} and absorption can also be
determined.

The description focuses on the mechanical layout, data acquisition and visualisation of the data. The equipment control with a UNIX
workstation and VME bus are detailed.

The theoretical part of this thesis gives a simple model for the material scattering characteristic \new{(BSDF)}.

\section{Notes on this translated English version }

{
\setlength{\parindent}{0mm}

It might be unusual to translate one's own work after over 30 years, but it seemed a worthwhile venture because this diploma thesis appears
to be unknown today even to close collaborators. It introduces some ideas that are still relevant in 2021. These are:
\begin{enumerate}
\itemsep0em
	\item	layout of the lamps, sample and detector of an out-of-plane gonio-photometer (section \ref{det-variants}),
		correlation of angular resolution, speed, data size and overall time (ch. ~\ref{general-layout})
	\item	measurements taken ''on-the-fly'' during detector movement, \ref{on-the-fly}
	\item	custom software and hardware for motor-control \ref{stepper-control-details}
	\item	data display by mapping $f(\theta,\phi)$ onto a plane (birth of {\tt mountain}) \ref{display-triangulation} \ref{mountain-hist}
	\item	foundation of my PhD thesis which was written five years later
\end{enumerate}

Any diploma work filed at a German university around 1990, i.e. prior to the Internet, is nearly impossible to reference.
When this English version is filed at \url{arXiv.org}, reference will at last become possible, - technically at least.

The purpose of this English version is to provide an accurate translation, without trying to paint over flaws in the concept or details:
For comparison, a scanned version of the original is \href{http://pab-opto.de/pers/phd/pab_dipl_1990_de.pdf}{available as PDF on my server}.
As reference, the bibliographic number at the FhG-ISE internal library is {\tt DIP6}.
\new{Additional notes in hindsight are printed in colour}.

The layout of paragraphs and pages was adapted to match the original singled-sided scheme, keeping numbers of figures and pages identical.
Technical details of the conversion are given in Annex \ref{conversion-technical}.

Happy reading.

\subsection{Licence of text and images}
Text and figures herein are distributed under Creative Common Licence \href{https://creativecommons.org/licenses/by-nd/4.0/}{\bf CC BY-ND 4.0},
{\em Attribution-NoDerivatives 4.0 International}

{\em This license requires that re-users give credit to the creator. It allows re-users to copy and distribute the material in any medium or
format in unadapted form only. The content may be used for commercial purposes.}
}

\vfill
\pagebreak[10]

\chapter{Introduction}

Since several years a new class of insulation materials is being examined: Transparent Insulation (TI) . This class of building
materials combines a {\em U-value} of $0.8 - 1.0 [W/m^2K]$ with a transmission of 60\% to 80\% in the visible spectrum (VIS).
TI has applications as facades insulation with solar gain, in solar collectors and increasingly as a way to illuminate indoor
spaces (daylighting). An overview and literature reference are given in \cite{wittwer:89} \new{, \cite{wittwer:92}}.

Some TI materials are already in commercial use, others are still in the development stage. The materials used most frequently are
TI-honeycomb structures and aerogel. Aerogel is an extremely porous material made of $SiO_2$. The air-filled cavities have dimensions that
are smaller than the average free path of air molecules at 1bar pressure. Combined with a very low density, the thermal conduction is lower
than $0.5 [W/m^2K]$ \cite{om-122}. Blocks of solid aerogel appear nearly transparent, while bulk aerogel granulate is translucent.
Stacks of TI honeycomb structures made of transparent plastic material appear translucent to clear and display a characteristic ring of
light when illuminated by direct light. Apart from theoretical work on TI, the following parameters are measured routinely at Fraunhofer ISE:

\begin{tabularx}{\textwidth}{|l|X|}
\hline
U-value		&heat flow between two plates of constant temperature \cite{pflueger:84}\\
g-value		&heat flow to plate of constant temperature when illuminated by direct light \cite{jacobs:89}\\
$\tau_{dh}$	&direct-hemispherical transmission depending on wavelength and incident angle \cite{platzer:86},\cite{wittwer:90}\\
$\tau$		&transmission with UV, VIS, IR spectrometers (spectral, direct and diffuse transmission)\\
\hline
\end{tabularx}

Up to now, it was not possible to measure angular dependent values of transmission and reflexion, however, there are four reasons why
measurements are important.  The first two are covered in this text:

\begin{enumerate}
\itemsep0em
\item	characterisation and comparison of materials
\item	indoor illumination levels on surfaces located behind the TI for varying incoming radiance (especially direct sunlight)
\item	calculation of absorption in the material
\item	simulation of multi-layered TI systems \cite{ta-233}
\end{enumerate}

Previously, the incident direction was designated with a single angle between the surface normal and incident direction. However, for TI materials,
which are not rotationally symmetric around their surface normal, the incident direction has to be specified by {\em two} angles,
$\theta_{in}$ and $\phi_{in}$. Consequently, a measurement device must allow all incident and outgoing angles (4 parameters) to be measured.
The aim of this work was the design and construction of a device that is flexible in its task, robust and precise.

Prior to this work, a similar device had been built at Lawrence Berkeley Laboratory (LBL) \cite{da-248}, which was originally conceived to measure
lamp distribution curves \cite{da-172}. This set-up differs from the "large scanning radiometer" developed at LBL, for example by a different
detector geometry. \footnote{In 2008, the follow-on to this work, my ''PG2'' gonio-photometer, was exported back to LBL. The great circle of life.}

\chapter{Theory}
\label{theory}

\section{Scattering in general}

Scattering is defined as the deflection of waves or particles by inhomogeneities in the media through which the waves/particles pass.
The energy of the particle (or wave) and the spatial distribution of the inhomogeneities (scattering centres) can vary over a wide range.
For example, photons of electromagnetic waves, from radio waves to gamma rays, can be scattered by elementary particles, atomic
nuclei, crystals, TI materials or TV-antennas. For the following, it suffices to restrict ourselves to classical electrodynamics, since
excitation of discrete energy levels in the materials can be neglected. In all energy ranges, the description of the scattering becomes more
complex when the wavelength ($\lambda$) of the incoming radiation is approximately equal to the size ($d$) of the scattering centres.\\

In case of the visible spectrum ($\lambda = 300 - 800nm$), there are three ways to describe scattering:

\begin{itemize}
\itemsep0em
\item $\lambda \gg d$ ~~Each scattering centre of size $d$ is sufficiently far away from other scattering centres, that each can be regarded
as an emitting dipole ({\em Rayleigh scattering}).
\item $\lambda \approx d$ ~~Each scattering centre becomes a multi-pole-radiator, for which the higher multi-pole terms can not be neglected
({\em Mie scattering}).
\item $\lambda \ll d$ ~~Scattering follows from geometrical considerations \new{Geometrical Optics, see \cite{luneburg:64}, \cite{kline:65}})
\end{itemize}
A material is typically made up of many individual scattering centres, whose contributions add up to the scattering amplitude. Depending on
the coherence length of the incoming radiation, the distance between scattering centres and their geometric arrangement (e.g. crystal lattice),
the sum is coherent or plainly the sum of each power contribution.

\section{Scattering in Transparent Insulation Material}

In this case, electromagnetic waves, $\lambda = 300-800nm$, are scattered at objects with a typical size given in meters and internal
structures of size $d$. The incoming radiation (sunlight) is parallel, incoherent and de-polarised. A detector averages over all polarisation
states.

The two TI groups, aerogel and honeycombs, exhibit very different scattering mechanisms:

Aerogel consists of pores with a diameter smaller than $\lambda$. Furthermore, density fluctuations occur in blocks of solid aerogel, which
has spatial dimensions in the order of the wavelength. Therefore, aerogel shows Rayleigh and Mie scattering. The scattering centres are dense,
causing multiple consecutive scattering events, which are not described by pure Rayleigh scattering. Nevertheless, the behaviour of many
aerogel samples can still be described by Rayleigh scattering, especially with good solid blocks of aerogel whose density fluctuations
are small \cite{hunt83}. The size of scattering centres is roughly $d=8..10 nm$ \cite{om-122}, the deviation from the Rayleigh theory is
described in \cite{om-194}. Chapter \ref{dipole} describes the deduction of dipole radiation pattern from the Maxwell equations and it is used
in \ref{rayleigh} to get the differential cross-section of Rayleigh scattering.

\pagebreak[3]

A quantitative verification of Rayleigh and Mie scattering with massive blocks of aerogel was not part of this work, since emphasis lay on
honeycomb structures. Chapter~\ref{meas1} contains a comparison between solid blocks of aerogel and bulk granular aerogel.

\begin{figure}[hbtp]
\begin{center}
\fbox{
\input{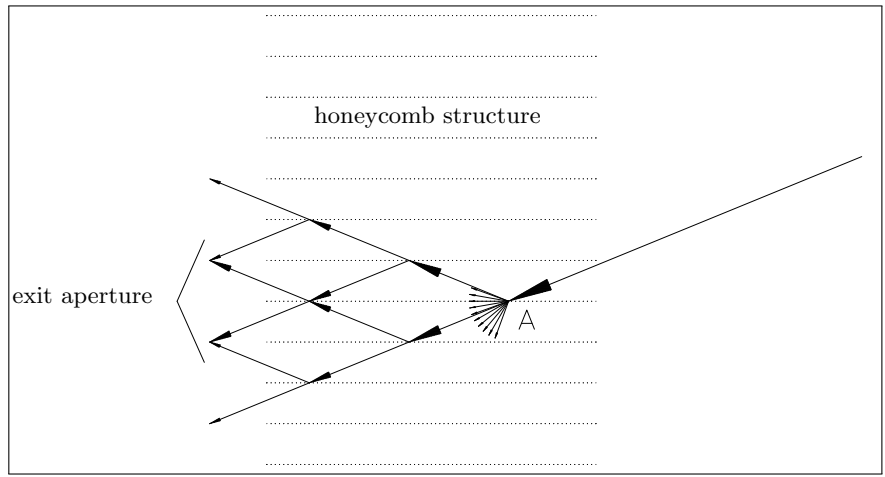}
}
\caption{ray-paths in an ideal honeycomb material, 2D cross-section \label{honeycomb-ray}}
\end{center}
\end{figure}

For TI honeycomb materials, the typical ray path is shown in Fig.~\ref{honeycomb-ray}, with a typical size of the honeycomb structure about
4mm. Each wall of the honeycomb divides the incoming ray in two rays. Diffuse scattering and absorption in the material add to the overall
transmission, as does the unevenness of the walls, distorting the ideal ray path. Chapter \ref{ti-model} gives a simple model for the
scattering mechanism of honeycomb materials.

\begin{figure}[hbtp]
\begin{center}
\includegraphics[height=11cm]{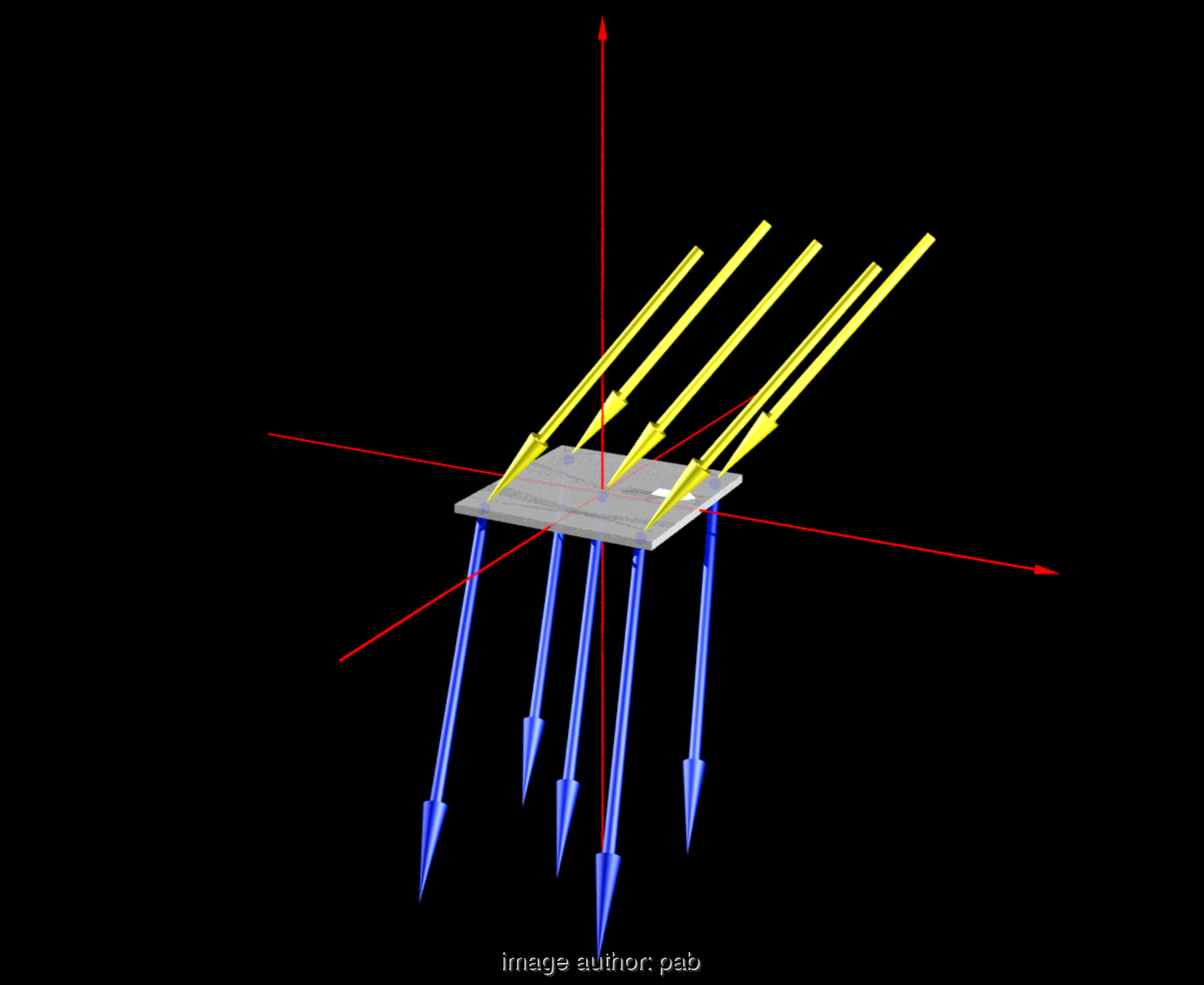}
\caption{Simplest beam path: Incoming light (yellow), outgoing light (blue) and sample (grey)\label{farfield}}
\end{center}
\end{figure}

The simplest case of scattering geometry on the scale of the sample is ''parallel-in / parallel-out'', meaning that the sample is
illuminated by ''parallel'' light, and the distance between the detector and the centre of the sample is substantially larger than the
sample size, as shown in Fig.\ref{farfield}. This beam path is used to derive the Rayleigh scattering. This geometry is also realised if,
with a larger sample, a lens of sufficient size is mounted in front of the detector, representing the measurement set-up of Fraunhofer
diffraction at a grid. At the LBL radiometer, considerations have been made to use a parabolic mirror to get more exact data for multi-layer
simulation \cite{klems:89}.

For practical daylighting applications with TI material, this simple case is not applicable: The sample size amounts to multiple square
meters when the material is applied to a facade, or to 40x40 cm as sample in the measurement device. For both cases, the distance between
the sample centre and a surface element behind the TI (e.g. the detector) is within the same order of magnitude as the emitting surface.
Contrary, the {\em incoming} light is ''parallel'' in the practice (sunlight) and in the device (by special illumination).
\new{See PhD work for a much refined analysis.}

Chapter~\ref{photometric-ti} describes how, with a converging scattering geometry and photometric methods, the material constants and light
distribution behind the TI can be derived for daylight applications.

\begin{figure}[hbtp]
\begin{center}
\includegraphics[height=11cm]{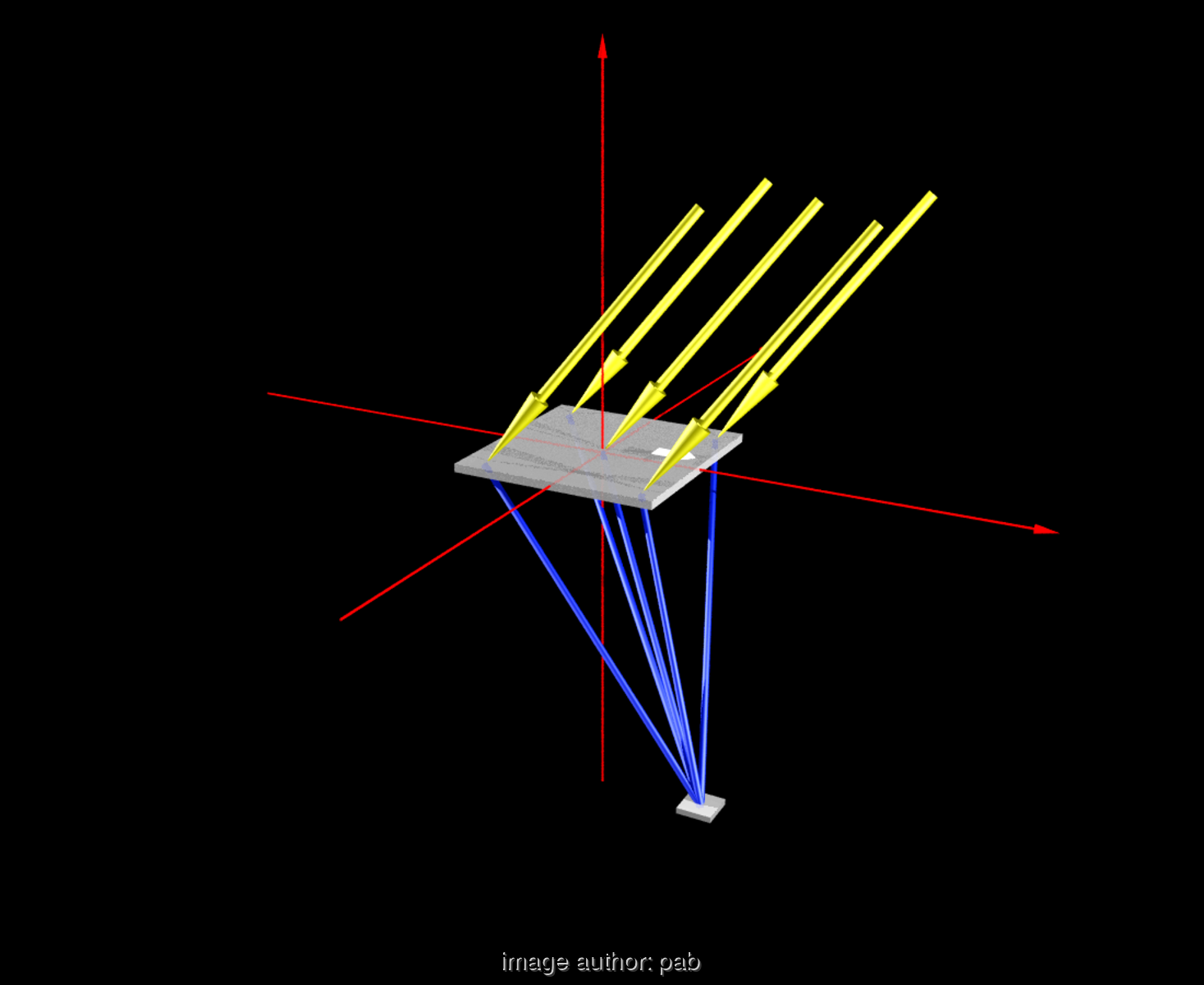}
\caption{Scattering beam path in the practice: The detector is not at infinity, and the outgoing beam paths are not parallel}
\end{center}
\end{figure}
\vfill

\clearpage

\section{Radiation patterns of dipoles}
\label{dipole}

The description of a harmonic dipole follow Chapters 6, 9 and 16 in \cite{jackson:75}. The four Maxwell equations can be written with vector
potential $\bf A$ and scalar potential $\Phi$ in a vacuum as:
\footnote{Vectors are printed in bold font in this and the following chapter.}
\begin{eqnarray}
\itemsep5em
	\nabla^2 {\bf A} \; - \; \frac{1}{c} \; \frac{\partial^2 {\bf A}}{\partial t^2} &=& - \frac{4\pi}{c} {\bf J}\label{a-def}\\
	\nabla^2 \Phi \; - \; \frac{1}{c} \; \frac{\partial^2 {\Phi}}{\partial t^2} &=& - 4\pi \rho\\
	\nabla {\bf A} \; - \; \frac{1}{c} \; \frac{\partial {\Phi}}{\partial t} &=& 0 \textrm{\hspace{4em}(Lorentz gauge)}
\end{eqnarray}
\begin{eqnarray}
	{\bf B} &=& \nabla \times {\bf A}\\
	{\bf E} &=& - \nabla \Phi \; - \; \frac{1}{c} \, \frac{\partial {\bf A}}{\partial t}\label{e-def}
\end{eqnarray}
With currency density $\bf J$ and charge density $\rho$.

A solution of $\bf A$ and $\Phi$ for a given $\bf J$ and $\rho$, without boundary conditions, is:
\begin{equation}
	{\bf A}({\bf x},t) = \frac{1}{c} \;\; \int\int \;\;
				\frac{ {\bf J}({\bf x'},t') } { \abs{{\bf x} - {\bf x'}} } \;\;
				{\delta \left( t'+ \frac{\abs{{\bf x} - {\bf x'}}}{c} -t \right) } \;\;
				d^3{\bf x}' \; dt'
\end{equation}
The Dirac function $\delta$ is the retarded Green's function, causally linking $\bf J$ and $\bf A$.

Assuming a harmonic dependency for $\bf J$ and $\rho$:
\begin{eqnarray}
	\rho({\bf x},t)	   &:=& \rho({\bf x}) \; \exp( -i \omega t )\\
\nonumber
	{\bf A}({\bf x},t) &:=& {\bf A}({\bf x}) \; \exp( -i \omega t )
\end{eqnarray}
then ${\bf A}$ follows with:
\begin{equation}
\label{solution-A}
	{\bf A}({\bf x},t) =  \exp( -i \omega t ) \; \frac{1}{c} \;  \int \; {\bf J}({\bf x'}) \;
				\frac{\exp( i k \abs{{\bf x} - {\bf x'}})} { \abs{{\bf x} - {\bf x'}} } \;\;
				d^3{\bf x}'
\end{equation}
with wave-number $k:=\omega/c$. Similar, $\Phi$ follows from $\rho$. Both ${\bf A}$ and $\Phi$ have a harmonic time dependency.

The $\exp$-term in \ref{solution-A} can be approximated by spherical harmonics, giving a spatial term:
\begin{equation}
\label{sum-A}
	{\bf A}({\bf x}) = \frac{4 \pi i k}{c} \; \sum_l \; \sum_{m=-l}^{m=+l} \; h_l^{(1)}(kr) \; Y_{lm}(\theta,\phi)
				\int {\bf J}({\bf x'}) \; j_l^{(1)}(kr) \; Y^*_{lm}(\theta',\phi') \;\;
				d^3{\bf x}'
\end{equation}
with $h_l^{(1)}$ Hankel-function, $j_l$ spherical Bessel-function, $Y_{lm}$ spherical harmonic. For $l=0$ :
\begin{displaymath}
	h_0^{(1)} = \frac{\exp(ix)}{ix} \;\;\;\; , \;\;\;\;
	Y_{00} = \frac{1}{\sqrt{4\pi}} \;\;\;\; , \;\;\;\;
	j_0 = \frac{\sin x}{x} 
\end{displaymath}

\vfill

\pagebreak[10]

for $kr' \leadsto 0$ (dimension of source is smaller than wavelength), the $l=0$ term in the sum \ref{sum-A} (spherical wave):
\begin{equation}
	{\bf A}({\bf x}) = \frac{\exp(i k r)}{c r} \; \int \; {\bf J}({\bf x'}) \; d^3{\bf x}'
\end{equation}
For a localised charge, and using continuity $i \omega \rho = \nabla {\bf J}$: %Jackson 9.15
\begin{equation}
	{\bf A}({\bf x}) = -i k \; {\bf p} \; \frac{\exp{i k r}}{r}
\end{equation}
with $r$ being the distance to the observation point and a dipole moment $\bf p$ defined as:
\begin{displaymath}
	\bf p := \int {\bf x}' \; \rho({\bf x}') \; d^3{\bf x}'
\end{displaymath}
$\bf B$ and $\bf E$ follow with:
\begin{eqnarray}
\nonumber
	{\bf B} &=& \nabla \times {\bf A}  \;\;\; \textrm{(definiton of $\bf A$)} \\
\nonumber
	{\bf E} &=& \frac{i}{k} \; \nabla \times {\bf B} \;\;\; \textrm{(outside of sources, harmonic, Lorentz gauge)}
\end{eqnarray}
this follows from \ref{e-def} with help of:
\begin{displaymath}
	- \nabla \phi = \frac{i}{k} \; \nabla(\nabla {\bf A}) \;\;\; \textrm{(Lorentz gauge)}
\end{displaymath}
\begin{displaymath}
	- \frac{\partial{\bf A}}{c\;\partial t} = \frac{-i}{k} \; \nabla^2 {\bf A} \;\;\; \textrm{(\ref{a-def} and ${\bf J}=0$)}
\end{displaymath}
then $\bf E$ and $\bf B$ fields of a harmonic dipole:
\begin{eqnarray}
	{\bf B} &=& k^2 \; ( {\bf n} \times {\bf p} ) \; \frac{\exp(i k r )}{r} \left( 1 - \frac{1}{i k r} \right)\\
	{\bf E} &=& k^2 \; ( {\bf n} \times {\bf p} ) \; \times {\bf n} \; \frac{\exp(i k r )}{r} +
			\left[ 3 {\bf n}({\bf n}{\bf p}) - {\bf p} \right]
			\left( \frac{1}{r^3} - \frac{i k}{r^2} \right) \exp(i k r) 
\end{eqnarray}
for near field zone ($kr \ll 1$) , $\bf E$ is approximated by the field of a static dipole multiplied with $\exp(i \omega t)$.
In the far field ($kr \gg 1$):
\begin{eqnarray}
	{\bf B} &=& k^2 \; ( {\bf n} \times {\bf p} ) \;\; \frac{\exp(i k r )}{r} \label{B-final}\\
	{\bf E} &=& k^2 \; ( {\bf n} \times {\bf p} ) \times {\bf n} \;\; \frac{\exp(i k r )}{r} ={\bf B} \times {\bf n}\label{E-final}
\end{eqnarray}
Which is the result we are looking for.

\pagebreak[10]

\section{Rayleigh scattering}
\label{rayleigh}

The derivation of Rayleigh's formula assumes the following: 1) the incident EM wave induces multi-poles in the material. 2) The wavelength
of the EM wavefront is much larger than the dimension of the scattering centres, and therefore it suffices to use the dipole terms in the
multi-pole approximation. The incident EM wave is assumed to be plane and monochromatic:
\begin{eqnarray}
	\label{EM-in}
	{\bf E}_{in} &=& {\bf e}_0 \; E_0 \; \exp( i k {\bf n} {\bf x} ) \; \exp(i \omega t)\\
\nonumber
	{\bf B}_{in} &=& {\bf n}_0 \times {\bf E}_{in}
\end{eqnarray}
with direction of polarisation ${\bf e}_0$ and direction of propagation ${\bf n}_0$ and $k=\omega/c$

The differential cross-section is defined as power re-radiated into a unit solid angle in direction $\bf n$ with polarisation $\bf e$, per
incident flux from direction ${\bf n}_0$ with polarisation ${\bf e}_0$:
\begin{equation}
	\frac{d\sigma}{d\Omega} ({\bf n},{\bf e},{\bf n}_0,{\bf e}_0) =
		\frac{r^2 \abs{{\bf e}^* {\bf E_{sc}}}^2}{\abs{{\bf e_0}^* {\bf E_{in}}}^2}
\end{equation}
with indices $sc$ for the scattered radiation. With \ref{B-final},\ref{E-final} and \ref{EM-in}:
\begin{equation}
	\frac{d\sigma}{d\Omega} ({\bf n},{\bf e},{\bf n}_0,{\bf e}_0) =
		\frac{k^4}{E^2_0} \abs{{\bf e}^*{\bf p} +  ({\bf n} \times {\bf e}^*){\bf m}}^2
\end{equation}
while assuming an electrical and magnetic dipole ${\bf m}$. ${\bf e}_0$ and ${\bf n}_0$ are hidden in ${\bf p}$ and ${\bf m}$.
The $k^4$ term is characteristic of Rayleigh scattering.  In a next step, a dielectric sphere is assumed to be the scattering centre (radius
$a$ and dielectric constant $\epsilon$). The incident field induces an electric dipole:
\begin{equation}
	{\bf p}  = \frac{\epsilon-1}{\epsilon+2} \; a^3 \; {\bf E}_{in}  \;\;\;\textrm{and}\;\;\;{\bf m}=0
\end{equation}
the differential cross-section follows with:
\begin{equation}
	\frac{d\sigma}{d\Omega} = k^4 \; a^6 \; \frac{(\epsilon-1)^2}{(\epsilon+2)^2} \; \abs{{\bf e}^* {\bf e}_0}^2
\end{equation}
by averaging over all ${\bf e}_0$ and a sum over all ${\bf e}$, the differential cross-section for scattering of an unpolarised, plane,
monochromatic wave at an electric dipole is:
\begin{equation}
\label{dipole-rayleigh}
	\frac{d\sigma}{d\Omega} = k^4 \; a^6 \; \frac{(\epsilon-1)^2}{(\epsilon+2)^2} \; (1+\cos^2\theta)/2
\end{equation}
with $\theta$ defined as angle between $\bf n$ and ${\bf n}_0$

For multiple dipoles, the scattered field is a coherent addition of each dipole radiation. If the distance between the scattering centres is
larger than the length of coherence, then it is just the incoherent sum (lack of multiple scattering and incoherent excitation of each dipole).

If $\epsilon$ is known, equation \ref{dipole-rayleigh} allows to conclude the size $a$ of the spherical scattering centres to be concluded
from the angular data of the scattering \cite{om-122}.

\vfill
\pagebreak[20]

\section{Basics of Photometry}
\label{photometry-basics}

\begin{figure}[hbtp]
\begin{center}
\includegraphics[height=10cm]{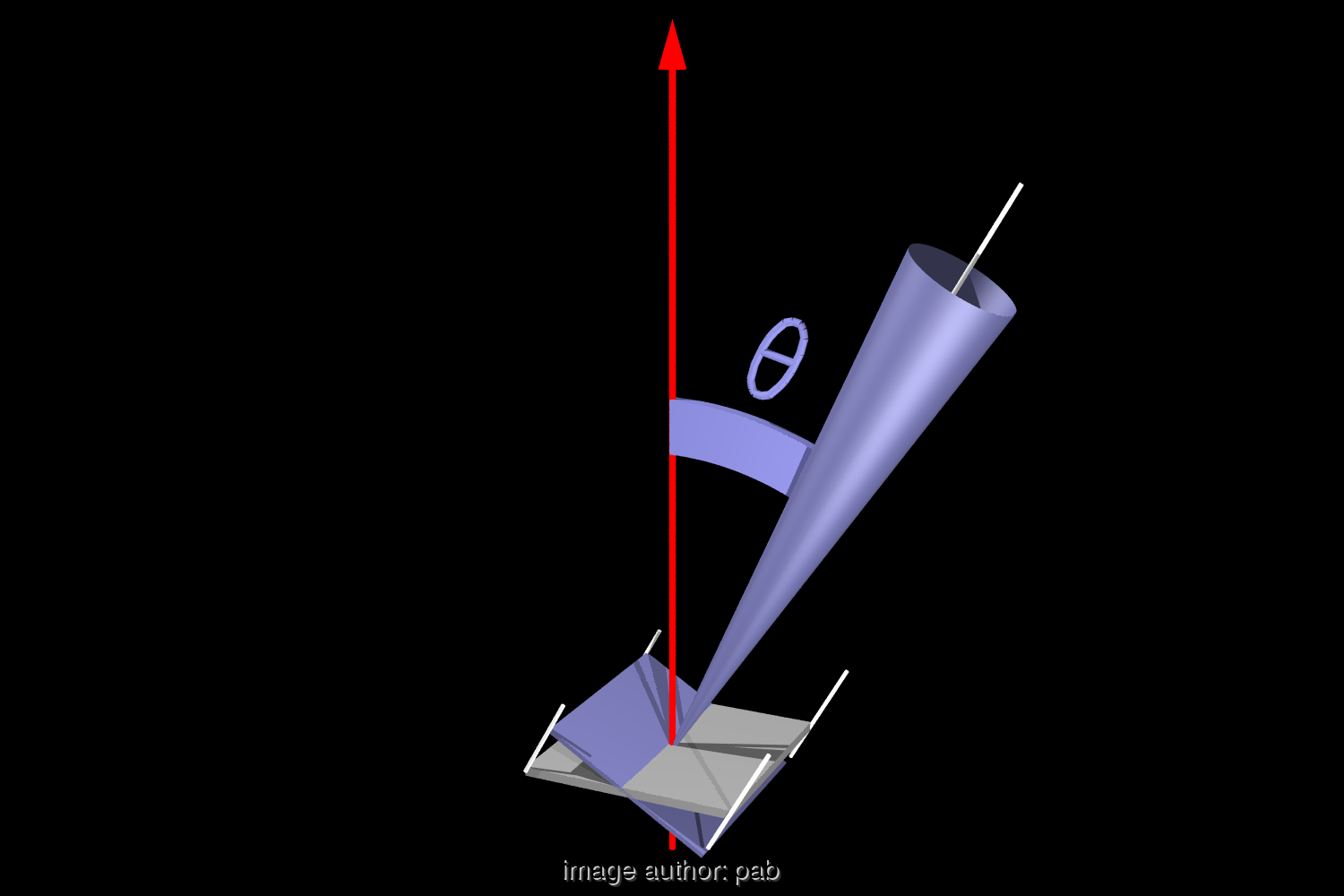}
\caption{The basic photometric units: A surface element \dA (grey surface) emits into a solid angle $d\Omega$ (blue cone) a power $dP$.
	The outgoing direction is given by $\theta$, and the projected surface element (blue area) is orthogonal to the outgoing direction.
	\label{radiance-def} }
\end{center}
\end{figure}

A small surface element \dA emits power $dP$ into a small solid angle $d\Omega$, integral over all wavelengths. The quantity ${\bf I}$
specifies the angular distribution of the emitted power:
\footnote{Chapter \ref{photometry-basics} and the following use bold font for relevant scalar quantities.}
\begin{equation}
	dP = {\bf I}(\theta) \; \cos \theta \; {\bf dA} \; d\Omega
\end{equation}
The term $\cos(\theta)$ is added so that an ideal black body radiator has ${\bf I}(\theta) = \const$ . Hence ${\bf I}$ specifies the energy
per time interval per solid angle per projected surface element, integral over all wavelengths (see Fig.~\ref{radiance-def}).

In the case of ''grey'' volume emitters with strong self-absorption, ${\bf I}$ is constant as well \cite{hottel-1967-radiative}[pp22]. The
surface of such an emitter is called a ''Lambertian'' or diffuse emitter. A ''grey'' material has an absorption and emittance which is independent
of wavelength. The naming of ${\bf I}$ varies: intensity \cite{siegel-1981-thermal} \cite{hottel-1967-radiative}, radiance
\cite{bs} \cite{malacara} or photometric brightness \cite{born-wolf}.
\footnote{In my own PhD, 5 years later, the symbol $\cal(L)$ is used, and the radiometric term is {\em Radiance}.}

In general, ${\bf I}$ is a function ${\bf I}(\theta,\phi,\eta,\xi)$, where the angles $\theta$ and $\phi$ define the direction and
$\eta$ and $\xi$ the coordinates of \dA on the surface A. Let $dP_{hem}$ define the infinitesimal power emitted into the hemisphere from one side
of the surface element, then:
\begin{displaymath}
	dP_{hem} = \int_\theta \; \int_\phi \; dP
\end{displaymath}

\vfill
\pagebreak[10]

Let \dA be a surface element emitting power towards a second surface element $\bf dA_2\:$ in the direction $(\theta_1,\phi_1)$ and distance $r$.
The angle $\theta_2$ is located between the surface normal of the second element and the line between the two surfaces. The solid of angle 
$\bf dA_2\:$ as seen from \dA is:
\begin{displaymath}
	d\Omega = {\bf dA_2} \; \frac{\cos( \theta_2 )}{r^2}
\end{displaymath}
the transferred radiative power is:
\begin{equation}
	dP = {\bf I}(\theta,\phi,\eta,\xi) \; \cos( \theta_1 ) {\bf dA_1} \; \cos( \theta_2 ) {\bf dA_2} \; r^{-2}
\end{equation}

If the dimension of a surface $\bf A$ is small compared to $r$, then $dP$ is sufficiently constant over the surface and the total
transmitted power from $\bf A_1$ to $\bf A_2$ is:
\begin{equation}
	\label{transport}
	P= {\bf A_2} \; \int_{\eta_1} \; \int_{\xi_1} \; {\bf I}(\theta_1,\phi_1,\eta_1,\xi_1) \;
			\cos( \theta_1 ) \cos( \theta_2 ) \; r^{-2} \; {\bf dA_1}
\end{equation}
with the angles depending on the integration variables $\eta_1, \xi_1$ over $\bf A_1$:
\begin{displaymath}
	\theta_1 = \theta_1( \eta_1, \xi_1 ) , \phi_1 = \phi_1( \eta_1, \xi_1 ), \theta_2 = \theta_2( \eta_1, \xi_1 ), r=r(\eta_1, \xi_1 )
\end{displaymath}

The following chapters are based on equation \ref{transport}.

\clearpage
\vfill
\pagebreak[10]

\section{Photometry of Transparent Insulation materials}
\label{photometric-ti}

\begin{figure}[hbtp]
\begin{center}
\includegraphics[height=90mm]{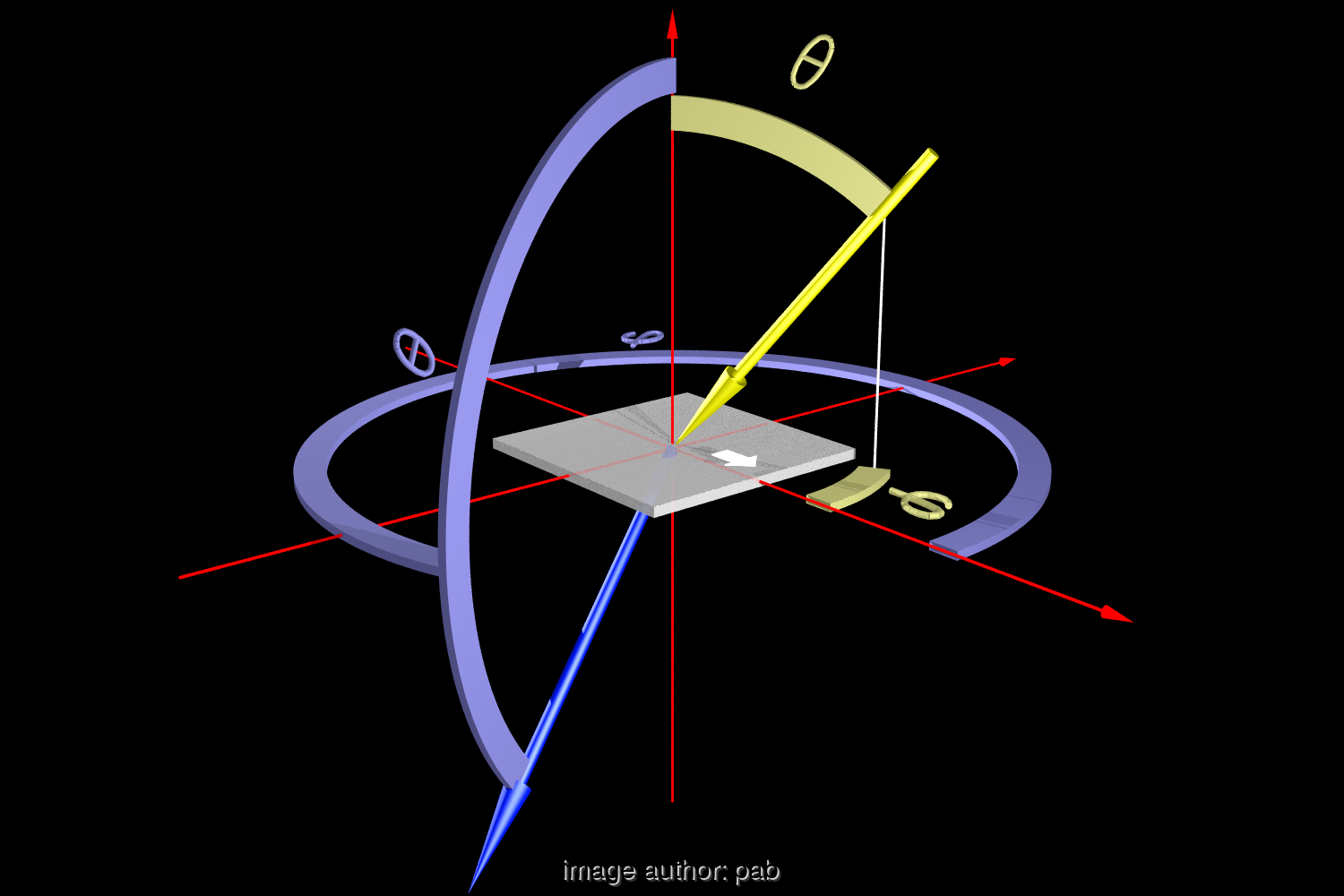}
\caption{Coordinate system of the sample with incident direction $(\theta_{in},\phi_{in})$ (yellow) and outgoing direction
$(\theta_{out},\phi_{out})$. The x-axis is from left-to-right, the y-axis front-to-back and the z-axis bottom-to-top. The sample is not rotationally
symmetric around the surface normal, symbolised by the white arrow. \label{coordinates}}
\end{center}
\end{figure}

This chapter describes the geometry of the measurement device. Fig.~\ref{coordinates} shows the coordinate system of a generic sample.
\footnote{Note in 2021: In 1989, this had been been different to the coordinates used at LBL, but later turned out identical to ASTM standard E2387
of BSDF measurements, see \cite{astm-E2387}.}

\begin{figure}[h!btp]
\begin{center}
\includegraphics[height=90mm]{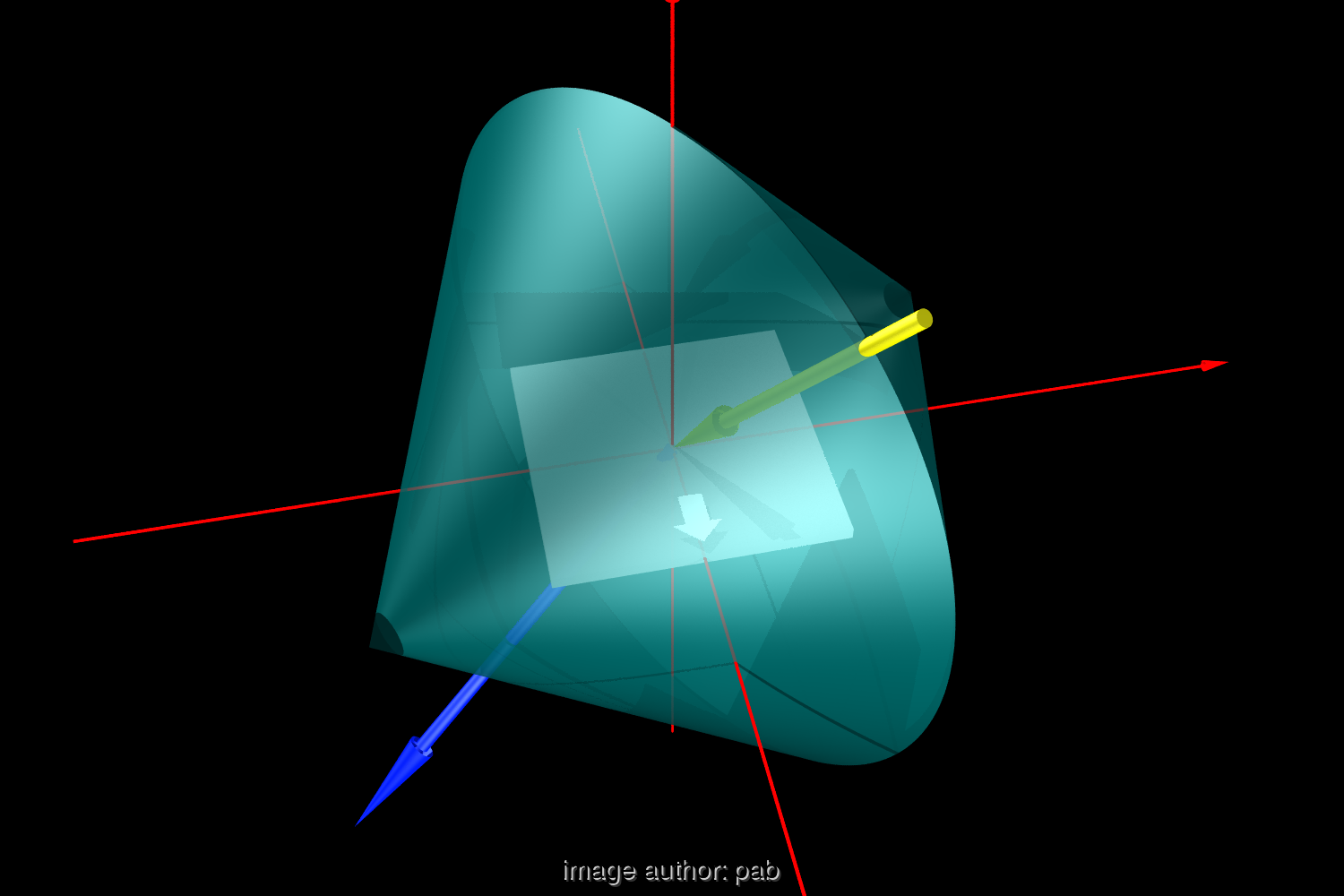}
\caption{The surface $S_2$ on which the detector moves (light blue), and the surface of the sample (light grey)\label{cones}}
\end{center}
\end{figure}

\clearpage

In general, the detector moves on a surface $S_2$ around the sample. In the simplest case, $S_2$ is spherical. For the device described
here, the surface is a combination of two cones for mechanical reasons (see Fig.~\ref{cones}).
The different distance between detector and the sample centre is compensated for by the inverse square of the distance.
\footnote{2021: This is a) only approximate and b) was avoided by a new circular rail right after completion of the diploma work.}

The overall energy balance of the sample is an important value for theories that model the inner scattering of a sample. However, to characterise a
sample \new{for indoor illumination}, it is sufficient to consider only the transmitted light and treat the surface facing away from the
lamp as self-luminous.

The coordinates of the detector position are $(\theta_{out}, \phi_{out})$, and the measured quantity is the power $P$ received by the detector.
Measurements are taken at specific, equidistant points $(\theta_{out}, \phi_{out})$.
\new{Actually, the device used a more sophisticated, adaptive process, see later.}

The calculation of the absorption is easy: All data values with $\theta_{out} = [0^o..90^o]$, the detector surface area, and the angular
distance from each point to neighbouring points, result in the total reflected power. Similarly, the transmission follows from all points with
$\theta_{out} = [90^o..180^o]$. A measurement without sample gives the total incident power, which allows calculation of the absorption.
Due to technical reasons, current data does not give absolute values for transmission and reflexion yet.
\new{See follow-on PhD for a full discussion.}

A comparison of different materials is a comparison of the luminous surface of $S_2$. Results are presented in chapter \ref{meas3}.

An interesting application of scattering data is the calculation of indoor illumination levels behind TI facades (''daylighting'').
Up to now, this simulation was only possible for overcast skies, whose near-diffuse radiation allows the rough assumption of a
diffuse illumination in the incident side of the TI, leading to a further approximation that the outgoing side of TI is also diffuse. Using
measurement data of this new device, it becomes possible to calculate illumination levels for direct sunlight.

If the TI (Transparent Insulation) is integrated in a facade, the area is much larger than the sample size of 40x40 [cm] when measured in the
device\footnote{The edges of the thick 40x40cm sample have reflective coating applied to approximate an infinite size.}, and the indoor
surfaces are not congruent with the surface $S_2$ of the detector positions. Nevertheless, data points are only available on $S_2$ and those
must be used for the calculation. We propose two methods to achieve this:

The first method consists of establishing a model for the scattering and fit its parameters to measured data. The model can than be applied
to larger TI surfaces as well, solving the problem. However, this works if the TI material is made of near-regular small structures,
whose dimensions are small compared to the detector distance. Most current TI materials fit this requirement.

The second method works for materials that have no regular structure, or whose structure is larger (e.g. glass bricks, shading devices).
Yet the method is technically more demanding, requires a special detector and generates substantially more data, taking up more storage
space.
\new{Commonly known as ''Near Field Radiometry''.}

The following describes both methods using the photometric basics of Chapter~\ref{photometry-basics}.

\vfill
\pagebreak[10]

To calculate the transmitted power from a large surface $\bf A_1$ to a small surface element $\bf dA_2$, the quantity $\bf I$ must be known
on $\bf A_1$ (see Equation.~\ref{transport}). In this context $\bf A_1$ is the surface of the sample opposite the lamp. However,
alternatively, $\bf I$ can be defined on {\em any} convex surface $S_2$ that encloses the luminous surface $\bf A_1$ . In this case, the
surface $S_2$ will be treated as luminous surface itself, and $P$ for any point outside $S_2$ is calculated in the same way.

The first method establishes, for each structural element of the TI surface, a physically plausible model of
${\bf I}(\theta,\phi,\eta,\xi) = {\bf I}(\theta,\phi)$. \new{Aka, constant over the surface.}
The model has 4-8 parameters and calculates the power received by a ''normal'' detector, depending on the position of the detector
(typically 80-800 different positions). These calculated values are fitted to measured data (see Chapter~\ref{ti-model}). A ''normal'' detector in
this sense is one that integrates over all incoming directions, e.g. a solar cell.

A model of $\bf I$ also allows for the interpolation of values for incident directions that have not been measured (Up to now it wasn't
mentioned that $\bf I$ depends on the incoming illumination as well). The disadvantage of this model is, that for each new sample, one has to
find a suitable physical model.

The second method measures ${\bf I}(\theta,\phi,\eta,\xi)$ on the surface $S_2$ \new{(surface of the detector paths)} at specific locations
$\theta_{out}, \phi_{out}$.  These two angles are effectively the coordinates $(\eta,\xi)$ on $S_2$. A large area made of TI can than be
assembled from smaller 40x40 cm samples, each with its own $S_2$. The total intensity is then the sum of each luminous surface element on
all $S_2$ surfaces.
\new{Near field photometry}

\label{camera-intro}
Measurement of ${\bf I}$ on $S_2$ requires a detector that resolves the received power by incoming direction. This should be possible using a
CCD camera. The precision with which illumination levels on arbitrary surfaces behind the TI can then be calculated, depends on the
positions of measurement points, which can not be arbitrarily close. This is subject to further investigation.
The mechanics to move the camera to a specific position is the same as for a normal detector. Thus adding a camera at a later time is possible,
should the need arise (depending on TI material types). Test runs with a camera had been concluded and showed problems with the calibration of
pixels, aperture control and higher requirements for computing power during measurements.
\new{For comparison, the workstation featured an MC68020 CPU in 1989}.

Typically for older setups, the detector takes measurements using positions on a given, regular $(\theta_{out},\phi_{out})$ grid, where the
detector path is independent of the sample characteristics. This is sub-optimal for samples with a strong directional transmission or
reflexion: Areas with diffuse transmission can be measured with a coarser angular resolution, whereas specular ''highlights'' require a
finer resolution. For this, a fast ''pre-scan'' is required that differentiates between angular areas with a ''diffuse'' background and areas
with high fluctuations. The required enhanced mechanical drive was built into this device from the start, and the control program is currently being
developed.

\vfill
\pagebreak[10]

\section{Simple model of scattering by TI honeycombs}
\label{ti-model}

This section describes a model for TI honeycombs and capillary structures, whose scattering pattern is independent of the angle $\phi_{out}$.

\begin{figure}[hbtp]
\begin{center}
\includegraphics[height=9cm]{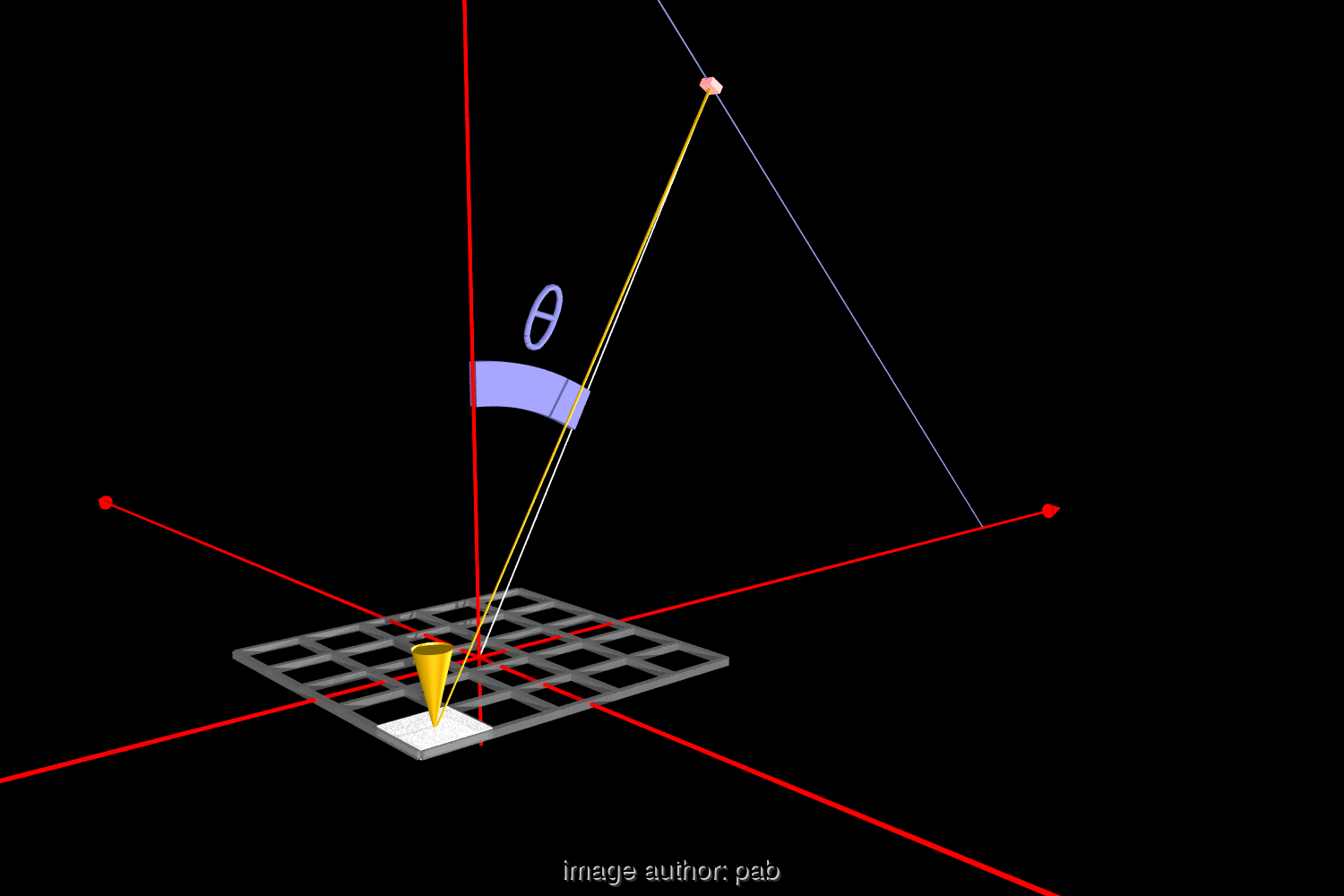}
\caption{Geometry of the sample and detector for modelling. Dimensions are to scale. For calculations, the grid on the sample is finer than
shown here. \label{ti-model-plot}}
\end{center}
\end{figure}

The sample is sub-divided into a checker-board of 20x20 patches, and $\bf I$ is considered to be the same for each patch
(Fig.~\ref{ti-model-plot}).  The sum of the radiated power from each patch towards the detector surface $\bf A_2$ is the total power measured
by the detector, according to Equation~\ref{transport}. The detector centre is moved along a diagonal line, representing the current mechanical
set-up in the measurement device.  The angle between the surface normal and the direction to the detector is labelled $\theta$.

$\bf I$ is assumed as ${\bf I}(\theta,\phi,\eta,\xi)={\bf I}(\theta)$, the model assumes a Gaussian distribution:
\begin{displaymath}
	{\bf I_{patch}}(\theta) = s  \; \frac{1}{\theta_{width} \sqrt{2\pi}} \; \exp( -0.5 ((\theta-\theta_{max}) / \theta_{width})^2 ) + o )
\end{displaymath}
$s$ simply scales the power, $\theta_{max}$ and $\theta_{width}$ characterise the cone shaped emittance of each patch and $o$ specifies a diffuse
background.

The total power received by the detector scales with the distance between sample centre and detector and is given as (\ref{transport}):
\begin{displaymath}
	{\bf I_{sum}}(\theta) = \frac{R(\theta)^2}{R_{max}^2} \sum_{patches} I_{patch}(\theta_{patch}) \; \cos(\theta_{patch}) \;
				{\bf A_{patch}} \; \frac{1}{r^2} \; \cos(\theta_{det}) \; {\bf A_{det}}
\end{displaymath}
\begin{tabular} {ll}
$R$			&distance between detector and sample centre at $\theta=0^o$\\
$R(\theta)$		&current distance between detector and sample centre\\
$\theta_{det}$		&angle at detector towards a patch\\
$\theta_{patch}$	&angle at patch towards detector\\
${\bf A}_{patch}$	&patch surface area\\
${\bf A}_{det}$		&detector surface area\\
$r$			&distance patch to detector\\
\end{tabular}

\vfill
\pagebreak[10]

\begin{figure}[hbtp]
\begin{center}
\fbox{
\input{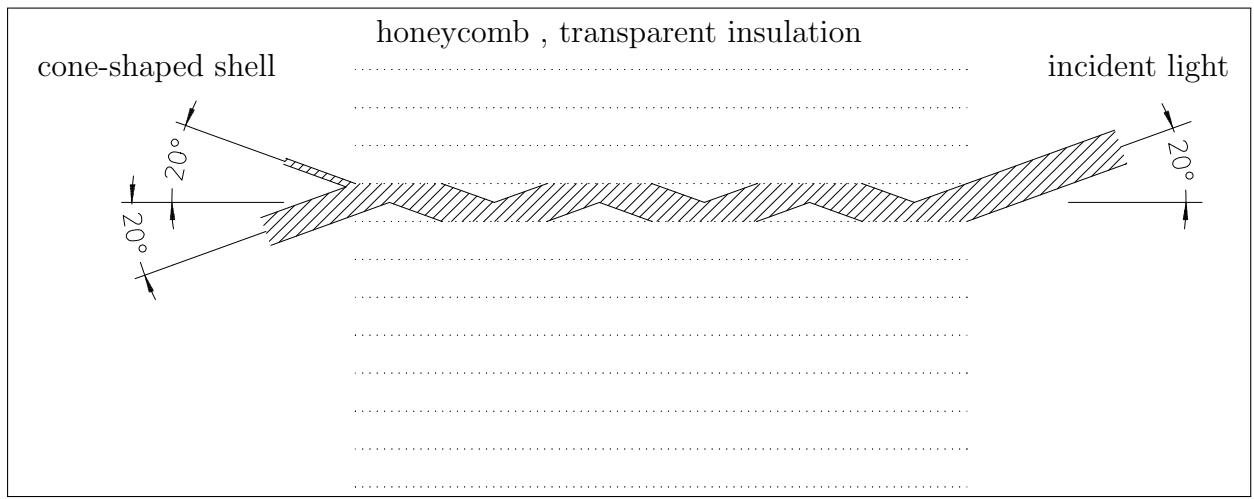}
}
\caption{motivation of the model for scattering at honeycomb TI\label{waben-mot}}
\end{center}
\end{figure}

$\theta_{max}$ depends on the incident direction that illuminates the TI material. As motivation, Fig.~\ref{waben-mot} shows a cross-section 
through the honeycomb material and a simplified beam path, which treats the honeycomb walls as ideal mirrors. Even such an ideal simple honeycomb
transmits light into two directions, depending on the incident direction. By random fluctuations of the walls and an incident direction that
is not parallel to one of the walls, the result is a cone-shaped transmittance of each honeycomb cell. The random nature of wall orientation suggests
a Gaussian distribution of the transmission cone, which is initially set as $\theta_{max} = \theta_{in}$ (see also Fig.~\ref{honeycomb-ray}).

A second fit, not described in detail here, can describe the functional dependency (e.g. polynomial) of parameters $\theta_{max}$,
$\theta_{width}$ and $o$ on $\theta_{in}$. This allows a description of the scattering for incident angles that have not been measured.

Future work is scheduled to improve the model with a dependency on $\phi_{in}$ and $\phi_{out}$, thereby allow asymmetric samples.
Computing time for one fit is currently 6 min on a SUN SPARC-STATION-1, and more complex models will take longer.

Preliminary tests using an IR thermal camera at $\lambda = 8\,{\mu}m$ show a ring with a less homogeneous distribution and two pronounced peaks. This
indicates that the ring is caused by random structures in the order of several micrometers.

\begin{figure}[hbtp]
\begin{center}
\input{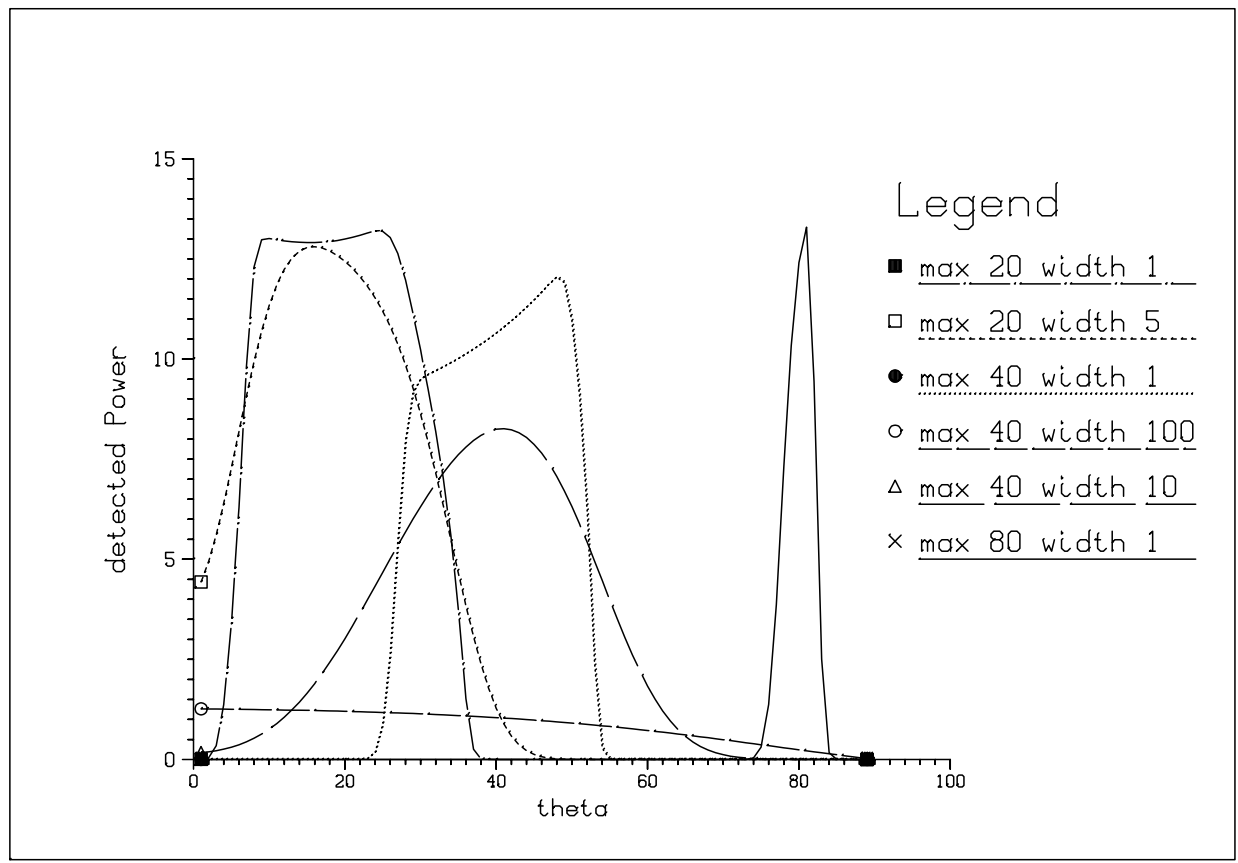}
\caption{Theoretical received power at the proposed detector \new{(near to the sample}), for different set of model parameters
	$\theta_{max}=20,40,80^o$ and $\theta_{width}=1,5,10,100^o$,
	With Lambert-Offset ($o=0$) and Scale ($s=1000$) as fixed parameters.\label{wsmodel}}
\end{center}
\end{figure}

Fig.\ref{wsmodel} shows the power received by the detector, multiplied with the square of the distance between detector and sample centre.
If the parameter $\theta_{width}$ is small (below $1^o$), the sample emits light in the shape of a cone-shaped shell, whose width depends on
the size of the sample and the emitting angle. The ''peak'' gets wider with increasing sample size and narrower with increasing detector
distance or greater emitting angle. If the parameter $\theta_{width}$ is large, the sample acts like a Lambertian emitter, and power
falls off with $\cos(\theta)$.

\begin{figure}[hbtp]
\begin{center}
\input{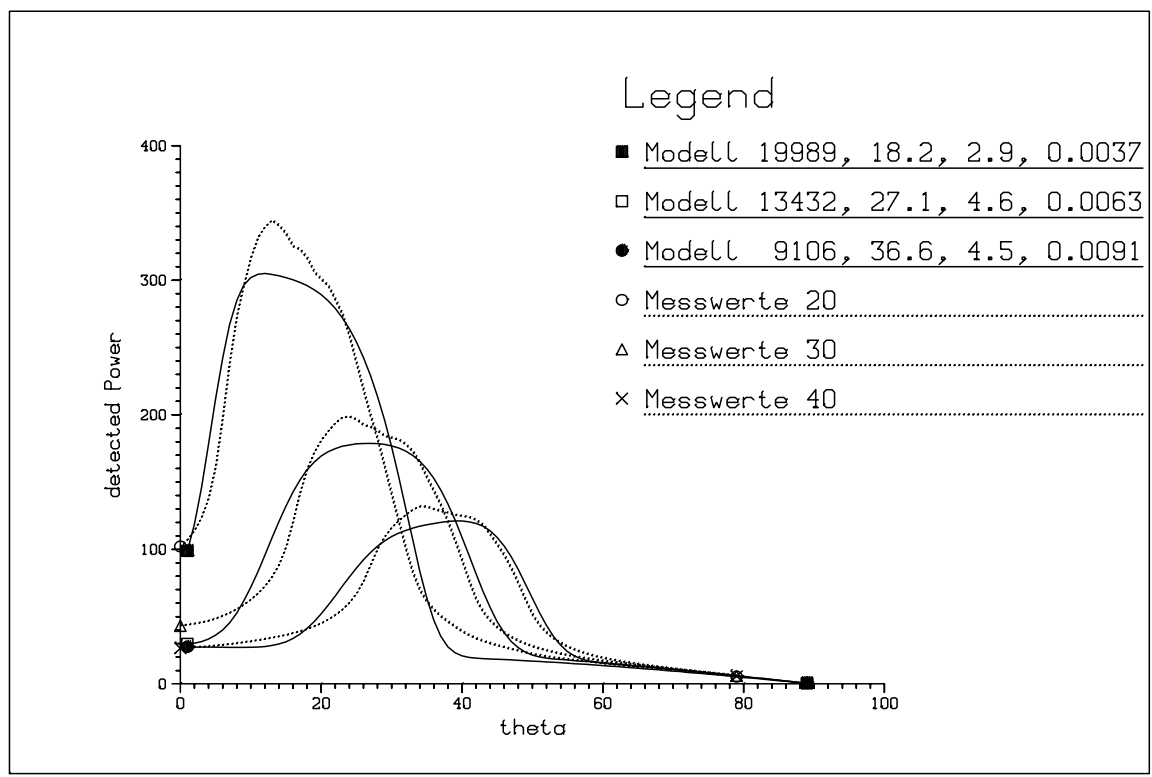}
\caption{Fit of the model to measured data of {\bf wa3101} and $\theta_{in}=20,30,40^o$,
	the legend of the model curves lists parameters $s$, $\theta_{max}$, $\theta_{width}$ and $o$\label{wsfit1}}
\end{center}
\end{figure}

\begin{figure}[h!]
\begin{center}
\input{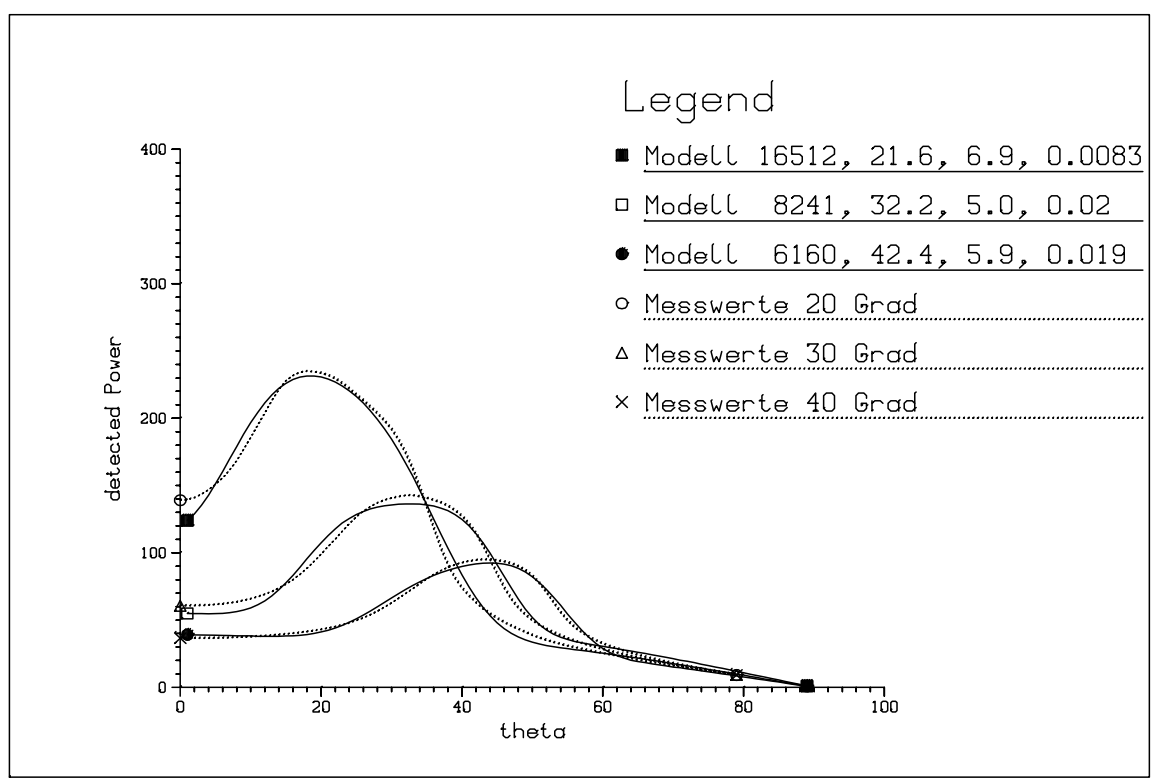}
\caption{Fit of the model to measured data of {\bf wa0301} and $\theta_{in}=20,30,40^o$\label{wsfit2}}
\end{center}
\end{figure}

For two types of Transparent Insulation (TI) materials, FhG-ISE sample numbers {\bf wa3101} and {\bf wa0301}, the parameters $\theta_{max}$
and $\theta_{width}$, $o$ and $s$ were fitted to the measured data using the Levenberg-Marquardt method \cite{num-recipes}[pp542].
A comparison between model and measurement shows the following (Fig.\ref{wsfit1} and Fig.\ref{wsfit2}):

\begin{enumerate}
\itemsep0em
\item	The maximum of {\bf wa3101} is shifted systematically towards smaller $\theta_{max}$ values.
	At incident angles of $\theta_{in} = 20^o, 30^o, 40^o$, $\theta_{max}= 18^o, 27^o, 36^o$.\\
	This contrasts with {\bf wa0301}, with shows a shift towards larger $\theta_{max}$. $\theta_{max}= 22^o ,32^o ,42^o$.\\
	These asymmetries are not yet understood.
\item	The diffuse offset increases with $\theta_{in}$, consistent with the explanation that this is caused by an increase in
	scattering {\em within} the material.
\item	The maxima get lower and wider with increasing $\theta_{in}$.
\item	Fits show a systematic misfit (lower and wider), indicating that the scattering mechanisms are not yet fully described by the model.
\item	The sample {\bf wa0301} shows a higher diffuse component and lower transparency.
\item	Overall, the model more accurately describes diffuse scattering. 
\end{enumerate}
\new{More generally, this works out the far-field pattern from photometric near field data using a model for the BSDF. Follow-on work at
FhG-ISE dealt with this problem more generally, see \ref{follow-on} for the diploma thesis by Matthias Braun in 1998.}

\vfill

\vfill
\pagebreak[10]

\chapter{General thoughts on device configuration}
\label{general-layout}

\section{Estimation of measurement time}
\label{mtime-estimate}

The simplest assumption is a constant angular resolution for all four degrees of freedom: $\theta_{in}, \phi_{in}, \theta_{out}, \phi_{out}$.
If each angle is subdivided into $d$ steps and each measurement takes $m$ microseconds, the total measurement $T$ follows with:
\begin{eqnarray}
	T &=& 360/d \; 360/d \; 180/d \; 180/d \; m   \;\;[{\mu}s]\\
	  &=& 1.16 \; m \; d^{-4} \;\;[h]
\end{eqnarray}
With the same assumptions, the required disk space in [MByte] is:
\begin{eqnarray}
	M &=& 360/d \; 360/d \; 180/d \; 180/d   \;\;\textrm{[Bytes per measured point]}\\
	  &=& 4190 \; d^{-4} \;\;[MByte]
\end{eqnarray}
This leads to practical limits for the angular resolution: As an extreme example, we assume a fast data acquisition device, for example, an
analog-digital converter with direct write access to a fast hard disk for storage. With a continuous acquisition rate of 0.5 MByte/s and 2 bytes
per data point (meaning a single value per position, not a CCD camera), $m$ is roughly $4 {\mu}s$ and $d$ follows as:

\begin{tabular}{lll}
$d=1^o$		&$T=4.6h$	&$M=8$ GByte\\
$d=2^o$		&$T=16m$	&$M=524$ MByte\\
$d=5^o$		&$T=1m$		&$M=32$ MByte\\
\end{tabular}

This combination requires a special, fast computer. If a commercial digital multimeter is used, $m= 100 ms$ (10 data points per second),
and the measurement time follows with:

\begin{tabular}{ll}
$d=1^o$		&$T=13a$\\
$d=2^o$		&$T=30d$\\
$d=5^o$		&$T=45h$\\
\end{tabular}

The long measurement times could be kept within reasonable limits by using special hardware and matching powerful mechanics, however, for a
fixed resolution of 1 or 2 degrees, the required storage space becomes too large. Using data compression during a measurement
has little advantage, since it increases the required computing power.

Hence, the assumption of sampling all four degrees of freedom with an equally spaced, fixed angular resolution makes measurements with a
resolution below 5 degrees impractical. This estimate is of interest to programs that calculate multi-layered TI materials, which require
data for {\em all} incident and outgoing angles. A practical choice of angular resolution takes into account the symmetries in the material and
interest in particular angles (e.g. positions along the path of the sun used as incident angles).

\vfill
\pagebreak[10]

\section{Light sources}

Incident illumination should be parallel and homogeneous over a surface area of 40 x 40 cm. Parabolic mirrors are the obvious compact choice to
achieve this.

Ideal for the task are off-axis mirrors, since they allow a homogeneous surface illumination without self-shadowing by the light source
and its mounts.
A drawback is the high price of a 40x40 cm off-axis, custom-made, parabolic mirror, especially since it is unclear whether the quality of
other components, like the small light-source or its isotropic radiation pattern would match the precision of the custom made mirror.

Suitable for our task proved an off-the-shelf concentric parabolic mirror with a diameter of 61cm, a focal length of 20cm, Al/SiO2 surface
coating and a surface precision suitable for illumination tasks. This model is used in marine lighthouses.

A 5 cm hole at the mirror centre allows two light source configurations: One directly at the position of the primary focus, two,
with a secondary mirror, in an geometric arrangement similar to astronomical telescopes with a Cassegrain or Gregory configuration.

The first configuration is easier to adjust, the second configuration allows for arbitrarily large lamp housings. The following primary
light sources have been considered: Xe-flash, Xe short-arc or halogen-lamp. A Xe-flash lamp wasn't considered further due to its problems with
adjustment, but it would offer possibilities that are mentioned in section~\ref{quarter-sphere}.

Criteria for the choice between Xe short-arc and halogen lamps are:

\begin{tabular}{lll}
			&Xe short-arc			&Halogen low voltage\\
size of emitter		&1.1 x 2.8mm (XBO1000W/HS)	&6mm\\
angular emittance	&$90^o$ (XBO1000W/HS)		&$120^o$\\
power			&150 W - 1 kW			&60 W\\
life time		&1000 h				&3000 h\\
interval operation	&no				&yes\\
mounting at primary focus	&no			&yes\\
\end{tabular}

Mounting the Xenon short-arc at the primary focus must be ruled out, since the lamp doesn't illuminate the full mirror. Additionally, the risk of
explosion of the Xenon bulb requires an enclosure which would shadow the beam.
\new{Actually, discussing the difference to usage of Xenon lamps in marine lighthouses is missing here.}

After estimating the detector's sensitivity and considering actual illumination levels, it was determined that a 60 W halogen lamp is sufficient.
Further advantages of a Halogen source include the possibility of switching a Halogen lamp on and off without long waiting times, and the
avoidance of the intense line spectra of a Xenon arc lamp.

Due its size and weight of the parabolic mirror, the light source remains at a fixed position in the measurement setup.

\vfill
\pagebreak[20]

\section{Variants of detector configuration}
\label{det-variants}

The choice of the detector configuration is based on pre-tests with typical TI materials: $\theta$ and $\phi$ should not be chosen in steps
of less than 5 degrees, in order not to overlook details in transmission. Two detector configurations have been considered:

Either a configuration consisting of many small detectors at fixed spatial positions that would offer an extremely fast measurement, or
a ''legacy'' configuration with a single moving detector. 

A configuration with fixed detectors could be realised, for example, by glueing small solar cells to the inside of a Plexiglas hemisphere,
with their active side facing inwards.  All outgoing angles could be measured without mechanical movement, so that the lower boundary of
the estimated measurement time (See \ref{mtime-estimate}) comes within reach.
Mechanical movement would only be required to adjust the incident angle. However, some of the incident light would be shadowed by the cells.

As an example, an arrangement of 5x5 mm solar cells in a 5x5 degree grid had been considered, using a sphere of 0.5 m radius.  In practice,
the wiring of 2592 solar cells and multiplexing analog signals pose problems. It would be possible to add an analog/digital converter to
each solar cell and use bus logic to keep wiring to minimum and connect the mini-detectors only for power and a serial, loop-through bus.
Similar set-ups are already used in detectors in high-energy physics \cite{tritsch:90}.

Another solution consists of using optical fibre instead of individual cells, with all fibres ending on a single CCD chip.  ''Pre-sorted''
fibre bundles, that are already optically connected to the CCD chip is possible, however this suffers from similar cable problems as the
electrical wiring concept. Furthermore, the fibres have to be enclosed in opaque material to avoid coupling, which conflicts with the
requirement to minimise self-shadowing of the incident illumination.

Finally an arrangement of multiple detectors was ruled out due the optional use of a CCD camera as detector: A grid arrangement of cameras
would, even for compact models, cast too much shadow.

\vfill

\section{Positioning of axis for scanning photo-goniometers}

Both the sample and the detector can rotate around two axis (four degrees of freedom in all). The angular ranges of each axis must be
arranged so that all incident and outgoing angles can be adjusted for. Closely coupled to this issue is the number of required lamps.
In the following, it is assumed that the detector moves around the sample with constant distance to the sample centre (path on the surface of
a sphere).

\begin{figure}[ht]
\begin{center}
\includegraphics[height=10cm]{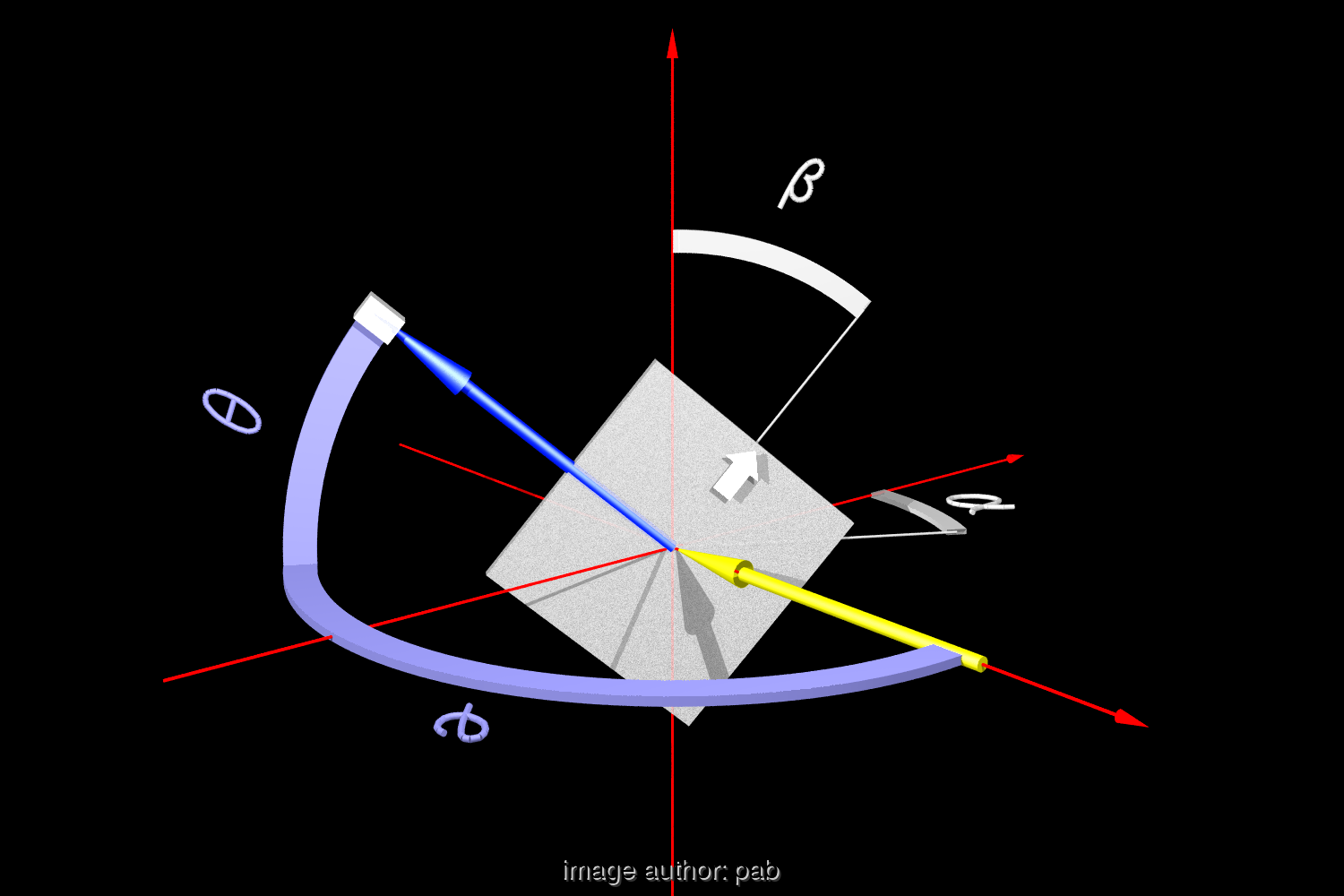}
\caption{Arrangement of sample (grey surface), detector (grey box) and light-source (yellow arrow) in lab coordinates.
	The x-axis runs left-to-right, the y-axis is front-to-back and the z-axis bottom-to-top. Included are the definitions of $\alpha$, $\beta$
	and $\theta_{det}$, $\phi_{det}$. \label{lab-coor}}
\end{center}
\end{figure}

Fig.\ref{lab-coor} shows the ''laboratory coordinate system'', used to specify the angles of the drive system.

\pagebreak[10]

There are multiple options for the geometrical configuration of the axes for sample and detector. The following text explains 
why this particular one is considered beneficial:

Using a right-handed coordinate system, in which the XY-plane is horizontal, the light source is assumed to be mounted on the positive
X-axis, shining onto the coordinate origin.

Due to the rotational symmetry of the incident light around the X-axis, the outer rotational axis of the sample mount can not point along the
same X-axis, which leaves this rotational axis to point either vertically along the Z-axis or horizontal along the Y-axis. With either arrangement,
any incident angle $\theta$ can already be catered for. The allowable angle $\alpha$ (Fig.\ref{lab-coor}) of this axis has to be at least
$180^o$ (illumination on front and rear side). The axis will be assumed vertical, parallel to the Z-axis, since it is technically simpler,
as will become clear later.

On this axis, a second, inner axis of rotation is orthogonal to the outer axis. A classical arrangement is a gimbal mount, as used in a
gyro compass, except that in this case, the mounting brackets would shadow the sample. An arrangement with 3 degrees of freedom for the
sample mount is given in \cite{davis:87}.

In order to measure samples that are not rotationally symmetric around their surface normal, the sample is therefore rotated around its
surface normal in the sample mount. The angle between the ''up''-direction of the sample and a vertical plummet is labelled $\beta$.

All incident directions are reachable with either $\alpha=[0..180]$ and $\beta=[0..360]$ or with $\alpha=[0..360]$ and $\beta=[0..180]$.  In the
following, a configuration of full scan angles, $\alpha=[0..360]$ {\em and} $\beta=[0..360]$ will be used. This already allows to set the same
incident direction twice, while using a single light source with two different combinations of $(\alpha,\beta)$. This fact offers choices
for the detector geometry:

\vfill
\pagebreak[10]

The detector should detect light in any outgoing direction, and therefore it must be possible to position it at any point on a sphere
centred on the sample, the so called ''scan area''.  Since each incident direction can be reached by two different combinations of
$(\alpha,\beta)$, the required scan path of the detector is reduced to one hemisphere. There are three possible orientations of this
hemisphere as follows:

The case ''symmetry axis of hemisphere equals the X-axis'' is not possible, since there would be no transmission measurements. 
The remaining cases are: The detector moves on the ''upper', ''lower'', ''front'' or ''back'' hemispheres. The detector is mounted on a
circular rail and this rail is free to rotate around its end-points as well. For technical reasons, the axis of the circular rail will be
vertical, and the detector scans the ''front'' hemisphere (negative Y-axis). The angle between X-axis and rail is labelled $\phi_{det}$ and
the angle between the horizontal plane and the detector is labelled $\theta_{det}$ (Fig.~\ref{lab-coor}).

This defines the standard arrangement with a single light source: Sample mount and detector rail swivel around the vertical axis, the sample 
can be rotated around its surface normal as well, and the detector moves along a rail. The range of the sample angles is 360 degrees each
and 180 degrees for $\theta_{det}$ and $\phi_{det}$. This configuration allows each incident/outgoing direction to be set with exactly one
unique combination of axis angles. This was the design concept of LBL's ''large scanning radiometer''.

Using more than one light source reduces the required range of the angular scan of one or more of the axis. Either $\alpha, \beta$ or
$\theta_{det}, \phi_{det}$ can be reduced. It is beneficial to keep the required angular scan range as small as possible for technical
reasons.

For a configuration with two lamps, the second lamp would be placed on the negative X-axis in Fig.\ref{lab-coor}. The sample mount angles
range from $\alpha=[0..360]$ and $\beta=[0..360]$ which reduces the scan-surface covered by the detector to a quarter sphere. Using the
angles in Fig.\ref{lab-coor}, with a detector rail swivelling around a vertical axis, this leads to two variants:

Either $\theta_{det}=[0..180]$ and $\phi_{det}=[0..90]$ or $\theta_{det}=[0..90]$ and $\phi_{det}=[0..180]$

\label{quarter-sphere}
The latter configuration scans the upper quarter sphere. This geometric configuration is used in this work. The scan area of the detector
(a solid angle of $\pi$) will be called ''quarter sphere'', even if the detector actually moves in the surface of a cone (see
\ref{cones}). Each incident and outgoing direction can either be specified in sample coordinates ($\theta_{in}, \phi_{in}, \theta_{out},
\phi_{out}$) or in ''lab coordinates'' ($\alpha, \beta, \theta_{det}, \phi_{det}, n$) , with $n$ specifying which lamp is switched on.

Finally, lets look at an extension with four or more lamps: Two more lamps can be positioned at the positive and negative Y-axis
(Fig.~\ref{lab-coor}), which reduces the required scan area to $1/8$ of a sphere. When using 72 lamps, placed on a circle in the XY plane around
the sample centre, the detector rail needs no swivelling movement at all, if outgoing directions are stepped in 5 degrees. And if this is
combined with pulsed flashers, the speed of measurement can be enhanced considerably. In that case, the footprint of the device would be
larger, due to the size of the parabolic mirrors used in the illumination system.

\chapter{Device with one detector and two lamps}

\section{Design overview}

The device consists of 3 subsystems, that have been designed and built as singular units with different manufacturing strategies:
\new{see also annex \ref{homework}}

\begin{figure}[h!]
\begin{center}
\input{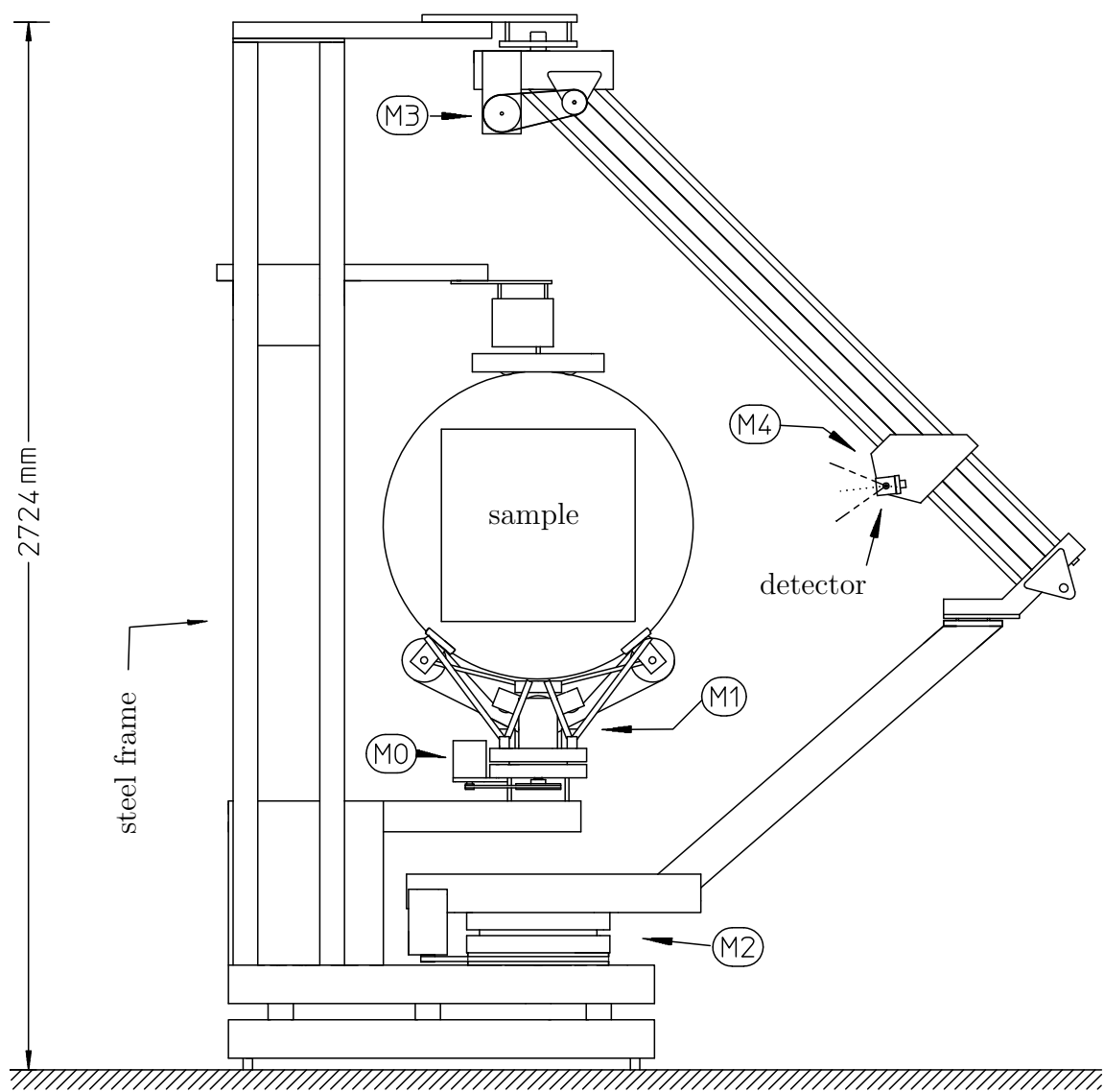}
\caption{\label{me10-design}Layout of my first FhG-ISE device in 1989, motor numbers explained in text}
\end{center}
\end{figure}

Fig.~\ref{me10-design} shows the overall plan of the device. The drives are labelled M0 to M4:

\begin{tabular}{ll}
M0	&vertical axis of sample mount\\
M1	&horizontal axis of sample mount, rotates sample around surface normal\\
M2	&vertical axis detector swivelling arm\\
M3	&drive motor for detector along rail\\
M4	&tilt motor of detector\\
\end{tabular}

\vspace{1cm}

\clearpage

\begin{figure}[hbtp]
\begin{center}
\includegraphics[width=16cm]{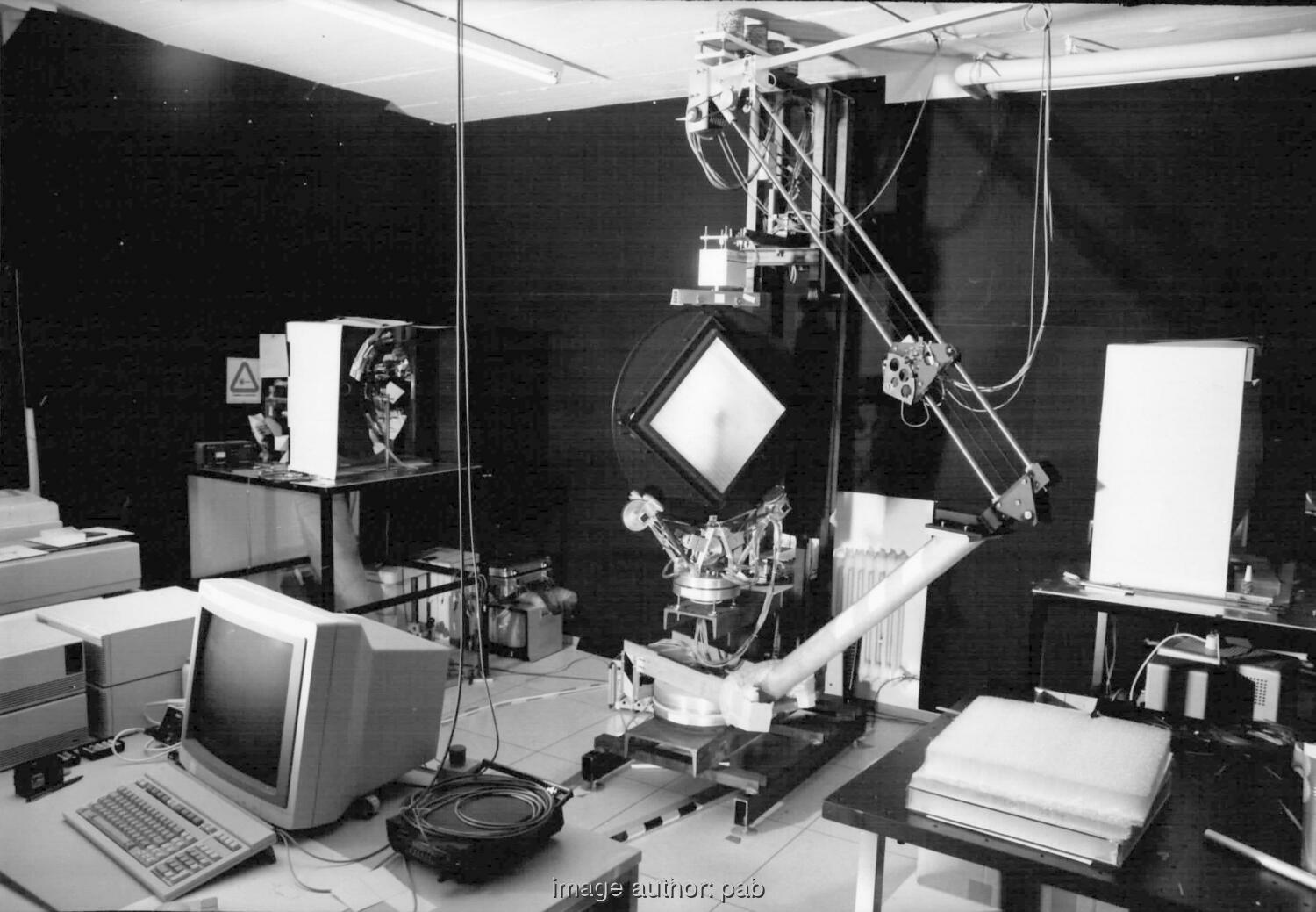}
\caption{\label{overall-photo}Photo of my first device in 1989: The detector mounted on a linear rail, first lamp to the left,
second lamp on right (with baffles) and the sample mount at centre}
\end{center}
\end{figure}

The detailed planning of the mechanics took up most of the time of this diploma work: Some aspects have been considered in detail. Solutions
were compared and the best ones were selected. Components had been stripped down to simple parts early on in the design process, in close
collaboration with the mechanical workshop at the institute. Modular manufacturing and an easy replacement of modules were thus ensured, in
case that changes are required later.  Special manufacturing procedures (large scale lathes, CNC grinders, etc.) and special materials
(titanium, carbon fibre composites) could be avoided.

There is an ink drawing for each of the 150 parts. When designing the second part, the detector rail systems, a 2D CAD program, ME10 by
Hewlett-Packard, sped up the construction process considerably. In can be said, with all care, that two months of flawless operation, without
requiring further re-work, proved that time spent on the design phase had been time well spent.

The three main parts of the device had to be adjusted themselves as well as relative to each other. The sample-mount and detector are
mounted in a steel frame, the ''backbone'' of the device. All flanges between the components and the steel frame are adjustable in six
directions (translation along 3 axis plus tilt around 3 axis). This compensates irregularities, e.g. caused by welding.

Based on the estimated measurement time, the device operates for hours autonomously and without the attention of an operator. Therefore, an
automatic detection of faults has been built into the mechanics and electronics.

\pagebreak[10]

A standard measurement procedure consists of scanning along one axis of the detector and then adjusting one or more of the other axes. This
is repeated until all scheduled combinations of incident and outgoing directions are completed. A future enhancement should include a
second procedure in which two or more axis move simultaneously during a measurement, to scan areas with a higher angular resolution. 
A very simple drive, which only drives one axis slowly, would not allow this second way of scanning. Hence, all axes are
laid out for fast movement, fulfilling the aim to reach any position of incident- and outgoing angles within 8 seconds from any other
position. This rule is matched for the sample mount. Only the main arm of the swivelling drive takes a little longer due to its inertia (15
seconds for a $180^o$ movement).

\section{Dimensioning stepper-drives and gear ratios}

The torque of a stepper motor is nearly constant up to a critical speed (RPM, revolutions-per-minute of the motor axis) and drops off
steeply beyond that (Fig.\ref{stepper-torque}):

\begin{figure}[hbtp]
\begin{center}
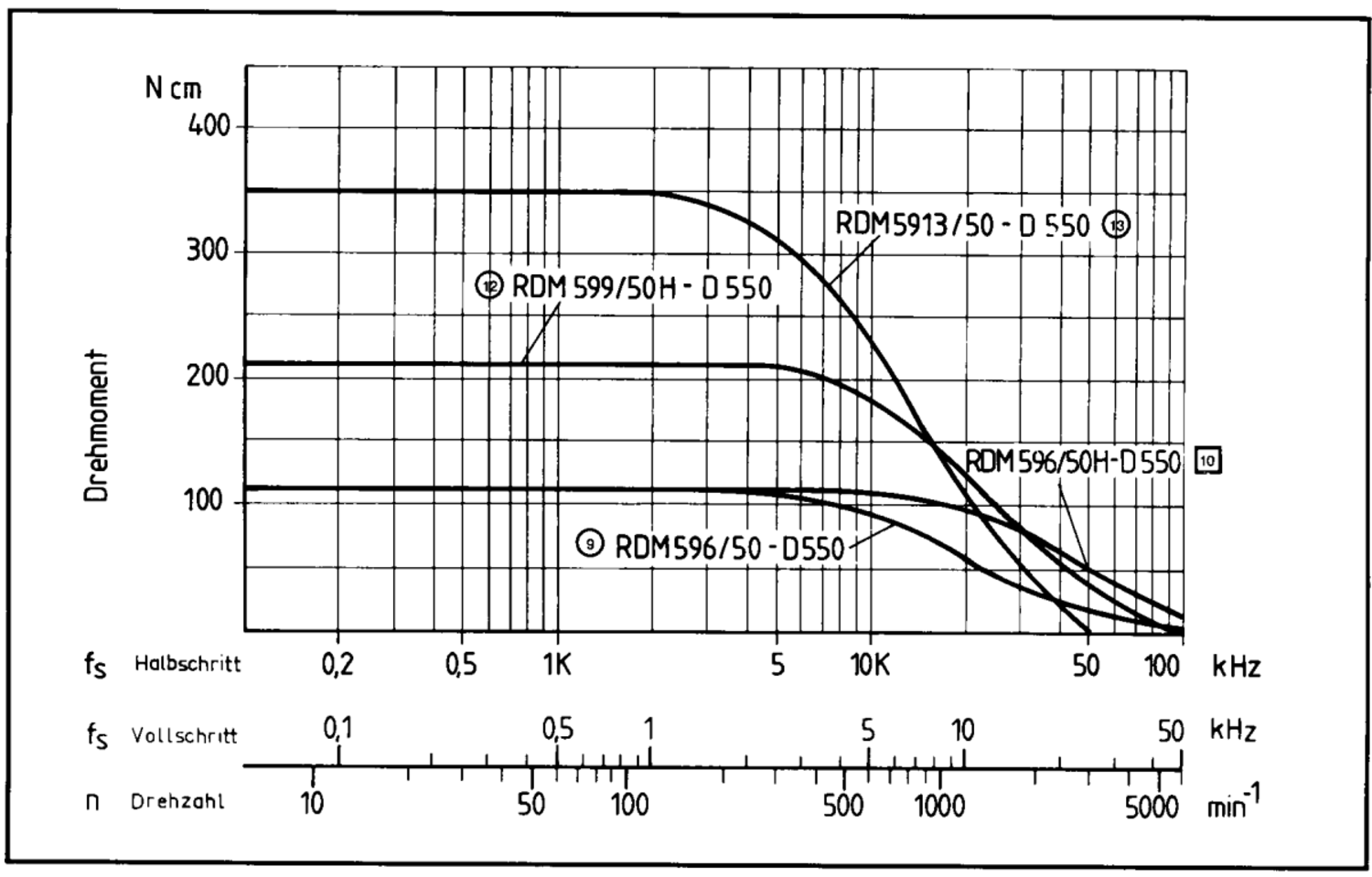
\caption{\label{stepper-torque}Torque plotted versus rotational speed of the motor axis and frequency of stepper pulses for a stepper-motor
		type RDM599 made by BergerLahr \cite{berger-lahr-man}{plot 12}}
\end{center}
\end{figure}

The critical speed increases with the supply voltage. For a RDM599 5-phase stepper-motor this is approximately 5 revs/second for a supply voltage
of 40 VDC. \footnote{Note of 2021: The 40V DC supply was caused de-facto by $V_{cc}$ of the electronic drive chips.}

For a precise movement, the reduction gear ratio between motor axis and detector axis should be large. However, since the
maximum RPM of the motor axis is limited, this also limits the speed of the driven axis, and this limits the maximum applicable gear ratio.
The trick is to chose the optimal ratio:
\vfill

\pagebreak[10]

The motor drives accelerate to the maximum angular speed for the axis at constant acceleration, and de-accelerate in time to reach the
desired position without exceeding maximum torque. Hence different speed ''profiles'' are possible, which all drive the angular position from
$\alpha_1$ to $\alpha_2$ within a time interval $T$: Either a high acceleration rate for a short time and low maximum speed, or a low 
acceleration rate for a longer time and higher maximum speed. The parameter for these profiles is the acceleration time $t_b$.
Our diagram with total time $T$ and path length $s$ shows if there exists a combination of gear-ratio and $t_b$.

From the moment of inertia $\Theta$ around an axis, the path length $s$ and time interval $T$ follows, for each $t_b$, an acceleration
$\tau_{max}$ and a maximum velocity $\omega_{max}$:
\begin{eqnarray}
	\tau_{max}	&=&	\Theta ( \alpha_1 - \alpha_2 ) \; / \; ( t_b ( T - t_b ))\\
	\omega_{max}	&=&	(\alpha_1 - \alpha_2 ) \; / \; ( T - t_b )\\
\end{eqnarray}

\begin{figure}[h!]
\begin{center}
\input{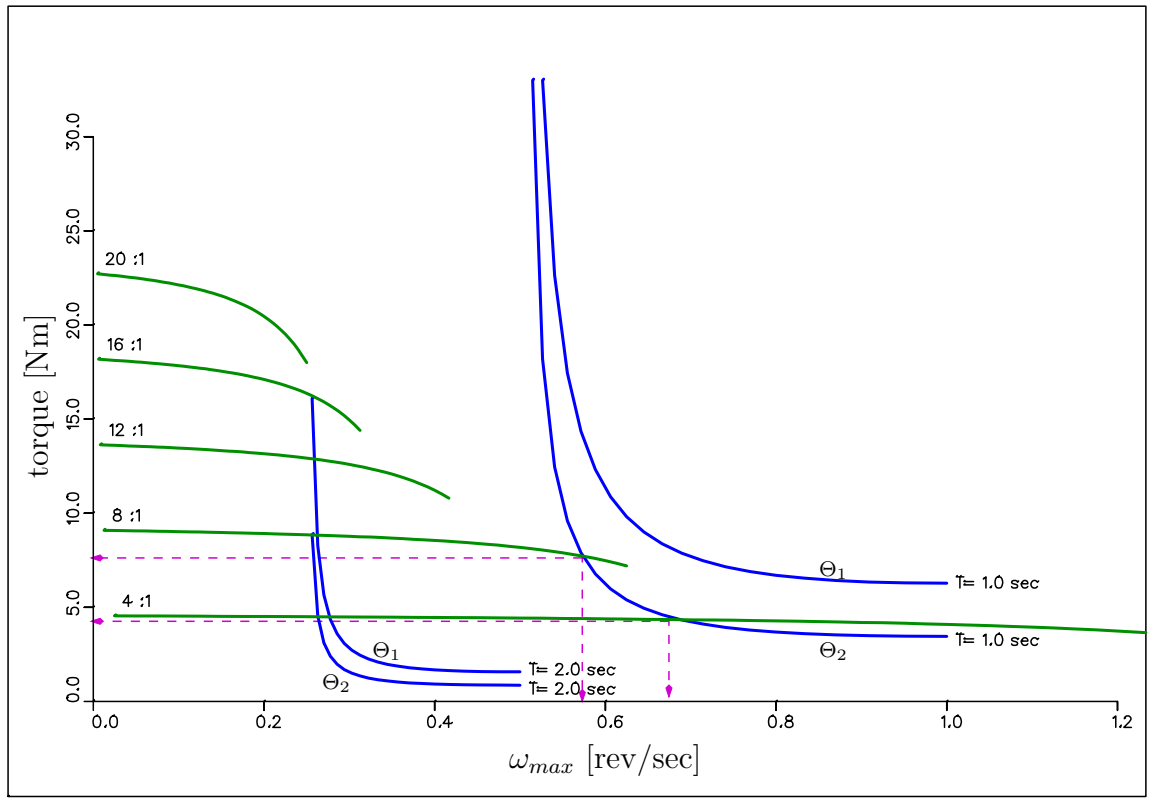}
\caption{\label{mot-hyp}Diagram to select a suitable gear ratio: Plots required torque $t_{max}$ on Y-axis and angular speed $\omega_{max}$
	on X-axis and the maximum motor toque $\tau_{mot}$ for different gear ratios.}
\end{center}
\end{figure}

The curve of different $t_b$ resembles a hyperbola. Fig.~\ref{mot-hyp} shows curves for two moments of inertia and two time intervals $T$ with
($ \alpha_1 - \alpha_2 = 2 \pi$). A second set of curves plots the torque curves for the stepper motor with different gears attached.

If a point on the hyperbola lies below a torque curve, this combination of torque and speed is plausible, and there exists a $t_b$ with which
the path $s$ can be accomplished in the given time interval $T$. Furthermore, it indicates which gear ratio is required to achieve this.

Examples of reading Fig.~\ref{mot-hyp}: For the longer time interval $T$ = 2 sec , any gear ratio between 16:1 and 4:1 works. The two curves of
$T=2$ differ in their $\Theta$ (assumed inertia around axis). The shorter time interval $T$ = 1 sec allows no gear ratio for $\Theta_2$,
while accomplishment of the movement in given time $T$ is feasible for $\Theta_1$ and a gear ratio of 4:1 and 8:1 .
The graphic proved valuable as a planning tool when laying out the mechanical drive system.

Practical aspects of the results: Using timing belts and gears the maximum practical gear ratio is roughly 10:1, higher ratios can be realised with
worm gears and custom-made gears. Worm gears require constant lubrication to work efficiently and require a precise alignment, and were not
considered. All gear ratios describe here have been made with timing belts.
\new{The PG2 Gonio-Photometer of the same author exceeded these limits with other drive types in 2005}.

\section{Details of parabolic mirror}

Building the mounts of the parabolic mirrors poses little problems: They remain static, no weight limit, all tolerances are within 1/10 mm and only
the lamp mount had be optimised to cast a small shadow. Currently it consists of a massive 18mm brass pole, which does cause a noticeable
shadowing in measurement results. An alternative has been considered already: A lamp mount suspended on steel wires will likely solve this problem.
The bottom part of the mirror rests in a CNC machined notch, adapted to the edge-profile of the mirror and cushioned with 5 mm cork.
In order to compensate temperature differences and to ease mounting of the mirror, cork is also used in the mounts at the side.
Mirror and lamp mount are placed on a 1x1m table with an aluminium top and a frame of welded steel pipes underneath, to achieve the required
stability.

The light source consists of a 12 VDC 60W halogen lamp, actually it is the full-beam filament of an H4 automotive head light.
Power supply and electrical control are described in Chapter \ref{electronics}.

To limit unwanted reflections off the unused mirror, an automated shading system, consisting of black lamellas, is considered to be mounted
in front of each mirror at a later stage.

\section{Details of sample mount}

\begin{figure}[h!]
\begin{center}
\input{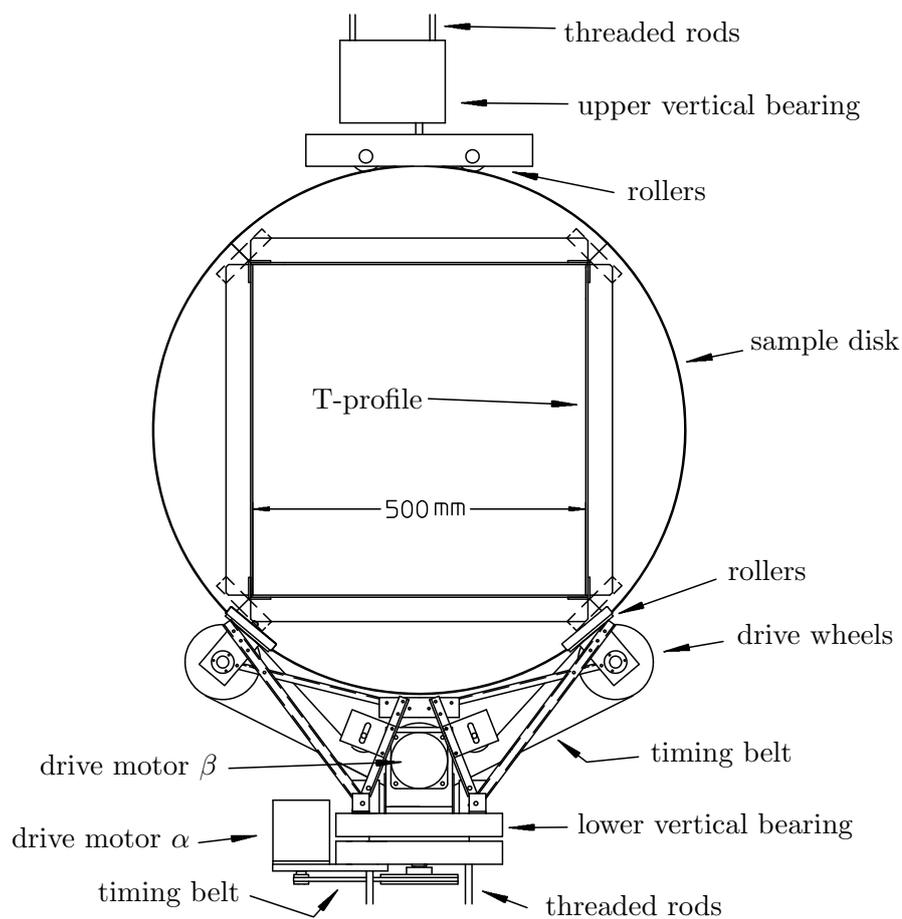}
\caption{\label{cad-sample}The sample mount}
\end{center}
\end{figure}

The sample (standard size 40 x 40 cm) is mounted in a quadratic opening of size 50 x 50 cm in the middle of an aluminium disc with 80 cm
outer diameter (Fig.~\ref{cad-sample}). Between the square opening of the disk and the sample, a frame is mounted that adapts to the sample
thickness, protects the sample from light shining onto the edges. the inner surfaces are selectable matte black or coated with a
mirroring adhesive tape. Technically the disk is assembled from four CNC machined segments, connected by plates. An additional T-profile at
the inside of the segments adds stability and mounting points. The width of the T-profile (40 mm) allows measurement of thin samples and
avoids shadowing by the sample mount. All aluminium parts are black anodised and painted with {\em 3M Ultra black} matte paint.

The aluminium disk is support at three points: It rests on two drive wheels, which are driven by the motor using timing-belts and rotate the
disk around its surface normal. At the top point, guiding rollers unsure the vertical orientation. At the lower wheels, secondary rollers
hold the disk in place.

\vfill
\pagebreak[10]

The load bearing support structure underneath the sample connects the drive wheels, supports sample weight and the disc, and supports the
forces in the timing belts. It is made of strutted, hollow aluminium extruded profiles ( [] and L profiles), giving it an extremely stable,
light weight, buckling resistant characteristic.
\new{For more details of its history, see the appendix.}

The strutted construction rests on a bearing with vertical axis that swivels the sample around the vertical axis. The bearing consists of two
aluminium ''cakes'' with two tapered roller bearings.

The electrical wiring (that feeds the motor $\beta$, which drives the sample around its surface-normal) limits rotation around the vertical axis.
Hence the motor that drives the vertical axis $\alpha$ has to have a safety loop that cuts power if the controlling computer drives the axis out of
range. Otherwise the cables may be subject to excessive forces. Therefore, a camshaft cuts power if the axis is turned out of range. It is
still possible to turn it full $360^o$, but not more than $5^o$ on either side. All incident angles are still fully adjustable.\\
\new{This led to the PG2 gonio-photometer having precision slip-rings and no swinging wires in 2005.}

\vfill
\pagebreak[10]

\section{Details detector mount}
\label{det-detmount}

\begin{figure}[h!]
\begin{center}
\input{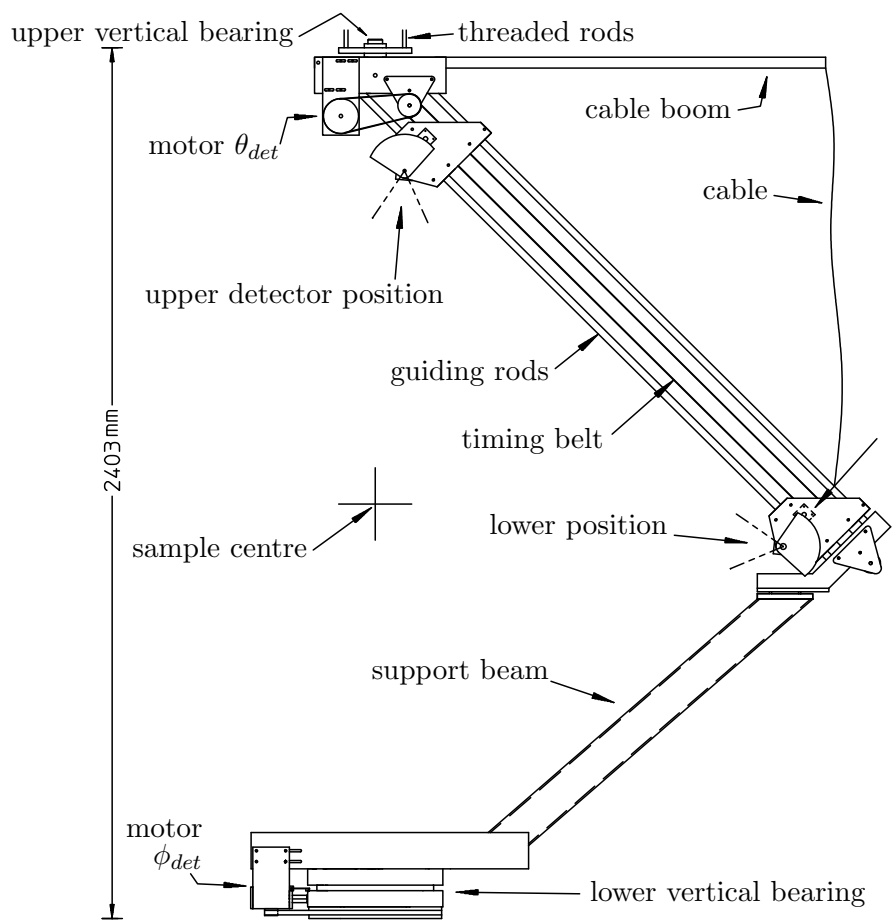}
\caption{\label{cad-det}The detector drive system}
\end{center}
\end{figure}

The detector swivels $180^o$ around the vertical axis and $90^o$ around the horizontal axis (Fig.~\ref{cad-det}). Two solutions for the first
rotation were considered: Firstly a half-circle rail on which a $90^o$ arc is mounted, and secondly a swivelling arm. A rail has some
advantages with the drive system, but the difficulty lies in manufacturing a horizontal flat half circle from segments, since milling this in
one piece is impossible in the institute's workshop. Hence a support beam was built that extends $45^o$ upwards, and is supported by an oversized
bearing. The stepper motors drives the arm with a 1:10 gear ratio.

The detector should move along a $90^o$ curved rail segment. However, such a segment of 1m radius would have to be custom-made. This would
be expensive since the surface must be hardened and ground to act as a support rail for the sled. Additional support structures would be
necessary to provide stability, e.g. to support the forces of a timing belt. This adds to self-shadowing if the detector is moved in front of
the lamp.
\new{Yet a commercial circular rail, newly available then, was replacing it a few years later, in an optimised version.}

Hence the $90^o$ circle segment was replaced by a linear segment. As a commercial product, this offers excellent stability and reliability 
(20 mm hardened hollow rust-free shafts as rails).

Therefore, the detector moves not on the surface of a sphere, but on the surface of a cone. An additional fifth motor tilts the detector to
keep it pointing at the sample centre.

\section{Control electronics}
\label{electronics}

Between computer and device, all signal lines are DC-isolated, de-coupled by opto-couplers. The following paragraph describes the
electronics between the stage of the opto-couplers and the drive system. Details of the connection between control-computer and
opto-coupler are given in section \ref{unix-workstation}.

The opto-couplers between device and computer fulfil two functions: Firstly they prohibit ground loops and noise spikes induced on long
DC-coupled loops spanning an area (See Maxwell's equations chapter~\ref{dipole}), and secondly they protect the workstation from high
voltages in case of malfunction.

An elegant solution would be a bidirectional fibre-optic connection between workstation and experiment. It would replace 40 opto-couplers
and the rather inelegant 6 m long cable of 45 strands with a sleek 2-fibre serial connection. Pre-tests with serial
multiplexers/demultiplexers have been carried out, outside of this diploma work, but sadly this nice idea wasn't used due to
time-constraints.

A 5-phase stepper motor features 5 separate coils in its stator and a permanent magnet in its rotor. One step consists of a slight angular
rotation of the rotor in one direction, whose angular step is a fixed value for a specific motor (here $0.36^o$). The coils have to
be switched on/off accordingly, to change the magnetic field and get the rotor to move to a new position. Each coil has three states: no
current, or one of two current directions. Power electronics of stepper motors, driving the coils, have at least three input signals:

\begin{tabular}{ll}
1.	&step impulse which rotates the motor by one angular step\\
2.	&direction (e.g. low=CW, high=CCW)\\
3.	&current reduction, which lowers the current during a pause between movements\\
\end{tabular}

Outside this diploma work, I have developed my own power electronics that control 5-phase stepper motors, with 8 units built. Commercially
available units do use up to 120V DC as supply voltage, which does offer higher torque at high speed (due to self-inductance of the coils
following {\em Lenz's law}), but at the speed required (4 rev/sec), the torque is the same as with 40V DC.
\new{A practical advantage is that 40V DC classify as ''Extra Low Voltage'' in electrical safety.}

After power-down of the motors, an initialisation is required on each axis, since the driving computer has no knowledge about the initial
position of the mechanical axis. Therefore' each axis either requires an absolute angular encoder, or each axis has a electronically
readable zero-mark. The latter is simpler and more exact.
\new{Depending on the absolute angular encoder, that is.}

The zero-mark is built with a small forked light barrier and a metal latch at each axis. The light barrier is mounted at the driven axis,
following the gears and not at the motor-axis.

These zero-marks don't have to be at ''physically zero'' for any angle, and they need not be an the end positions of the swivel movement, in
fact they are better placed in the middle of the scan area. A calibration procedure determines where the ''physical zero'' is relative to the
zero-mark.

While searching for the zero-mark during initialisation of an axis, the computer has to know in which direction the zero-mark is to be found. This
is helped by two reflective-light-detectors and a coarse angular encoding on each axis that tells the driving software if a zero-mark is
CW or CCW from the current actual position.

\vfill
\pagebreak[5]

Power supply of the stepper-motors is a lavishly dimensioned, custom, three-phase 480W 40V DC supply (internal resistance 0.4 Ohm) that
ensures that all 4 motors can be ramped up concurrently with maximum acceleration and very little drop in the supply voltage. An output
with TTL levels indicates to the controlling workstation if the 40V DC supply is ''on''.

Intrinsically built into the power supply is a hardware circuit for an emergency-stop loop: A power MOSfet interrupts the 40V supply if a
signal line is opened. This happens if the scan-area of an axis is exceeded or if a stop-button is pressed manually.

The 12V/60W lamps are fed through batteries (car battery 63Ah), which are charged by commercial charging units. Alternatively, the lamps can
be operated on battery alone, which ensures a 100\% ripple-free output. This is to avoid possible beat-frequencies between the 50Hz power
supply and a CCD camera running at 25Hz or 30Hz. Power to the lamps is controlled by power MOSfets, switched with TTL levels.

\new{With the availability of high-frequency (20-90kHz) AC/DC power supplies, there is no detectable ripple in light from a Halogen lamp due
to the thermal mass of the filament.}

\begin{figure}[h!]
\begin{center}
\input{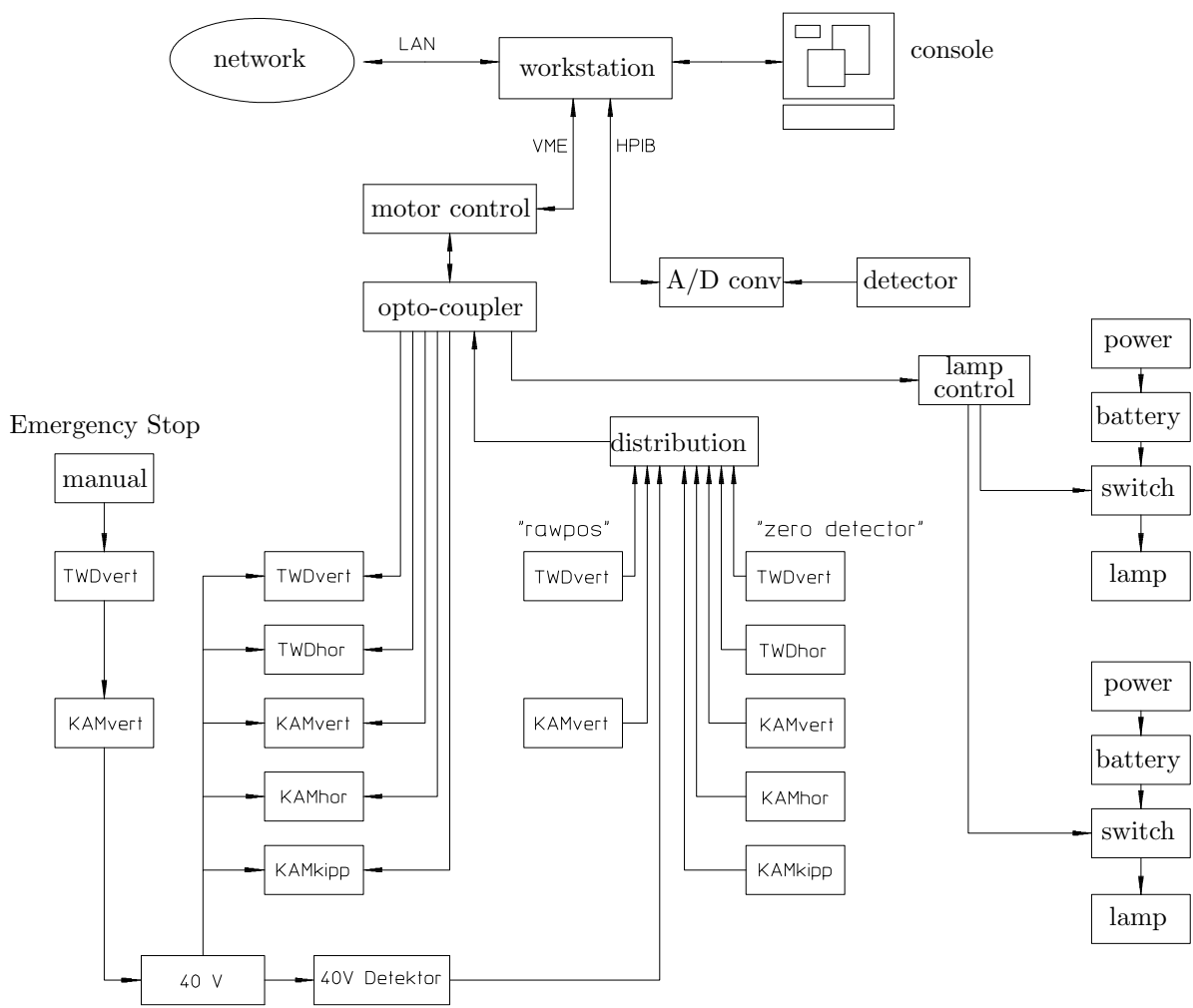}
\caption{\label{scheme-e}Electrical schematic. Not shown is the wiring for 5V and 12V power.}
\end{center}
\end{figure}

Fig.~\ref{scheme-e} shows signals between the workstation and the device. The emergency stop loop is shown on the left, power control of the
stepper motors in a column (''TWDvert'' etc), positional control in the middle (''rawpos'' and ''zero detector'') and lamp control on the right
side.

\section{Assembly and alignment}

Alignment of the components requires very basic tools only: Plummet, hose water level, level, and a laser. The sequence of alignment steps is
shown in Fig.~\ref{just}:

\begin{figure}[h!]
\begin{center}
\input{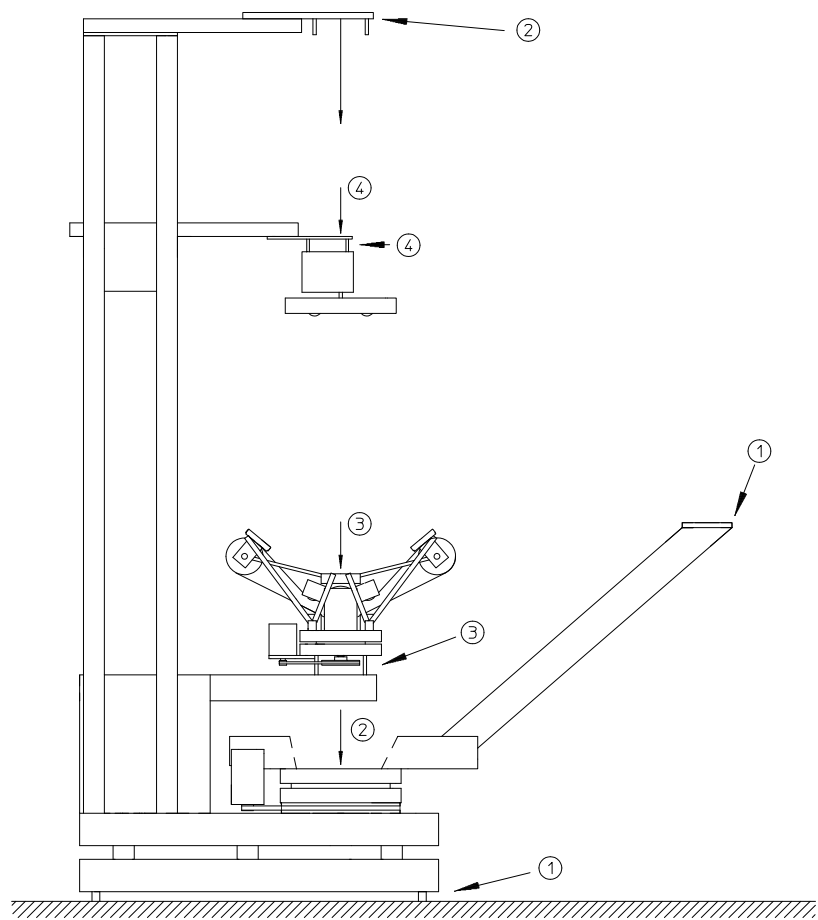}
\caption{\label{just}The sequence of alignment}
\end{center}
\end{figure}

The mechanical connection between the steel ''backbone'', the lower bearing and the swivelling arm is fixed. Conclusively, the end of the
swivelling arm has to put into a horizontal plane first, by adjusting screws mounted at the ''feet'' of the steel ''backbone'' (1).
To achieve this, a four-way hose water level is used, and the vertical position of the end of the swivelling arm is checked with the meniscus
at each arm of the hose water level. A difference smaller than 2 mm was achieved, indicating that the difference between the axis of rotation
and the vertical is less than $0.2^o$.

A plummet is used to position the upper bearing of the detector and the sample mount: Before the sample mount is installed, and before the
upper bearing of the detector is added, the top plate (2) is positioned vertically above the lower bearing by using a plummet on a long
string. The deviation between the tip of the plummet and the centre of the lower bearing is less than 1 mm. The exact vertical distance
between the two bearings of the detector is adjusted with threaded rods later.

The lower bearing of the sample mount is adjusted next, by using a shorter plummet (3). Additionally required is the horizontal levelling of
the mount with a standard level (1mm/1m precision) and its vertical position relative to the lower detector bearing ($\pm 2mm$).

\vfill
\pagebreak[10]

The upper bearing of the sample mount is adjusted next (4). And finally, the upper mount of the detector sled is installed and the vertical
orientation of the rails is cross-checked.

The alignment of the zero positions for detector, sample mount and lamps is facilitated with a HeNe laser that is installed at a fixed
position at the wall. Its beam is adjusted to be horizontal (with a hose water level, distance 7m, deviation $\pm 5mm$, or $2'$ arc minutes),
and to centre on the sample centre ($\pm 3mm$). Then the mirror centre and the filament of the bulb can be adjusted to the HeNe beam.
A second alignment step of the mirrors adjusts their tilt so that each beam hit the other mirror dead-centre, and on the laser-beam. With the lamp
bulb in focus, the size of the shadow of an abject must be independent from its distance to the mirror. A calibration by auto-collimation was
scheduled for a later time.

\chapter{Control of Measurement and Data Acquisition}

\section{CCD camera}

A CCD camera offers, apart from fundamental aspects (see \ref{camera-intro}), some practical advantages: Connecting even a simple analog
monitor identifies stray-light signals quickly (e.g. control lamps, reflecting tape at the sample mount, light barriers emitting IR).
During testing, the software control of the tilt motor at the detector was verified with the camera as well: The marked sample centre had to
be at the centre of the image for all detector positions.

Furthermore, a detector with a lens is less susceptible to stray light than a detector with a simple baffle (assuming a sufficient quality of the
lens and its internal glass surfaces).

Using a camera with a synchronised Xenon-flash source could speed up measurements considerably: Each frame of the CCD is illuminated by
one flash (25Hz frame rate). With a multi-lamp setup, lamps could be fired in succession, giving a high data rate with low speed mechanical
movement of the detector. Blur caused by the detector moving would be avoided. Before a CCD camera can be considered further,
these problems need addressing:
\begin{enumerate}
\itemsep0em
\item	The usable range of intensity is smaller than anticipated: The difference between the dynamic range published by the manufacturer of
the chip (Thomson TH7863 \cite{thomson:89}) of 1:5000 and the measured value of 1:40 requires further investigation (e.g. on the electronics).
\footnote{2021: There's more detailed text on this from 1990, which was not included in this work.}
\item	The aperture at the lens is not controllable. It probably requires removal of the automatic aperture control circuit in this compact lens.
\end{enumerate}

\section{Estimation of required processing speed}

Data acquisition should allow a CCD camera. With a pixel resolution of the CCD camera of 60x60 pixel (after averaging with a hardware image
processor) and 25 frames-per-second (FPS) 90000 pixel/sec must be processed. Each pixel undergoes a 3x3 transformation matrix to transform
from pixel coordinate to sample coordinate (9 multiplications and 6 additions). The require compute speed is around 1.4 MFLOPS (mega
floating point operations per second), matched by a fast workstation, and excluding a PC.

\section{UNIX workstation and VME bus}
\label{unix-workstation}

The {\em VME bus} is an international standard for hardware that allows to combine hardware modules of different manufacturers.
The ''host computer'' is the basis with the required ''infrastructure'': video card, keyboard, monitor, disks and operating system.
In this case: A Hewlett Packard (HP)  9000/360 workstation (MC68030 CPU + MC68882 floating-point processor at 25MHz), running the HP variant
of UNIX (HP-UX). An VME ''bus expander'' HP 98577A (the workstation uses a different internal bus) offers slots for 4 VME cards: Two are used for
generic extra CPU cards (SAC700, MC68070 CPU, 128kB RAM, 3 timers and parallel, serial I/O), made by Eltec (Mainz,Germany).
\vfill

\pagebreak[3]

The generic CPU cards run a custom written assembler program to control the stepper-motors. Each card runs up to 3 motors, since each motor
requires its own timer, and there are 3 timers per card (unless hardware is added).

Signals for lamp control and all TTL signals are routed via the generic CPU cards and the motor-control I/O PCB as well.

Two more VME slots are taken up by the video digitiser: They digitise a TV signal (CCIR standard) in real-time to 744x600 pixel and write to
an on-board 512kB RAM.

Access to the address space of the VME cards is easy: The VME addresses are mapped 1:1 into the address space of the HP. The 32bit address bus
offers ample space for additional cards.

A user program gets the VME addresses mapped on request (trough the Memory-Management-Unit MMU) see \ref{software}. Thereby, a program running under
UNIX can directly access the RAM of the motor control and the RAM of the video digitisers, without requiring new device drivers on system
level (kernel). Access conflicts between the UNIX side and the additional processors are resolved by hardware on the VME cards, 
making the whole control extremely transparent and easy for the UNIX program.

The HP workstation is integrated via Local Area Network (LAN) (Ethernet, TCP/IP) in the network of the institute.

\section{UNIX workstation, VME software}
\label{software}

The software written for this diploma work is divided into 3 groups. The following table gives an overview, and adds the size of the source
files in kByte:

\begin{tabular}{lll}
1.	&measurement program			&8\\
2.	&library of subroutines			&\\
	&motor control, UNIX side		&15\\
	&mathematical helpers			&16\\
	&e.g. data sorting, min/max, etc	&\\
	&coordinate transformations		&10\\
	&data interpolation			&9\\
	&e.g. linear and 2D			&\\
	&control video digitiser 		&3\\
	&control of Keithley 485		&1\\
3.	&motor control, assembler on SAC700	&18\\
\end{tabular}

There are 82 subroutines written for this project. All software to control the device was written in the C language and administered with
the Revision Control System (RCS), which archives each new version together with a description or comment. This allows the specific
version of the software to be recorded in the measured data-file, allowing forensics later.
\vfill
\pagebreak[3]

\subsection{Control of drives and lamps}

\subsubsection{Assembler program on the SAC700 CPU cards}
\label{stepper-control-details}

\begin{figure}[h!]
\begin{center}
\input{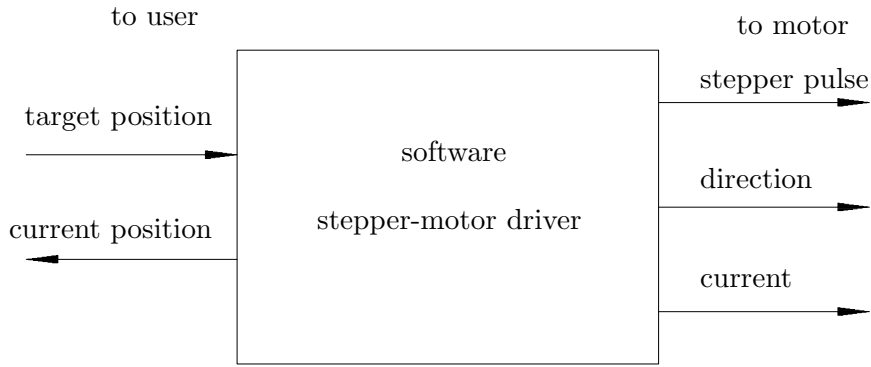}
\caption{\label{stepper-control}in- and output signals of a classical stepper-control software}
\end{center}
\end{figure}

The stepper control module generates a frequency modulated series of pulses to be fed to the motor driver power hardware, so that the drives
speed up and slows down with constant acceleration and the final position matches the intended position. Fig.~\ref{stepper-control} shows the
logical input and output of the software that handles the stepper control.

The ''current position'' is a 2 byte value of the actual position of the motor axis in real-time. All positions are given in unit ''step''.
As an example: If the axis is at position 0, a ''to-go'' position of 500 represents a half turn at the motor axis, 1000 is a full turn and
2000 are two full revolutions (for this particular 5-phase stepper motor in half-step mode).

The ''to-go'' position is another 2 byte word, meaning that the motor-axis can be programmed for a maximum of 65 revolutions. This word is
kept in the RAM of the SAC700 cards, and therefore writable by the UNIX side. Handshake between UNIX side and assembler program is
''write-and-go'': As soon as UNIX writes the 2 bytes, the drive will start (and by writing the 2-byte word in a single 16byte cycle, this is
safe without ''race conditions''). Further updates to the ''to-go'' position are ignored while this motor is moving.

A third 2 byte word resets the current position to a new value, without movement. This is required to initialise a position.

The assembler program supports a 'search'' mode: It drives the motor at low speed until a zero mark is reached, and loads the ''true''
physical position as new ''current position''.

\vfill
\pagebreak[3]

To check the ''current position'' during normal operations, a ''zero feedback'' counter triggers if the zero-mark is crossed during a normal
movement: it starts a step-counter whose value is compared to what it should be at the end of the movement.\\
\new{
To check this from the UNIX program is beneficial, since the major drawback of ''pure'' stepper drives is their lack of a feedback loop:
They either reach the intended position, or they miss steps (e.g. due to accidental torque overload) and hence miss the intended position
considerably.
}

During speed-up and slow-down, the frequency ramp of motor-steps has to be generated correctly for the resulting acceleration to be
constant. There's one timer per motor that generates a pulse after a preset time interval. Synchronously, the stepper control program 
updates the ''current position'' counter and then checks if the axis has to be speed-up further, or slowed
down. Namely if the next value loaded into the timer has to be a smaller or a larger value. 

This works with a table of pulse intervals: At the start of movement, the assembler program loads the first entry of the table into the
timer and starts the timer. After the first step, provided further acceleration is required, it loads the second entry of the table. During
acceleration the table is worked up with increasing index (and in reverse during de-acceleration). There are tables for each acceleration
and for each motor (since the inertia they are driving is different, so is the acceleration).
The simplest way to pre-generate the table is, by straight forward classical mechanics: 
\begin{eqnarray}
	s_i &=& i \; \Delta s 	\textrm{\hspace{4em}$s_i$ Position after step $i$, $\Delta$ step size} \\
	s_i &=& 0.5 \; a \; T_i	\textrm{\hspace{3em}$T_i$ time at step $i$, $a$ acceleration} 
\end{eqnarray}
therefore
\begin{displaymath}
	i \; \Delta s = 0.5 \; a \; ( \sum_{j=0}^i t_j )^2 \;\;\; \forall i \in N \;\;,\;\; t_j: = T_{j-1} - T_j
\end{displaymath}
and hence, for $v_0=0$ :
\begin{displaymath}
	t_n = ( \sqrt{n} - \sqrt{n-1} ) \; \frac{2 \Delta_s}{a}
\end{displaymath}
However, this theoretical result is not ideal in the practical world of motors: There is a maximum time interval (16bit timer: 65535 timer
units) and the real step angle is not infinitely small, so the start speed $v_0$ is actually {\em not} zero. Hence the better approach is:
\begin{displaymath}
	i \; \Delta s = v_0 \sum t_k + 0.5 \; a \; ( \sum_j t_j )^2
\end{displaymath}
leads to:
\begin{displaymath}
	t_n = \sqrt{ ( \frac{v_0}{a} + \sum_{k=0}^{n-1} t_k)^2 + 2 \frac{\Delta_s}{a}} \;\; - \frac{v_0}{a} \;\; + \sum_{k=0}^{n-1} t_k 
\end{displaymath}
No explicit formula was found for this recursive definition, - but in fact this isn't required, since the values are generated numerically
in a loop anyway.

\vfill
\pagebreak[5]

$v_0$ is given by the maximum time interval $t_{max}$ that can be programmed in the timer. The resulting acceleration ramps are nicely smooth.
This leads to a precise movement of the axis, without over-swing at the end, even for axis driven by a timing-belt.

\new{All timing-belt reduction gear have a residue elasticity, which tends to over-swing the driven axis. So a smooth de-acceleration
helps to reach a position precisely.}

The assembler program can control multiple motors in parallel, however the maximum frequency gets lower if multiple motors have to be cared
for. The written algorithm ensures that motors move independently and correctly, unless there is an ''overrun'' condition, which is detected.
This happens during acceleration if the frequency gets too high. By configuration of maximum frequency for each axis, an ''overrun'' can be
avoided during operations.

The maximum frequency has other limits as well: A motor can actually follow the steps up to a maximum frequency only

Safety-checks are implemented on the UNIX side, in case the assembler program gets stuck or breaks for unforeseen reasons. There's a
''dead-man'' switch: The assembler program, besides its other tasks, resets a memory location to zero every 10ms (''alive byte''). The UNIX
side sets it to 1 , and if there's still a 1 in this byte after 50ms, the assembler program obviously stalled. No such problem occurred in
half a year of operation.

It is possible for the user (UNIX) side to stop the assembler program. The SAC700 processor board then enters a firmware monitor in EPROM
(Erasable Programmable Read Only Memory), supplied by the manufacturer, which allows debugging through an RS232 terminal connection.

The assembler program itself resides in RAM and is downloaded from the UNIX side (''firmware download''). This benefits from compatibility
within the family of MS68000 CPUs: The assembler program is compiled with the HP compiler on the UNIX workstation (on a Motorola MC68030
CPU, using the MC68010 option) and executes on a 68070 CPU (an MC68000 derivative made by Valvo).

\subsubsection{Library of subroutines on UNIX side}

The software subroutines are the interface between the measurement program (on UNIX side) and the assembler program (on the SAC700 boards). These are
simple calls for positioning an axis in the device, that hide minuscule details of the motor drives from application programs:
\begin{verbatim}
	motor_position( deg, motor, MOTOR_WAIT )
\end{verbatim}
Will move the axis with name {\tt motor} to an angle of {\tt deg} degrees and waits for completion of the movement.
Multiple axis can be moved at once:
\begin{verbatim}
	motor_position( 20.0, TWD_VERT, NO_MOTOR_WAIT );
	motor_position( 45.0, TWD_HOR, NO_MOTOR_WAIT );
	motor_wait_to_finish();
\end{verbatim}
\new{This moves the two axis of the sample mount at once and waits for completion of both movements.}

\vfill
\pagebreak[10]

Other frequently used subroutines are {\tt read\_position(axis)} to read the actual position. even if moving, {\tt motor\_axis\_init(axis)}
to initialise (zero-find) an axis and {\tt motor\_init()} to initialise the SAC700 CPUs and the assembler program.

Parameters used in the assembler program (maximum acceleration, maximum frequency, zero-position) are read from file by {\tt motor\_init()}.
All status information is logged to a file.

For movements during which measurement points are taken, the speed for an axis can be reduced to any value.

\section{Measurement program and measurement sequence}

While measurements are taken {\new{(measurements-on-the-fly})}, only one axis is moved (typically the detector). The coordinates of the detector
are read in the ''lab coordinate'' system and then transformed to sample coordinates. This conversion can also be done later with a second program.
The main measurement program plans the detector path, reads and processes the data, and writes it to file.

\begin{figure}[h!]
\begin{center}
\includegraphics[height=10cm]{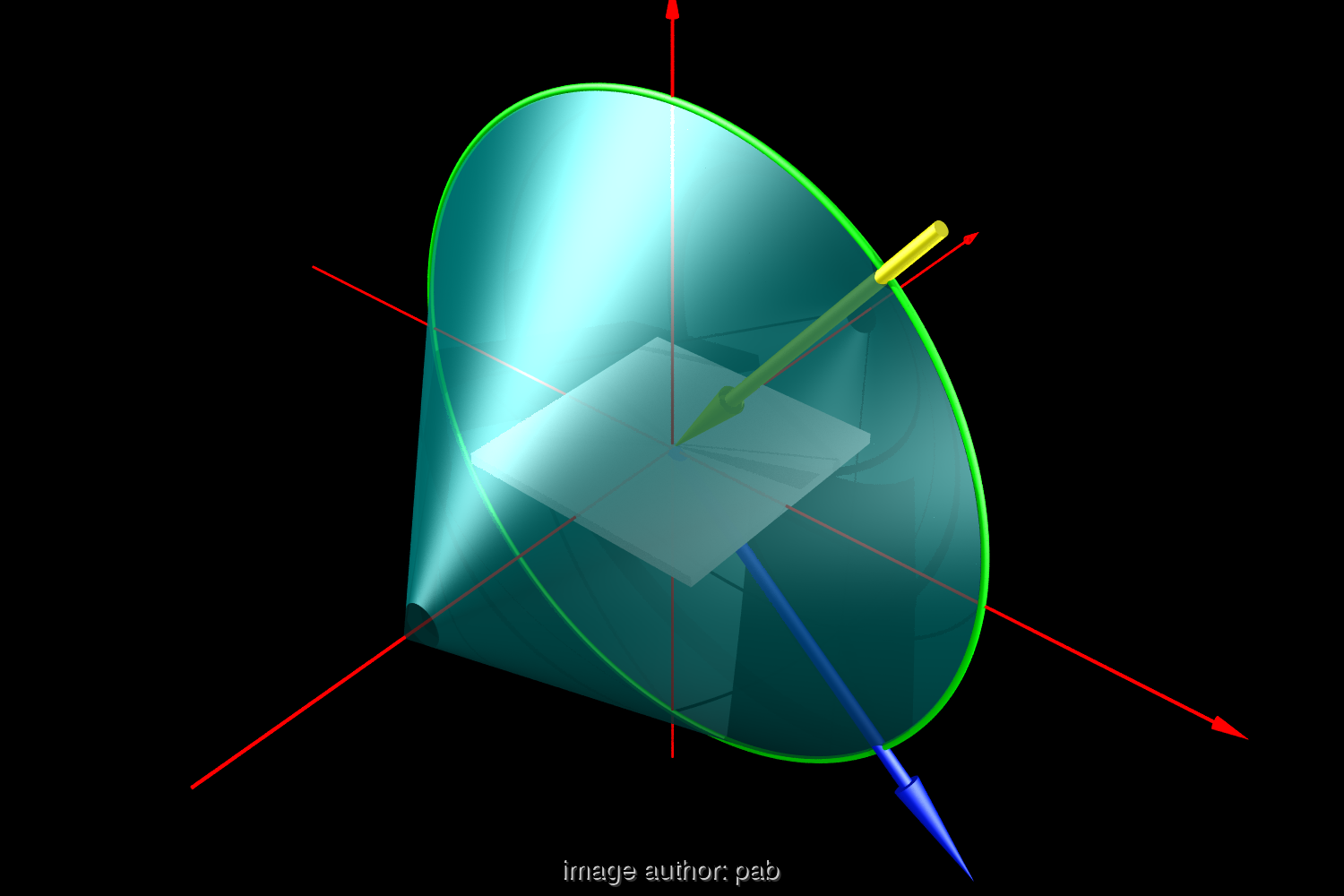}
\caption{\label{mess1}
	In-plane scans use only one outgoing angle and the detector moves on the circle shown in circle.
	All possible detector positions are on the surface of the green-blue sphere.}
\end{center}
\end{figure}

\begin{figure}[h!]
\begin{center}
\includegraphics[height=10cm]{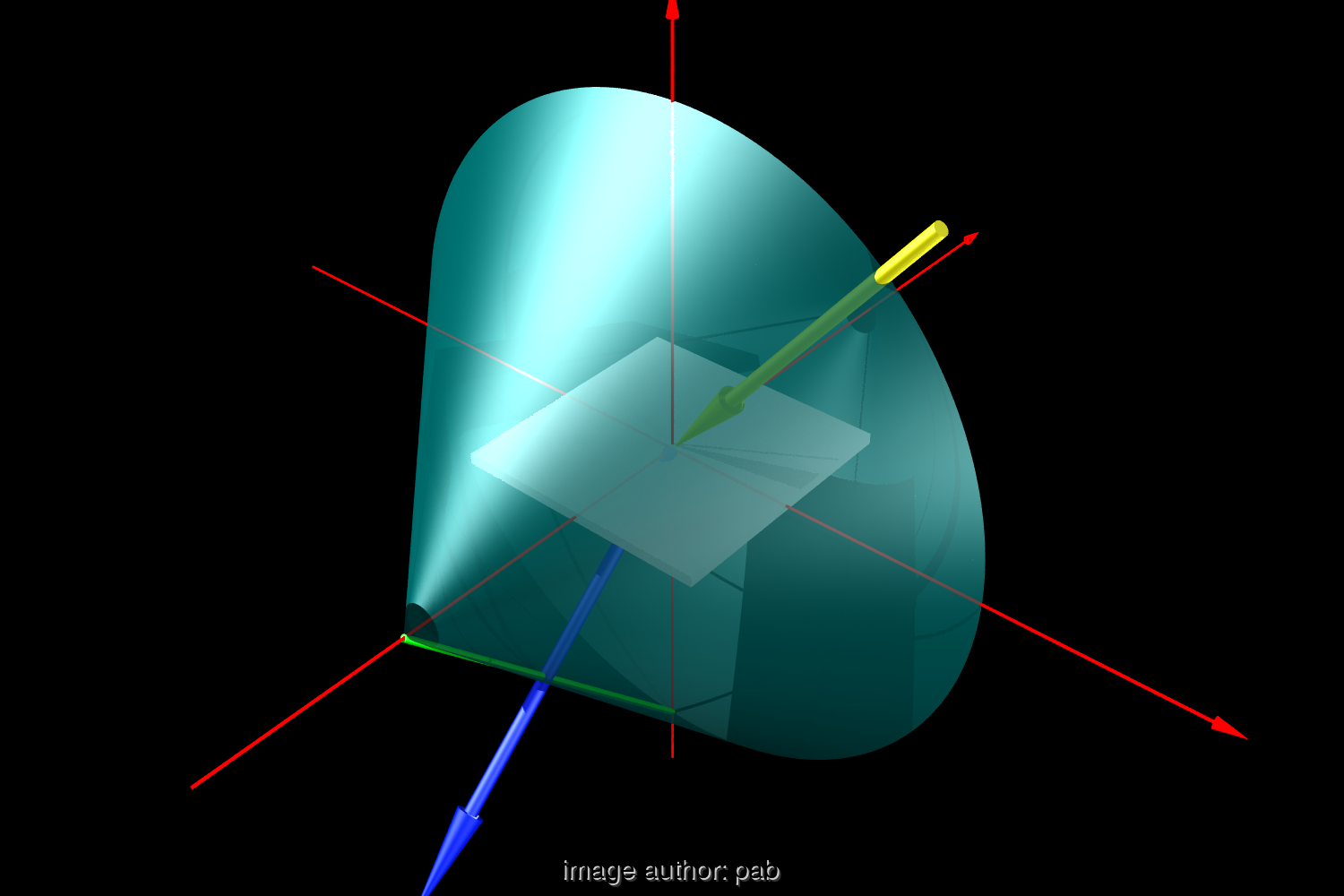}
\caption{\label{mess2}For the fit of TI data, the data scans along the line (green), see also Fig.~\ref{cones}}
\end{center}
\end{figure}

For this diploma work, there are three measurement methods that use the aforementioned library of subroutines:

\begin{enumerate}

\item Measurements with 2 degrees of freedom use just one incident angle and one outgoing angle, the detector moves on a circle
(Fig.\ref{mess1}). The outgoing angles in sample coordinates are calculated and written to file. The 2D plot uses a program with calls to
the DISSPLA graphics library (see \ref{disspla}).\\
{\em Note 2021: The original text states that is done for {\em rotationally symmetric samples}, which is not correct: Even symmetric samples
require an out-of-plane scan. A fact that even some folks at ISE higher up the ranks weren't aware of in 1990.}

\item Measurements with 1 degree of freedom for a special model of honeycomb structures. This takes measurements along a linear movement on
the surface of detector paths (Fig.~\ref{mess2}).

\item Measurements with a full set of degrees of freedom: For each incident direction, all outgoing directions are scanned. The detector
moves on the surface shown in Fig.~\ref{mess1}. This standard procedure is described in more detail in the following text.
\end{enumerate}

As explained in section \ref{quarter-sphere} (see the definition of scanning a quarter sphere and variables $\alpha, \beta, \theta_{det},
\phi_{det}, n$) there are 4 combinations of $\alpha, \beta$ that set all the same incident angle (using 2 lamps). For each, the detector
scans a solid angle of $\pi$ (quarter sphere) and writes the measurement data to file. The detector scans each quarter sphere by swivelling
around $180^o$ with a target resolution of $2^o$ ($\phi_{det}$). After each swivel, the detector is moved up $4^o$ along the rail ($\theta_{det}$).
All data is stored using sample coordinates. Each measurement of a quarter sphere can thus be verified on its own.

\clearpage

\label{on-the-fly}
While the detector moves, the control program reads two values: The incident power onto the detector (by a pico-ammeter) and the actual
position of the detector. Angular resolution depends hence on the selected scan speed. Since the scan speed is not constant along a swivel
(due to the acceleration ramps), the position of data-points is interpolated linearly. With a maximum angular step of $2.2^o$ between two
original points, an interpolation to $2^o$ is plausible.

Already during measurement, the variable distance is corrected for, by multiplication with $R^2$ ($R$ is the distance of the detector to the
sample centre).\\
\new{Of course, this is only a crude approximation, since the sample is not a point source.}

With UNIX being a multi-tasking system, there's a chance that a running program is stopped momentarily (in the order of milli-seconds) to allow
another program to execute. With the measurement program, this would cause gaps in the data, since the detector movement continues
undisturbed nevertheless.

The solution is to give the measurement program ''real-time priority'', so it gets top priority amongst the running processes. There exists
a second problem similar to this: The pico-ammeter is connected with an IEEE488 bus, which it shares with other I/O units (e.g. plotter).
Under extreme conditions, this could make the IEEE488 bus unavailable for access to the pico-ammeter if its used heavily otherwise (bus
congestion). While this rarely of concern, one would consider a second exclusive IEEE488 bus for the pico-ammeter if timing gets more
critical. It is not considered to be of serious concern, since the pico-ammeter is scheduled to be replaced by a faster VME-bus card anyway.

When using the Keithley pico-ammeter, the measurement program spends most of its time waiting for results from the pico-ammeter. In fact the
measurement program uses about 10\% of available CPU time, which makes working interactively at the workstation still possible, even during
measurements.

\section{Data processing}

The 4 files with data in ''lab'' coordinates must be transferred to sample coordinates and assembled. The transformation consists of 4
rotational matrices and concludes the incident and outgoing directions from the position of the device. The 4 quarter spheres should be
checked for continuity and smoothness at the boundary between two quarter-spheres, and might need fine-tuning (e.g. when using two lamps of
slightly different intensity), but this isn't implemented yet. After transformation to sample coordinates data files follow this naming
scheme
\begin{verbatim}
	<sample-name>-<phi>-<theta>
\end{verbatim}
With {\tt <phi>-<theta>} specifying the incident direction.
Each line in the file is formatted as:
\begin{verbatim}
	<theta> <phi> value
\end{verbatim}
With {\tt <theta> <phi>} specifying the outgoing direction in [degrees] for each data-point. Further processing is done at a later stage,
e.g. during visualisation.

\chapter{Data usage and Requirements for Visualisation}

Data processing depends on the intended usage.
The following applications for the measured data are currently planned, with the state of measurement and display software:

\begin{tabularx}{\textwidth}{lllX}
	&							&measurement	&display\\
1.	&checks of device during assembly and test phase	&o.k.		&o.k.\\
2.	&qualitative display for comparison of materials	&o.k.		&o.k.\\
3.	&illumination levels in arbitrary surfaces		&limited	&limited\\
4.	&absorption measurements				&limited	&-\\
5.	&simulation of multi-layer TI				&no		&-\\
\end{tabularx}

Measured data doesn't allow exact absolute values yet and hence no values for the absorption in the material.
The main problems hindering quantitative measurements are:

\begin{itemize}
\itemsep0em 
	\item	shadowing by the lamp post
	\item	stray light from second mirror
	\item	''peaks'' in transmission by light shining past the sample mount
	\item	light source not a point source
	\item	reflexions off the sample mount
\end{itemize}

Solutions to these problems consists of small changes to the device (automatic blocking of the unused mirror, anti-reflex paint for the
sample mount) and refined methods of data-processing (half-automatic elimination of false peaks in transmission before integrating measured
data, purchase of a dedicated graphic-workstation to handle complex, multi-variable measured data).

For point 1 in above list plots are generated by a commercial graphic package (''DISSPLA'' by Computer Associates, see fig~\ref{srd}), for  2
the graphic is rendered by a public domain ray-tracer 
\href{https://graphics.stanford.edu/~cek/rayshade/rayshade.html}{{\tt rayshade} from Yale University} and my own software for scene
generation from measurement data. The rest of this chapter describes plotting a function $f(x,y)$ with continuous codomain for $x,y$.

\section{Raw data, visualising $f(x,y)$ on a grid (DISSPLA program)}
\label{disspla}

\begin{figure}[h!]
\begin{center}
\fbox{
\input{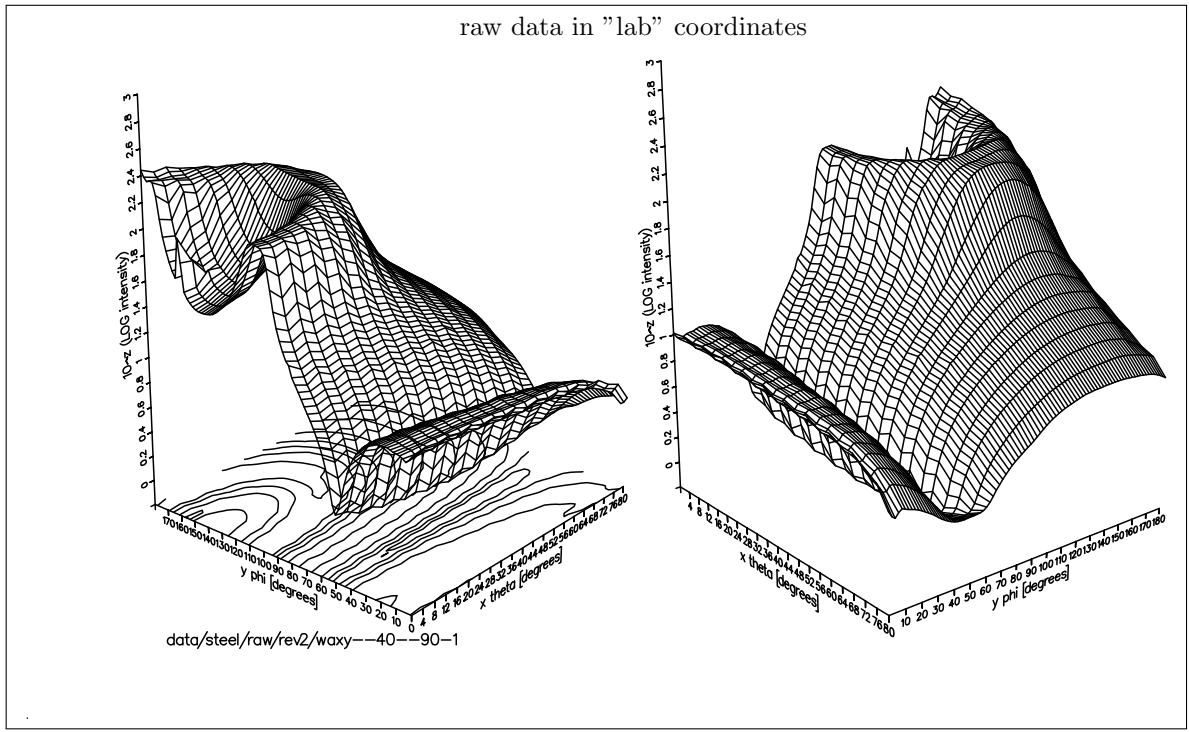}
}
\caption{\label{srd}Measurement results of a quarter sphere (half of the transmission ''ring'' and part of reflection in 
	''lab coordinates'', seen from two sides. The drawing is scaled from an A3 print.}
\end{center}
\end{figure}

The ''raw'' data can be displayed  in ''lab'' coordinates after each measurement of the quarter sphere in a 3D Cartesian coordinate system
(Fig.~\ref{srd}). This offers two advantages:

\begin{enumerate}

\item The signal depends on two angular parameter, whose values are on equi-distant grid in the XY plane. Codomains are
$\theta_{det}= [0..85^o]$ and $\phi_{det}= [0..180^o]$. The intensity is plotted as z-coordinate over each x-y point. Each point of the
resulting grid correspondents exactly to one measured data point.

\vfill
\pagebreak[30]

\item Each measured data point is taken at a easily checked position of the detector in ''lab'' coordinates. If the data show a suspicious
singular peak, the source is easily verified by moving the detector to that position and identifying the cause of the peak.

\end{enumerate}

This type of 3D plot, using a grid, is only possible if two assumptions are met: Firstly, the domain of $x$ and $y$ makes a plot in Cartesian
coordinates meaningful and, secondly, the points lie on an equidistant xy-grid. If $x$ and $y$ are not on a regular grid, problems arise,
which are described in annex \ref{interpol-problem}.

\section{Data in sample coordinates and visualisation by triangulation of $f(x,y)$}
\label{display-triangulation}

In computer-graphics, ''triangulation'' is the tessellation of a surface into triangles. The surface is given here by the measured
data points, one z-coordinate for each x-y point. However, the resulting triangulation is {\em not defined uniquely} (see annex
\ref{interpol-problem}.

\begin{figure}[h!]
\begin{center}
\fbox{
\input{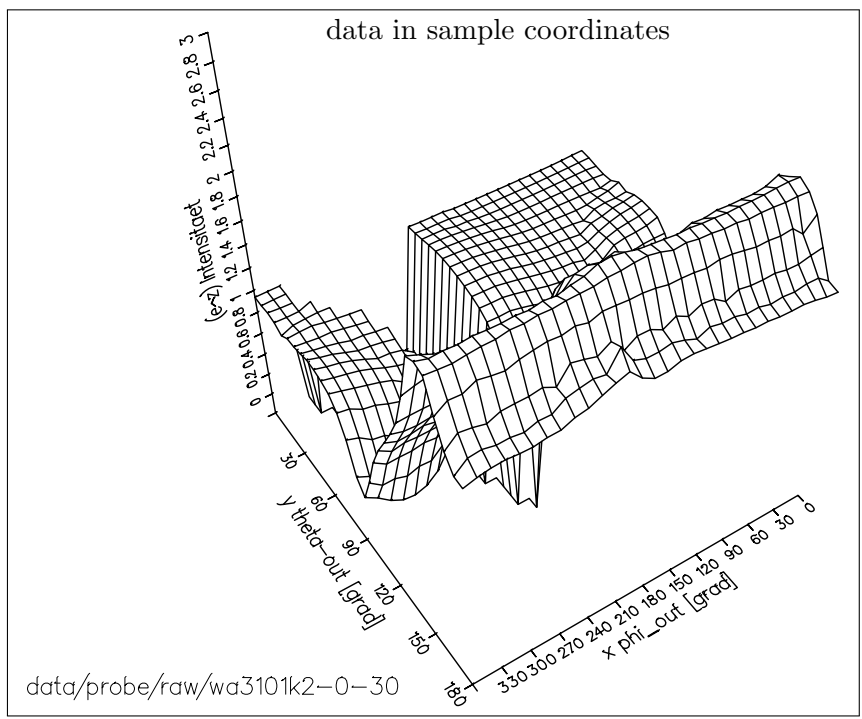}
}
\caption{\label{prd}Power received by the detector for all outgoing angles, for one incident direction, in sample coordinates ($\theta_{out},
\phi_{out}$), as Cartesian 3D plot}
\end{center}
\end{figure}

After coordinate transformation to sample coordinates and assembly of the 4 quarter-spheres, measurement points do not lie on an equi-spaced
grid anymore, and the co-domain is larger: $\phi=[0..360]$ and $\theta=[0..180]$. Displaying this in a Cartesian coordinate system results in
a hard-to-read, suboptimal plot, Fig.~\ref{prd}).

\clearpage

\begin{figure}[h!]
\begin{center}
\input{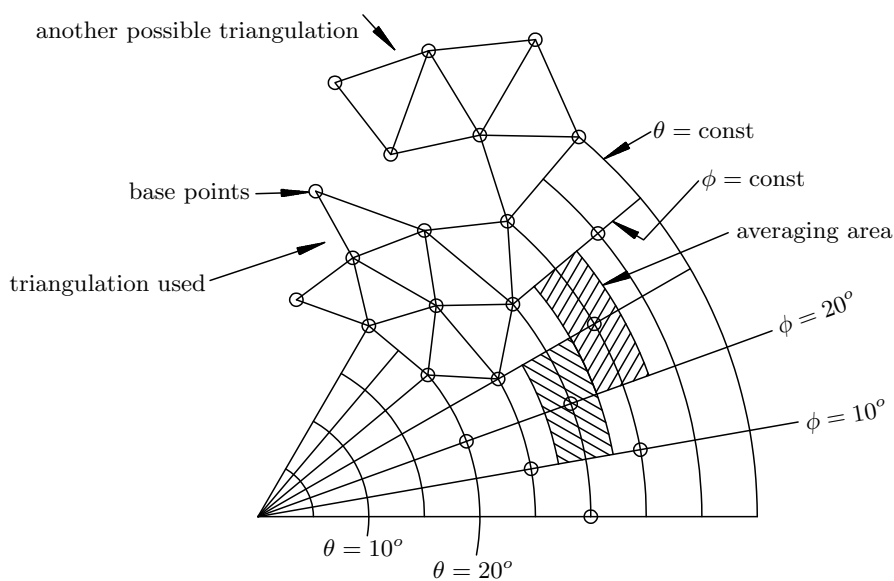}
\caption{\label{triangle}Averaging of measurement data: The edges of the triangle (small circle), the area for averages
	(hatched area) and the resulting triangles (top) }
\end{center}
\end{figure}

A better way to visualise this:
A function $f=f(\theta,\phi)$ is defined on a sphere. A display similar to a globe, with exaggerated mountain ridges, is not sufficiently
easy to interpret. Hence we suggest to project a hemisphere onto the plane: {\em Polar Azimuthal Equidistant}, not true to length and angle, but
sufficient for our purpose. Lines of constant $\phi$ are shown as lines, loci of constant $\theta$ are concentric circles. The measured
value is plotted as height above the XY-plane. The surface is defined by triangles whose corners are generated by averaging over measured
data.
\footnote{2021: This method was obsoleted a year later by switching to Delaunay triangulation. Furthermore, when using {\em Lambert
Azimuthal} Projection, the area of solid angles is preserved.}
Fig.~\ref{triangle} shows the area that is used per corner.

\clearpage

This particular triangulation generates a well-defined surface, since it not only depends on the data points, but also on the algorithm
building the triangles: For every triangle, 2 points lie on a circle of $\theta=\const$. This is not the only possible triangulation, see
Fig.~\ref{triangle}.

The graphical plot must be a clear and exact visualisation of the function $f$: Each triangle is coloured depending on the z-values at its
corners. With increasing average value, a rainbow colour is used, starting at blue, green, yellow to red, a colour scale that works for
grey-scale images as well (grey-scale transformation according to NTSC standard). It was found that mapping a $\theta,\phi$ grid onto the
surface helps to localise a position in $\theta,\phi$. Additional information on sample name, incident direction and z-scale is included in
the image.  More information could be added, but was postponed for time constraints. See also chapter~\ref{meas-intro}.

\begin{figure}[h!]
\begin{center}
\includegraphics[height=11cm]{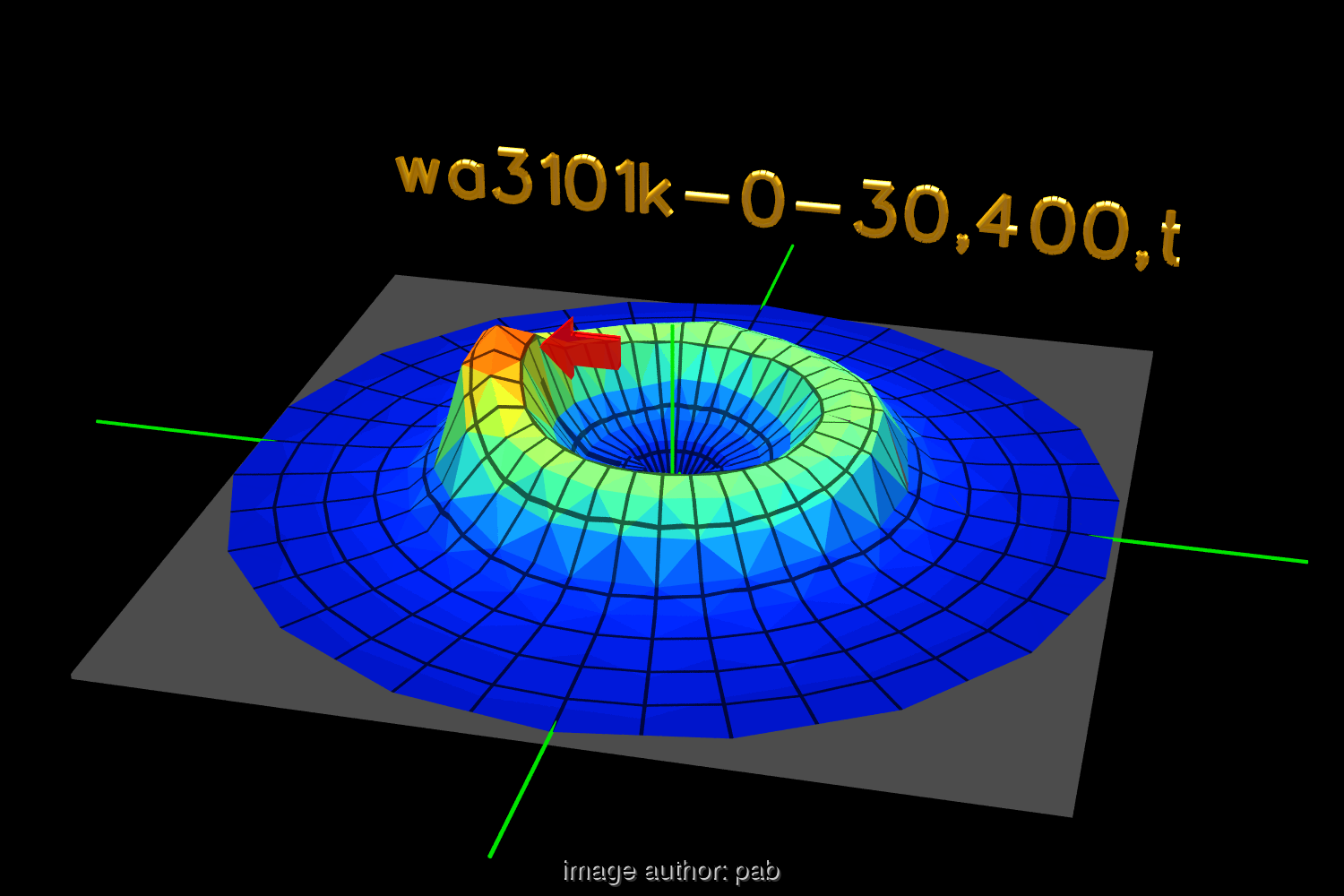}
\caption{\label{rendering1}The final visualisation,
The arrow demonstrates that extra elements can be added easily}
\end{center}
\end{figure}

While visual localisation is sufficient in ($\theta,\phi$), to read an explicit estimate of the z-value is hard. This is helped by adding a
transparent z-plane (Fig.~\ref{rendering1}).

When displayed on a workstation screen, moving the z-plane interactively allows a precise analysis of the data-value at each point.

H. Akima describes a triangulation without averaging \cite{akima:78}: Triangles defined in the XY-plane by the data points are the base to
fit a 3rd-order polynomial to its corners. Neighbouring polynomials are continuous and differentiable and the surface contains each
data point exactly. This method wasn't implemented yet.

Each of the images is generated in a two step process:

Firstly, triangles are generated from data points, as described, and their coordinates, plus a colour triple (RGB) of the triangle is written
to file. The 3D text is added as cylinders. This describes a 3D scene.

In a second step the program {\tt rayshade} renders an image of the fictitious 3D scene as would be seen from a specific eye-point
(''photo'' resolution 1200x1000 pixel). The scene is made out of simple objects (sphere, cone, cylinder, triangle, cube, plane). The
position, colour and surface parameter (shininess) for each object are defined in a standard ASCII text file. {\tt rayshade} allows multiple
coloured light-sources, shadows can be generated, as well as fog, depth-of-focus and focal length. The underlying method is called
''raytracing'' \cite{thalmann:87}. Rendering time was about 45min on a SUN SPARCSTATION-1 . Images were output on a Post-Script laser printer
or on a graphic monitor (255 colours on an X11-display). Apart from measurement results, rayshade was also used to render the images in
chapter~\ref{theory}.

\chapter{Measurement Results}

\label{meas-intro}
Four honeycomb samples of Transparent Insulation (TI) material and two aerogel samples have been measured.
(For photos see annex~\ref{ti-photos}.)

\begin{description}
\itemsep0em 
\item[wa0301]
plastic tube structure (2mm diameter, 15mm length) with the front and back surface melted to get a stable package

\item[waxy]
irregularly shaped transparent cylinders (4mm diameter, 30mm length) glued together in a dense package

\item[wa0325]
glass cylinders with very thin walls (4mm diameter, 40mm length) stacked together, with glass plane at front and back

\item[wa3101]
legacy material with quadratic honeycombs (4mm edge length, 80mm length), with periodic thicker walls along one direction due its
manufacturing process

\item[aerogel-I]
bulk granular aerogel spheres (diameter about 4mm) packaged between two glass panes of 15mm distance

\item[aerogel-II]
massive block of aerogel, thickness 20mm

\end{description}

\begin{figure}[h!]
\begin{center}
\fbox{
\input{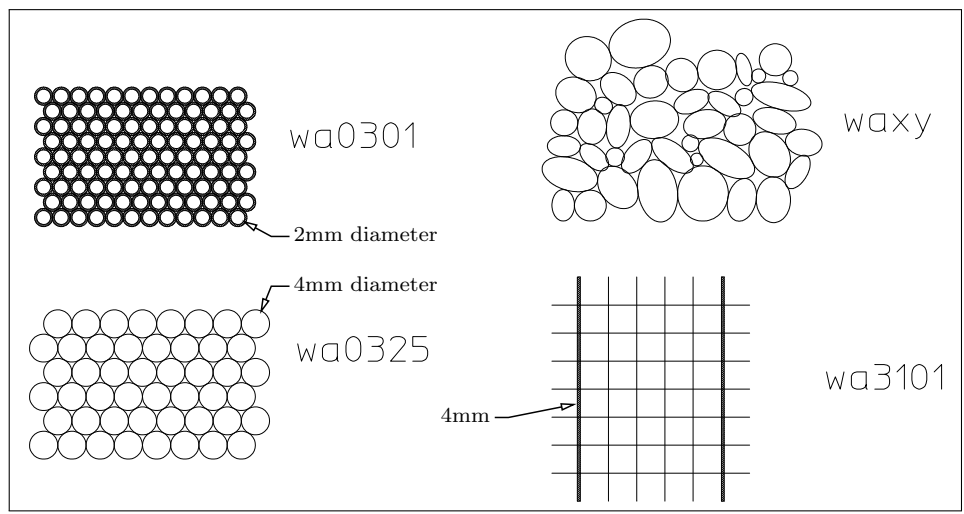}
}
\caption{\label{waty}The four tested samples of TI materials, shown to scale. The surface incident to the light is shown.}
\end{center}
\end{figure}

Measurement parameters: sample material, incident directions ($\theta_{in},\phi_{in}$), range of outgoing directions
($\theta_{out},\phi_{out}$). A measurement sequence consists of measurements with variation of one parameter:

\begin{tabularx}{\textwidth}{llllX}
		&TI type		&outgoing					&parameter	&constant\\
\hline
\ref{meas1}	&aerogel		&$\theta_{out}=[0..360^o]$			&material	&$\theta_{in},\phi_{in}$\\
\ref{meas2}	&honeycomb model fit	&$\theta_{out}=[0..90^o]$			&$\theta_{in}$	&sample\\
\ref{meas3}	&honeycomb model fit	&$\theta_{out}=[90..180^o], \phi=[0..360]$	&$\theta_{in}$	&$\phi_{in}$, sample\\
		&honeycomb transmission	&$\theta_{out}=[90..180^o], \phi=[0..360]$	&material	&material\\
		&honeycomb transmission	&$\theta_{out}=[90..180^o], \phi=[0..360]$	&$\phi_{in}$	&$\theta_{in}$, sample\\
		&honeycomb reflexion	&$\theta_{out}=[0..90^o], \phi=[0..360]$	&$\theta_{in}$	&sample\\
		&honeycomb reflexion	&$\theta_{out}=[0..90^o], \phi=[0..360]$	&material	&$\theta_{in},\phi_{in}$
\end{tabularx}

\section{Measurement 2D, in-plane, aerogel}
\label{meas1}

\begin{figure}[h!]
\begin{center}
\input{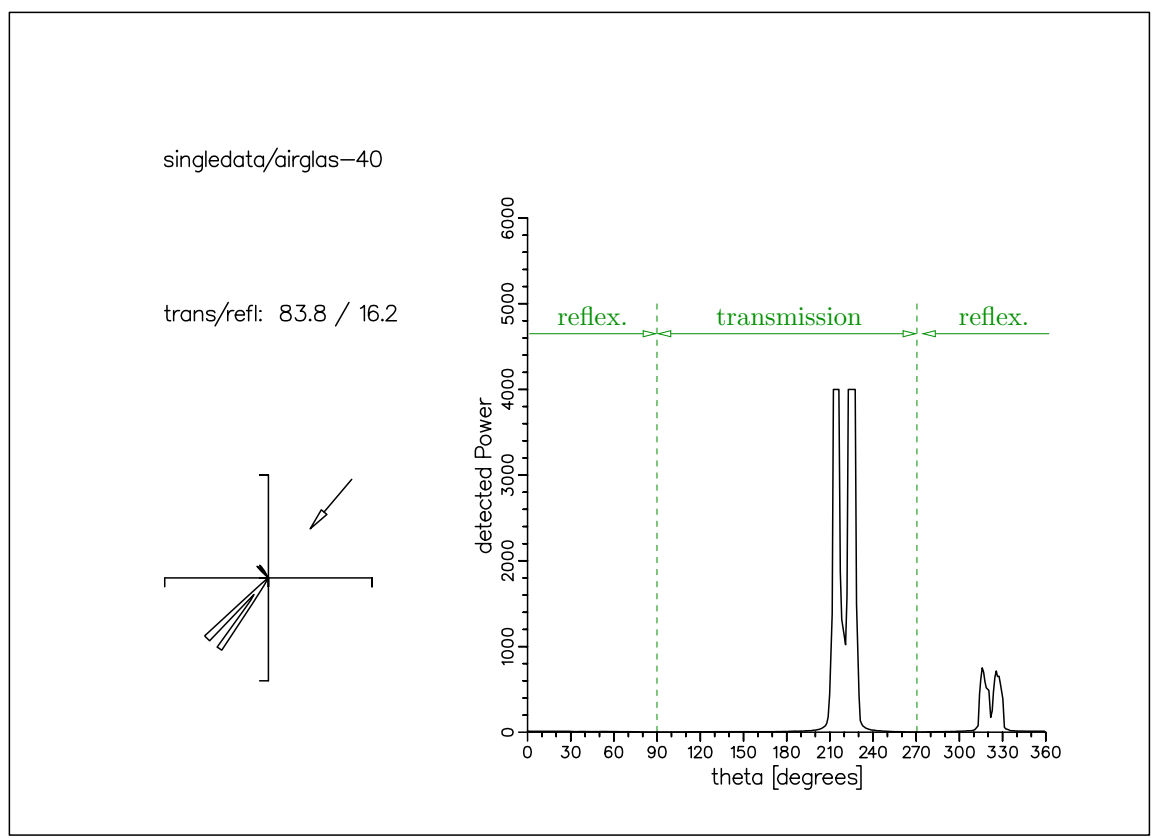}
\caption{\label{airgla40}Scattering of  massive block of aerogel, incident direction $\theta_{in}=40^o$}
\end{center}
\end{figure}

\begin{figure}[h!]
\begin{center}
\input{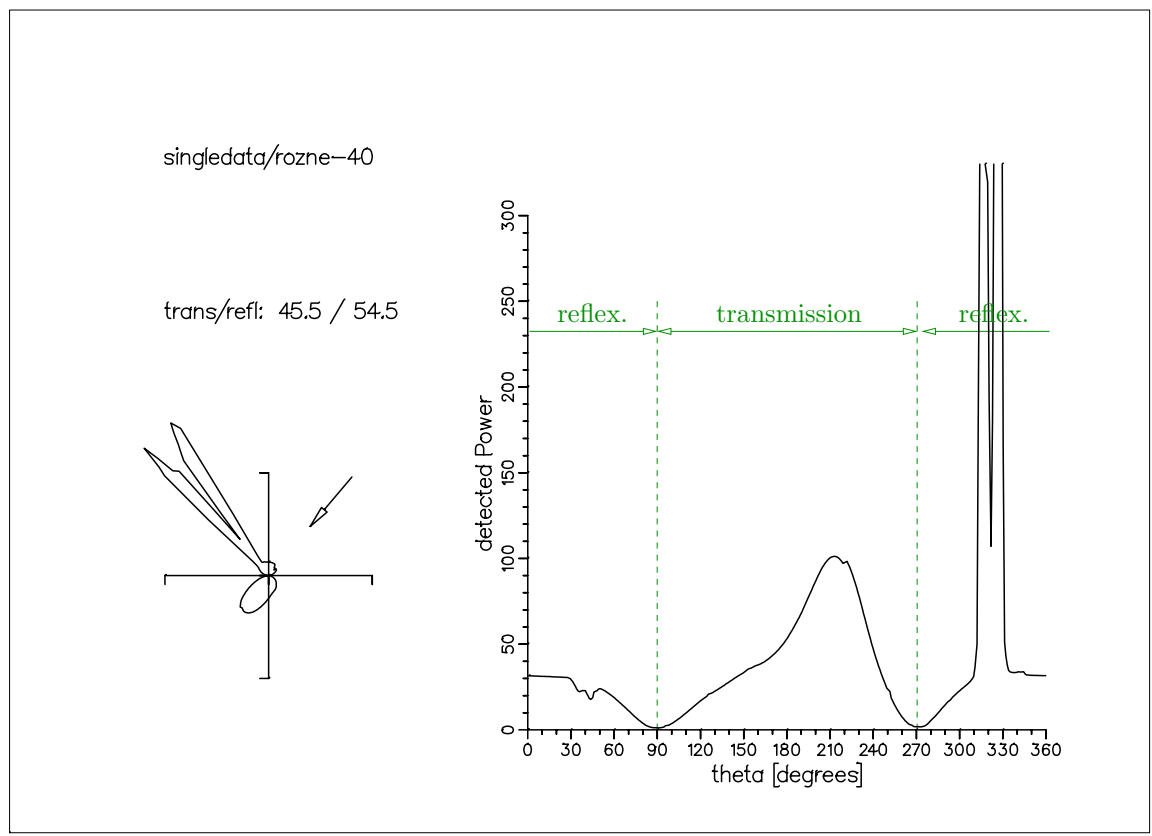}
\caption{\label{rozne40}Scattering of granular aerogel, $\theta_{in}=40^o$}
\end{center}
\end{figure}

Measurements of {\bf aerogel-I} and {\bf aerogel-II} showed an expected higher reflexion at the glass pane and much broader scattering of
the aerogel bulk material. The first is due to the higher index of refraction of solid glass compared to en-bloc aerogel (aerogel density is 1.02 to
1.05); the latter is due to surface scattering of the individual spheres. Fig.~\ref{airgla40} and \ref{rozne40} also include a ratio
of integrated transmission / reflexion.

\begin{figure}[h!]
\begin{center}
\fbox{
\input{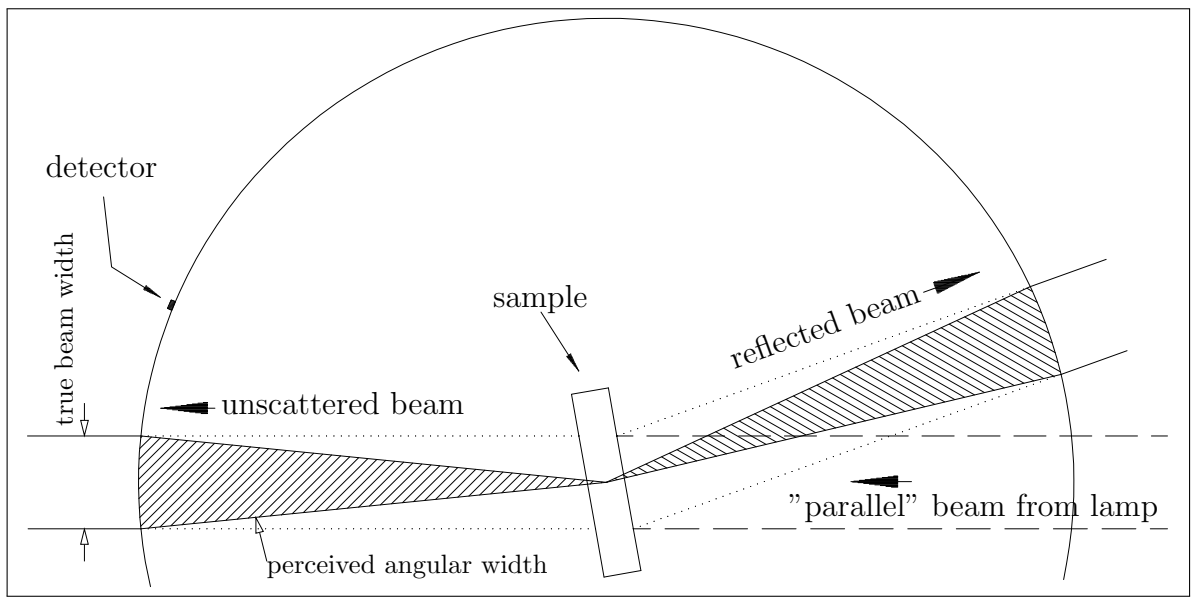}
}
\caption{\label{widebeam}The detector signal is broad even for a perfectly transparent sample, due to the width of the beam (hatched angular
area). Similar, the reflection peak is broad even for ideal reflection}
\end{center}
\end{figure}

The peak width shown in the plots is due to the beam width of the incident light (Fig.~\ref{widebeam}). The two spikes are actually caused
by shadowing of the mirror by the lamp support rod. Compare with \ref{meas2} to see how this is corrected for TI materials.  \footnote{2021:
Good side of this: At least the shadow of the rod after a 4m beam path proves the beam to be nicely ''parallel''.}

\begin{figure}[h!]
\begin{center}
\input{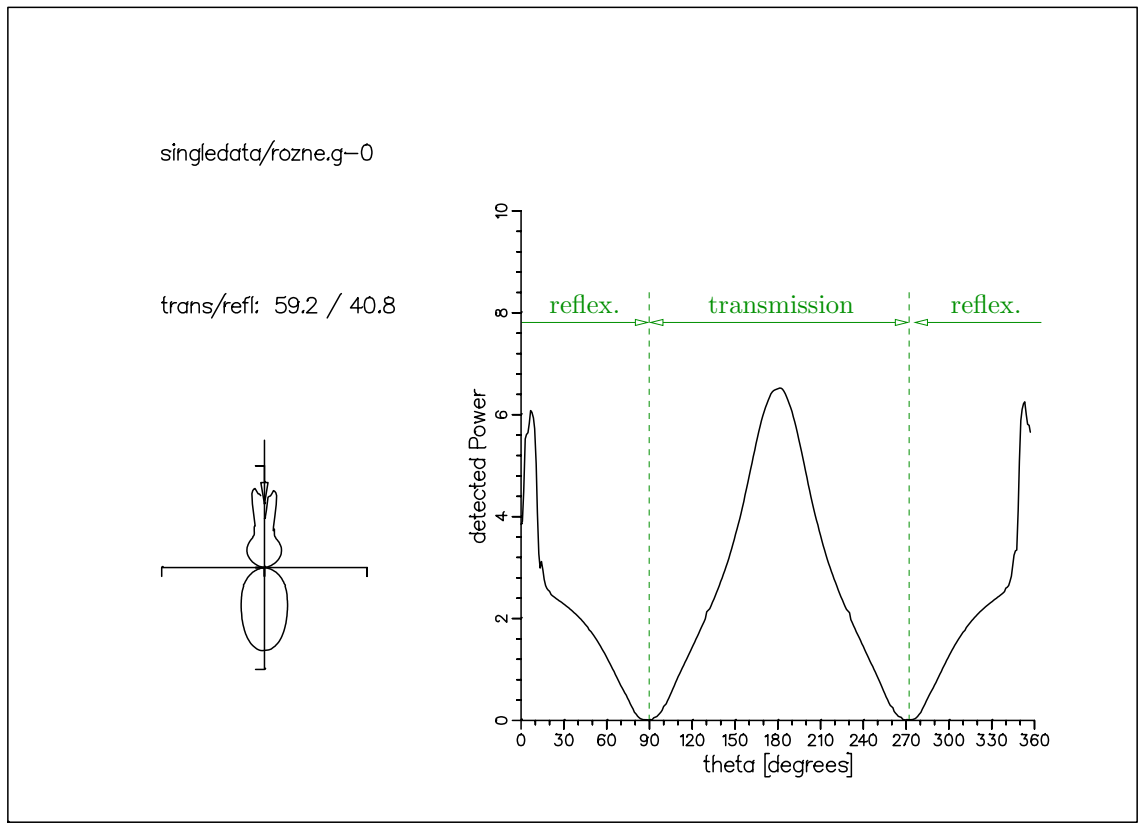}
\caption{\label{rozneg0}Scattering of granular aerogel, $\theta_{in}=0^o$, short wavelength}
\end{center}
\end{figure}

\begin{figure}[h!]
\begin{center}
\input{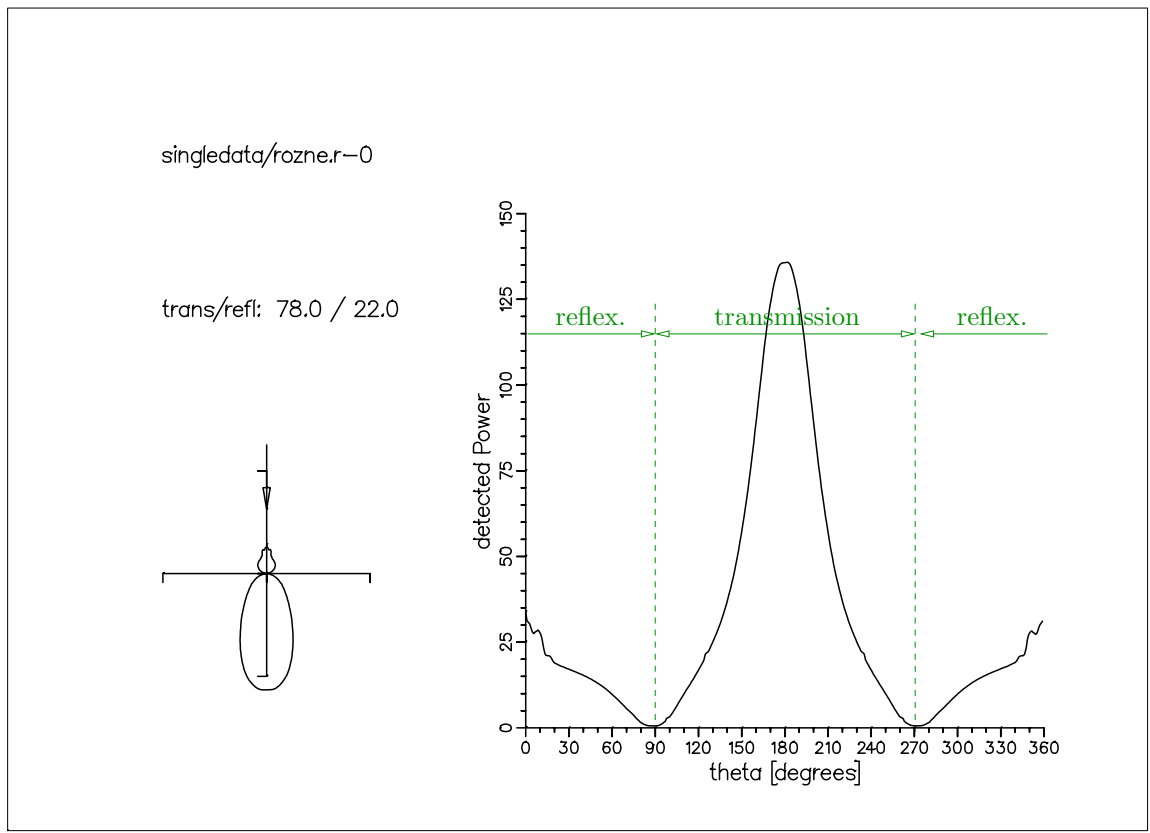}
\caption{\label{rozner0}Scattering of granular aerogel, $\theta_{in}=0^o$, $\lambda>650nm$}
\end{center}
\end{figure}

Additional test on spectral behaviour:

By adding a spectral filter at the detector (either greenish $V_{\lambda}$ filter or a red cut-off, Schott RG715), the assumed
wavelength dependency of aerogel scattering could be shown (Fig.~\ref{rozneg0}, \ref{rozner0}):

Scattering of bulk aerogel is caused by Rayleigh scattering at the aerogel pores and Mie scattering at the sphere surfaces (asymmetric in
forward/backward direction).  At shorter wavelengths, backscatter is larger. The spectral sensitivity of the Si detector material causes the
different scale of Fig.~\ref{rozneg0}, \ref{rozner0}. Nevertheless, the difference in scattering behaviour is obvious.

The current state of the device can not resolve the small angle scattering of a block of aerogel, and hence the difference in scattering of
red or green light.

\clearpage

\section{Measurement to model Transparent Insulation}
\label{meas2}

These measurements supply data for the modelling of TI materials described in \ref{ti-model}. This model assumes a rotational symmetric
distribution around the surface normal, so the light distribution depends solely on $\theta_{out}$. For sample this, the detector moves
along its rail in a straight line (See sections \ref{det-detmount} and \ref{ti-model}).

\begin{figure}[h!]
\begin{center}
\input{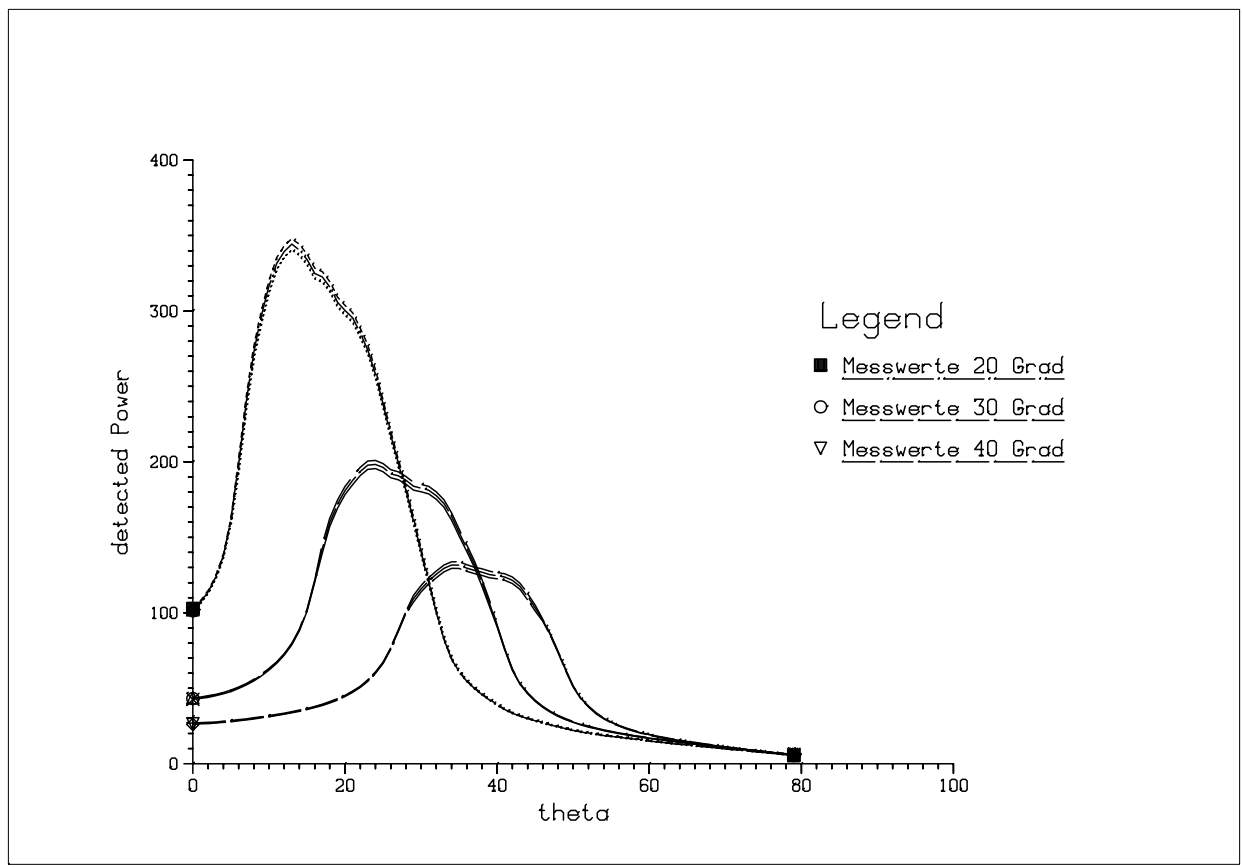}
\caption{\label{wsm1}Measurement data for a set of incident angles, sample {\bf wa3101}}
\end{center}
\end{figure}

Three honeycomb materials ({\bf wa3101}, {\bf wa0301}, {\bf waxy}) have been modelled for three incident angles of $20,30,40^o$. All data
was multiplied with the distance detector to sample centre. Also included in Fig.~\ref{wsm1} are the error bounds as given in the manual of the
Keithley pico-ammeter. 

\begin{figure}[h!]
\begin{center}
\input{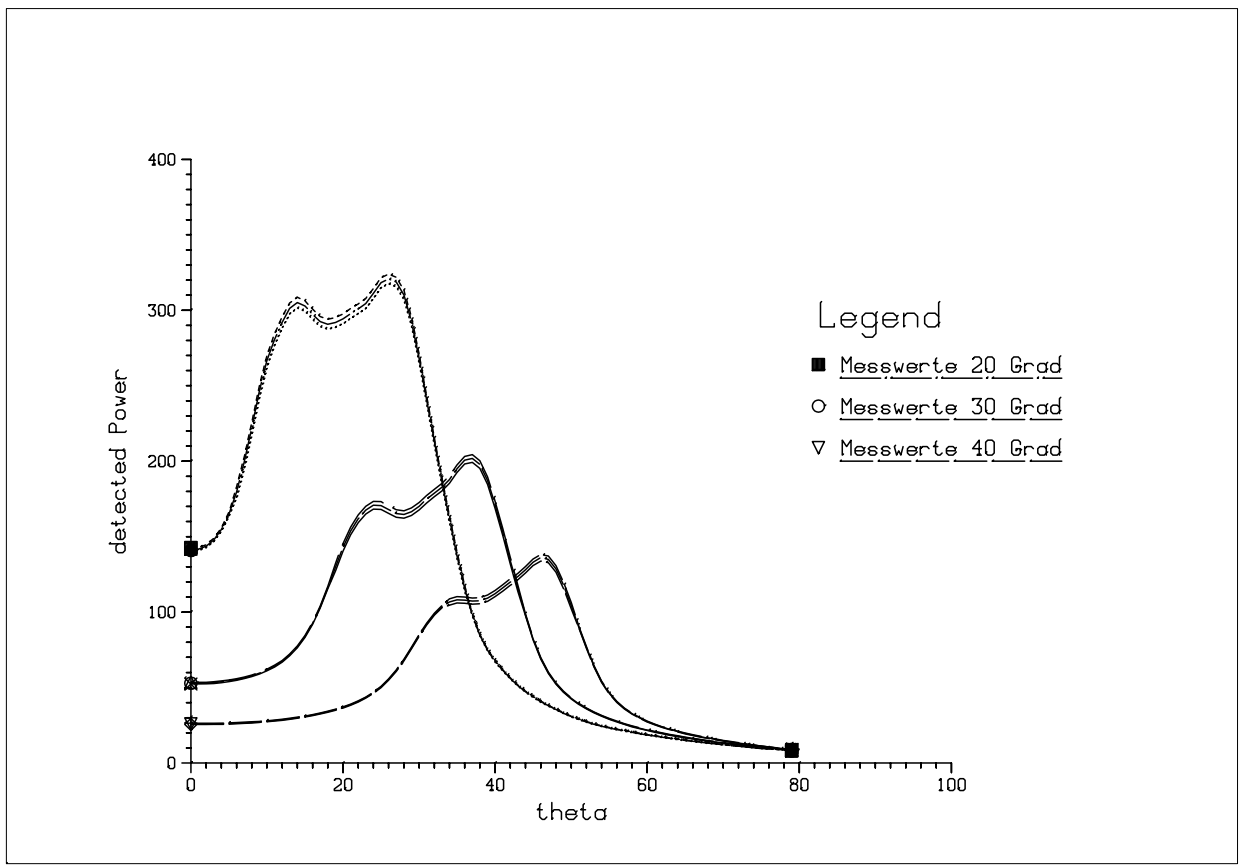}
\caption{\label{wsm3}Measurement data for a set of incident angles, sample {\bf waxy}}
\end{center}
\end{figure}

Sample {\bf wa0301} scatters more diffusely. The transmission is 80\% of {\bf wa3101}. These two measurement curves were
used to fit the TI honeycomb model to (See \ref{ti-model-plot}). Data for {\bf waxy} contains the shadow from the lamp rod (Fig.~\ref{wsm3}), and
therefore this data set was not used for fitting.

These plots represent a 2D cut through the 3D mountain (See \ref{meas3}). The 2D way of data representation allows easy reading of
values, but does not show the dependency on the outgoing angle $\phi$.

For a comparison between the fitted model and measured data, please refer to Fig.~\ref{wsfit1}, \ref{wsfit2}.

\begin{figure}[h!]
\begin{center}
\input{results/wsm2.pdftex_t}
\caption{\label{wsm2}Measurement data for a set of incident angles, sample {\bf wa0301}}
\end{center}
\end{figure}

\clearpage

\section{Measurement out-of-plane}
\label{meas3}

\subsection{Transmission data}

\label{tdata}
\begin{figure}[h!]
\begin{center}
\includegraphics[height=90mm]{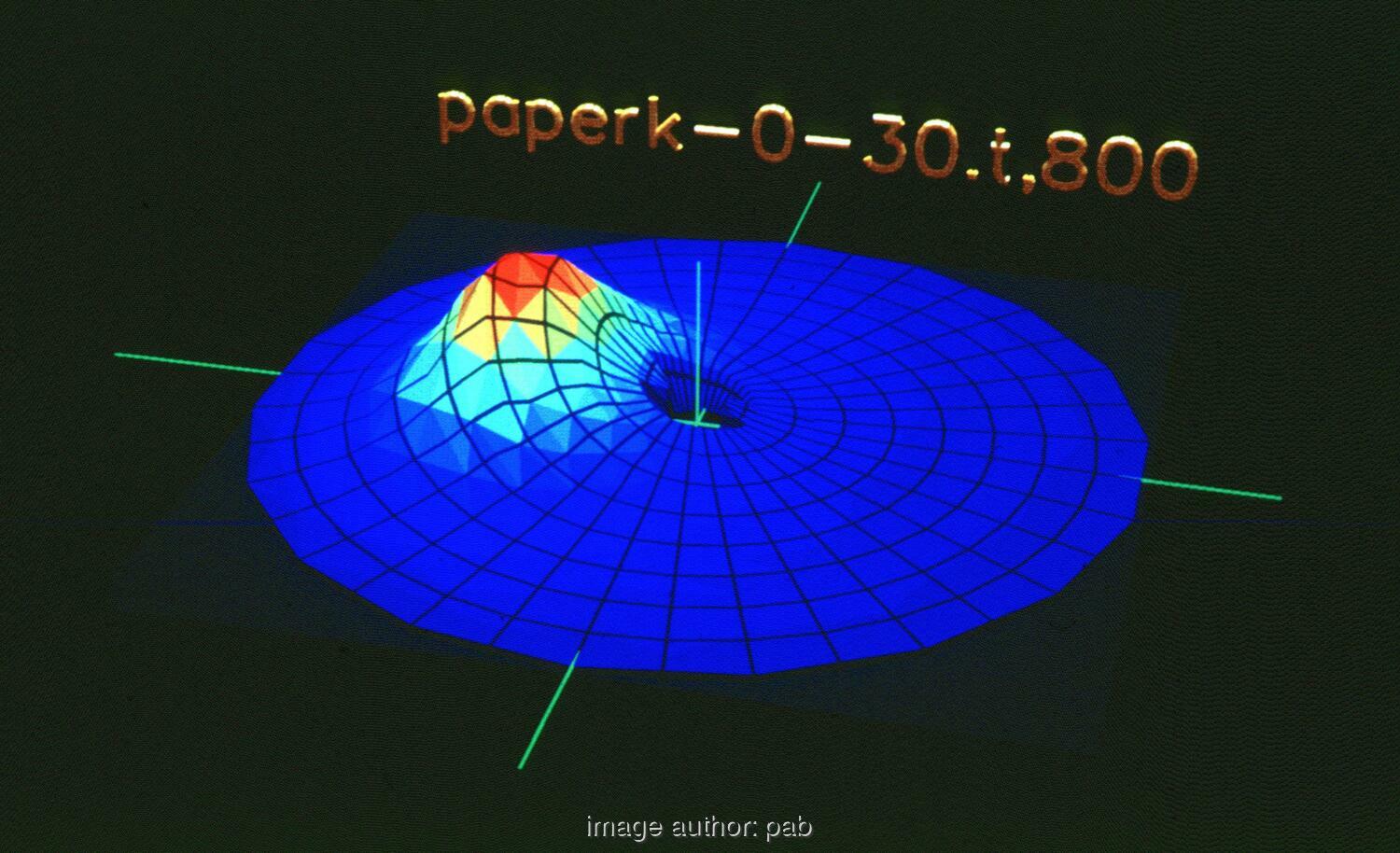}
\caption{\label{rsmountain-paper}Angular transmission of tracing paper, $\theta_{in}=30^o , \phi_{in}=0^o, Z_{max}=800$}
\end{center}
\end{figure}

\begin{figure}[h!p]
\begin{center}
\includegraphics[height=75mm]{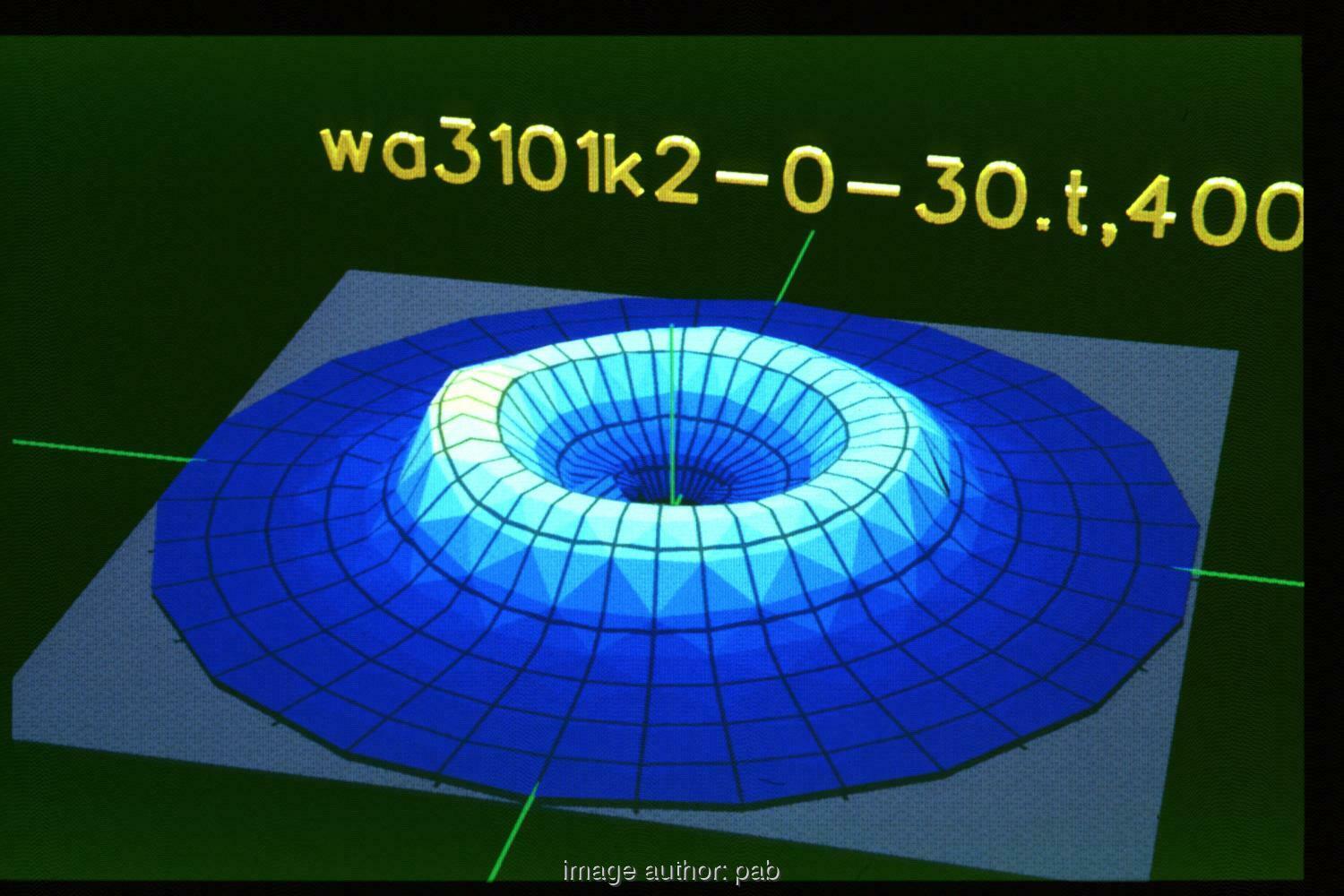}
\caption{\label{rsm-wa3101-30}{\bf wa3101} incident angle $\theta_{in}=30^o, \phi_{in}=0^o$}
\end{center}
\end{figure}

\begin{figure}[h!p]
\begin{center}
\includegraphics[height=75mm]{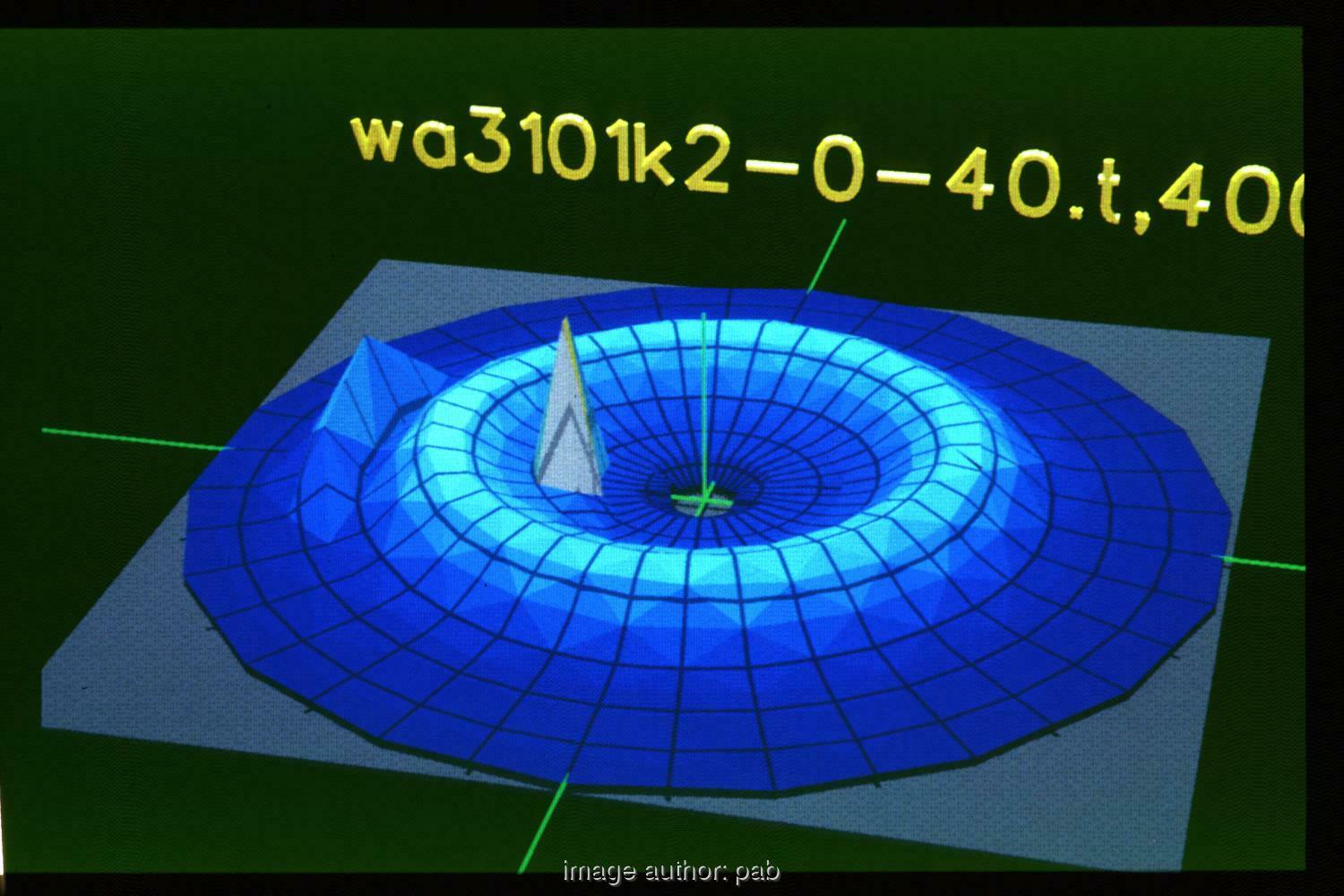}
\caption{\label{rsm-wa3101-40}{\bf wa3101} incident angle $\theta_{in}=40^o, \phi_{in}=0^o$}
\end{center}
\end{figure}

\begin{figure}[h!p]
\begin{center}
\includegraphics[height=75mm]{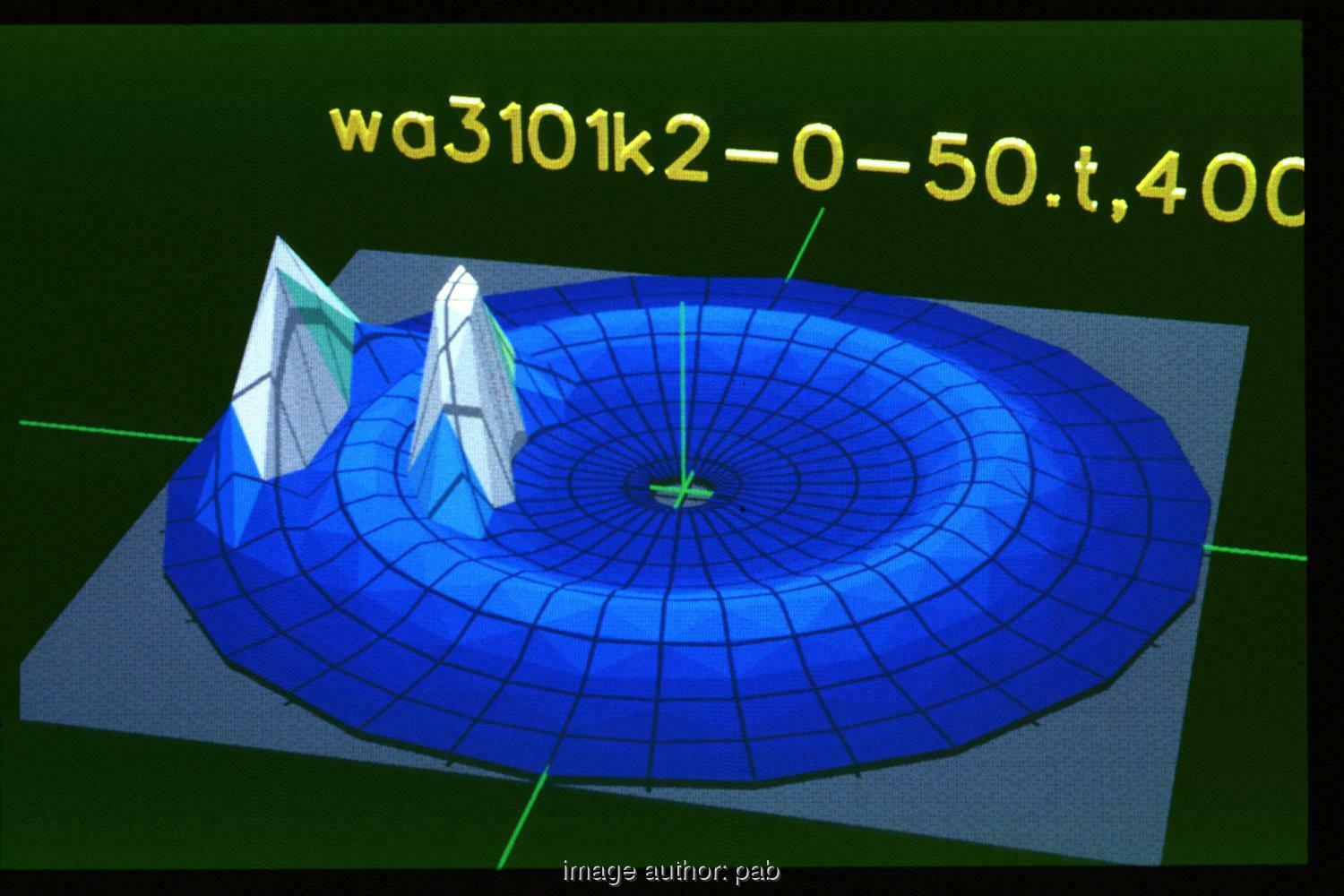}
\caption{\label{rsm-wa3101-50}{\bf wa3101} incident angle $\theta_{in}=50^o, \phi_{in}=0^o$}
\end{center}
\end{figure}

\begin{figure}[h!]
\begin{center}
\includegraphics[height=75mm]{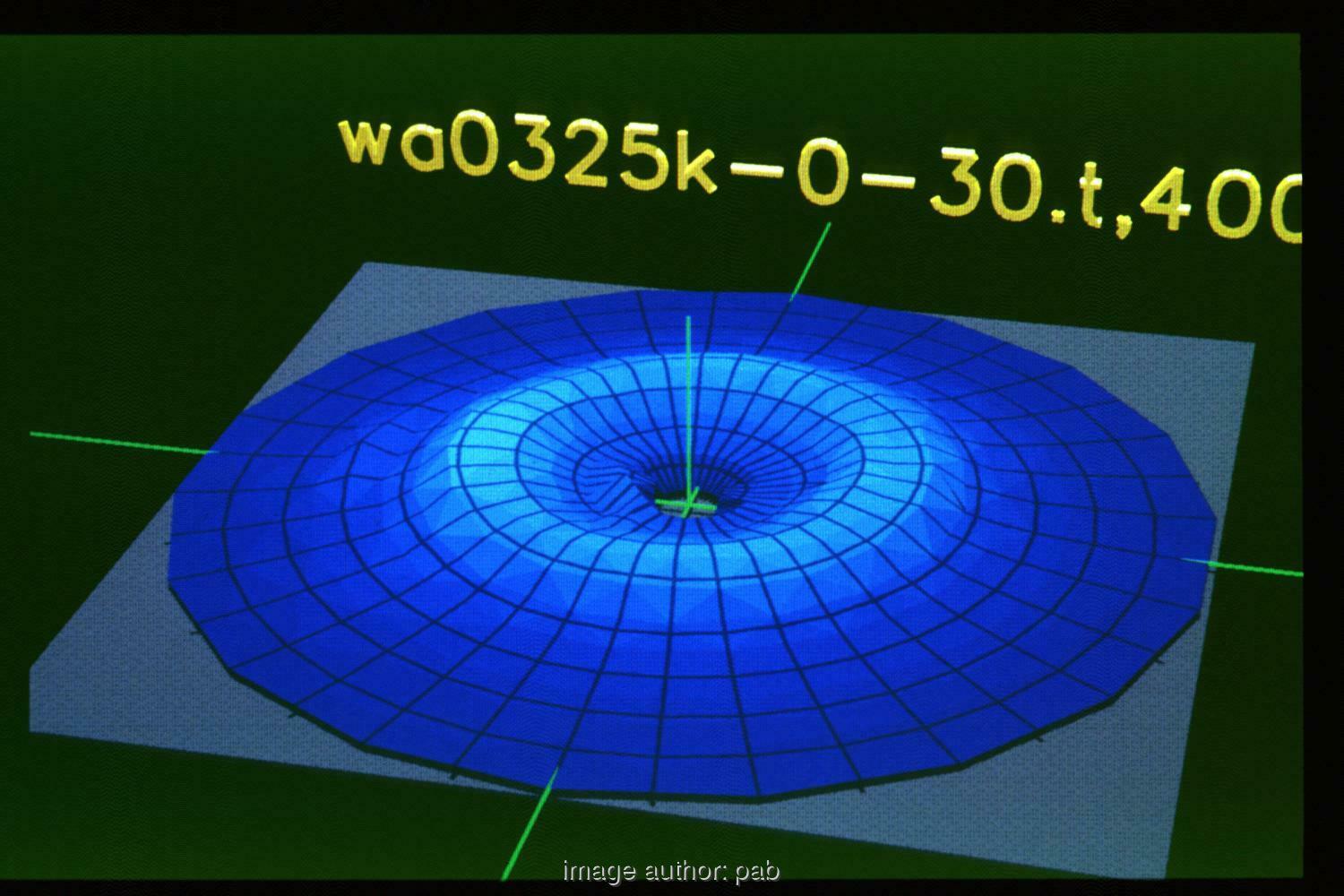}
\caption{\label{rsm-wa0325-30}{\bf wa0325} (glass tubes), less transmission due to reflection off outer panes}
\end{center}
\end{figure}

\begin{figure}[h!]
\begin{center}
\includegraphics[height=75mm]{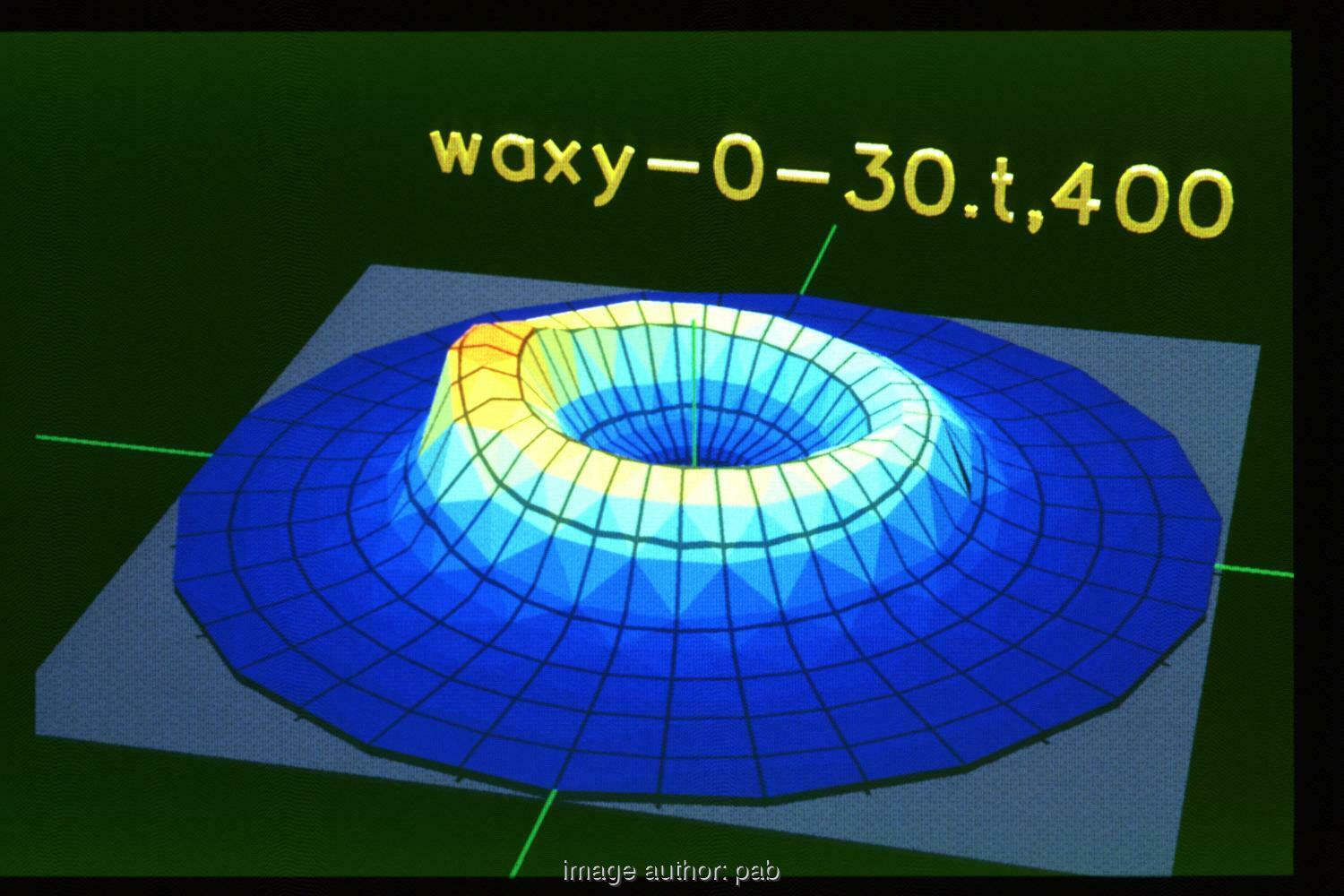}
\caption{\label{rsm-waxy-30}{\bf waxy} irregular plastic cylinders, forward peak due to unscattered part}
\end{center}
\end{figure}

\begin{figure}[h!]
\begin{center}
\includegraphics[height=75mm]{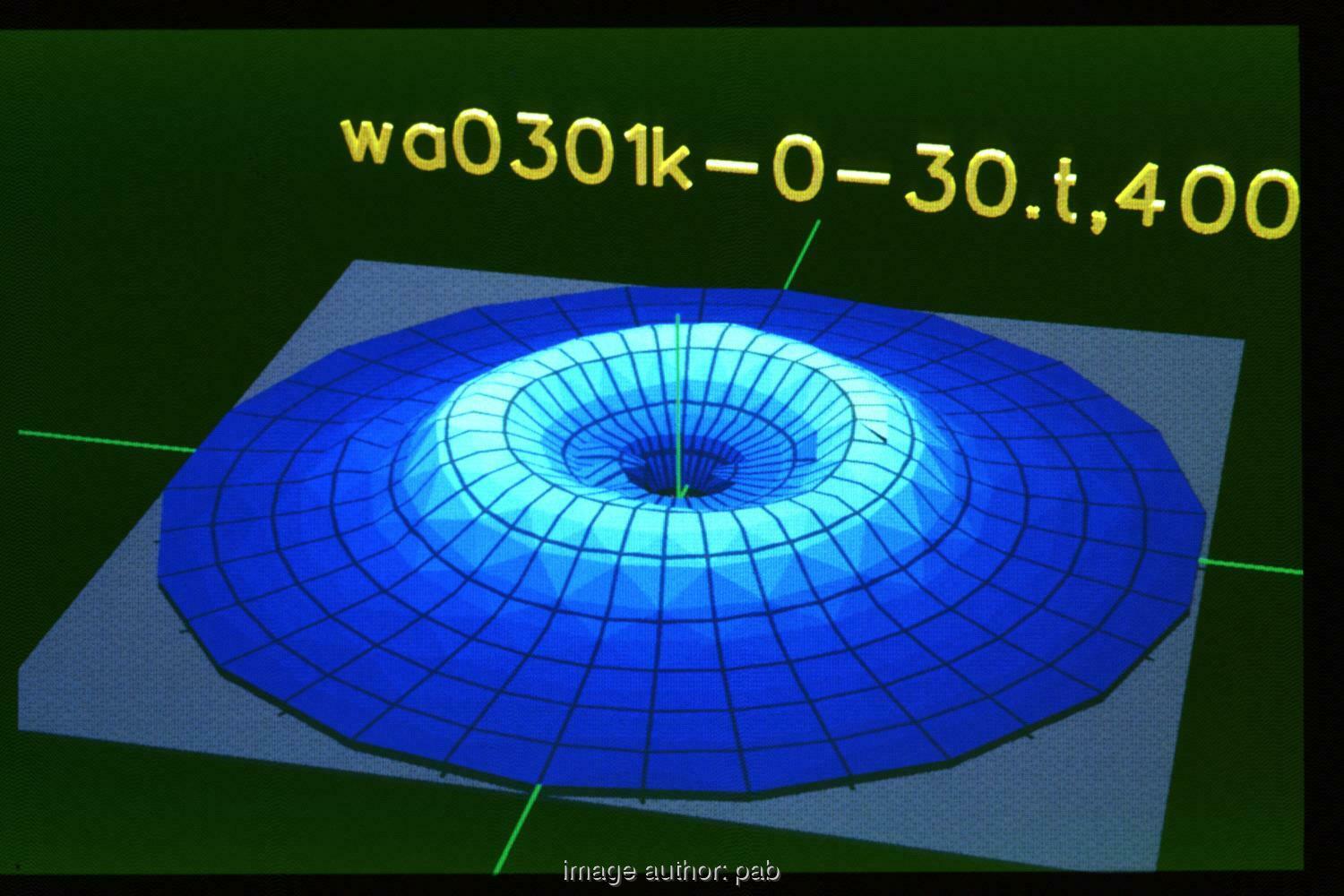}
\caption{\label{rsm-wa0301-30}{\bf wa0301} plastic cylinders, melted together at the surfaces}
\end{center}
\end{figure}

\begin{figure}[h!]
\begin{center}
\includegraphics[height=75mm]{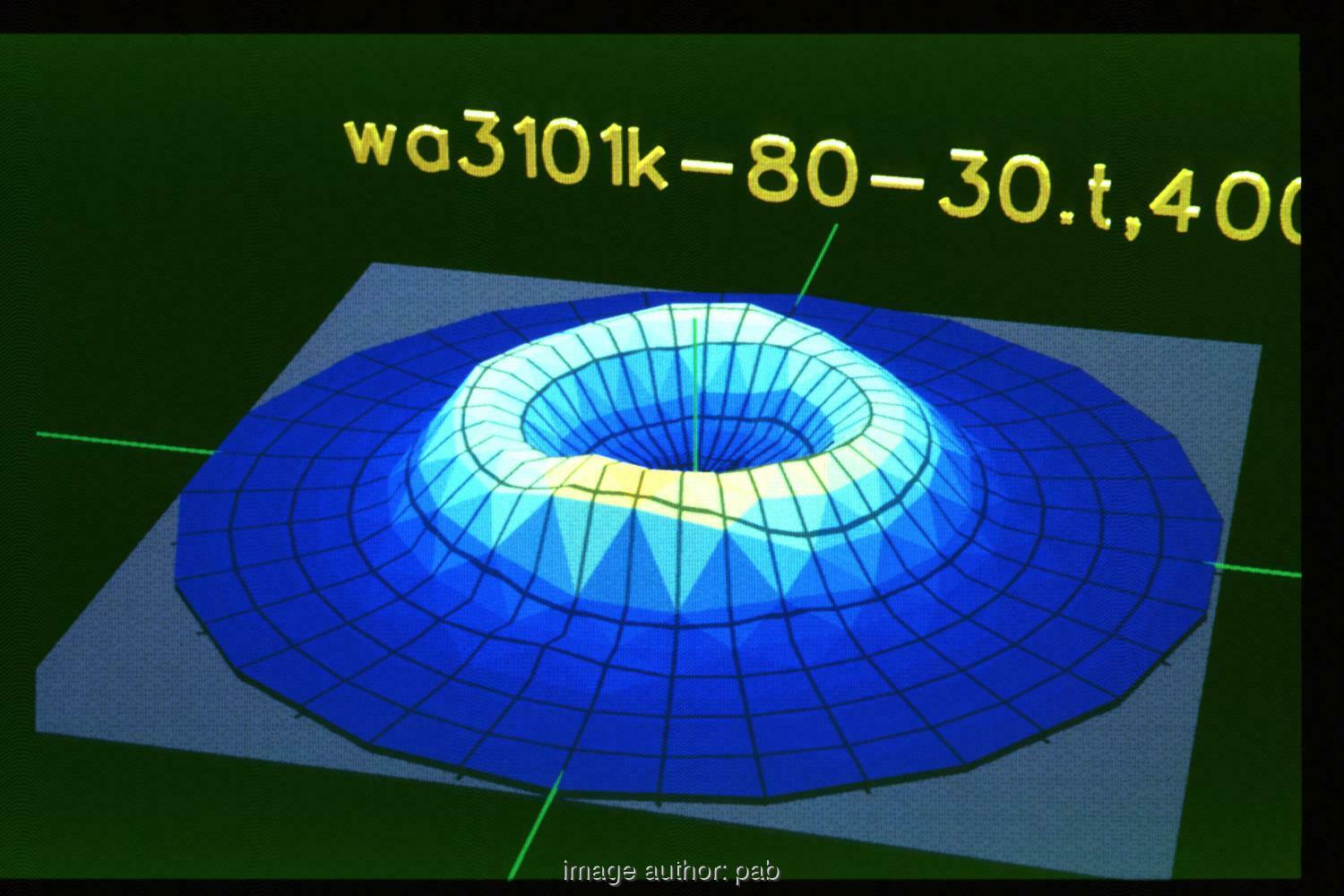}
\caption{\label{rsm-wa3101-p80}Sensitivity on $\phi_{in}$: {\bf wa3101}, $\phi_{in}=80^o$, ring + peak}
\end{center}
\end{figure}

\begin{figure}[h!]
\begin{center}
\includegraphics[height=75mm]{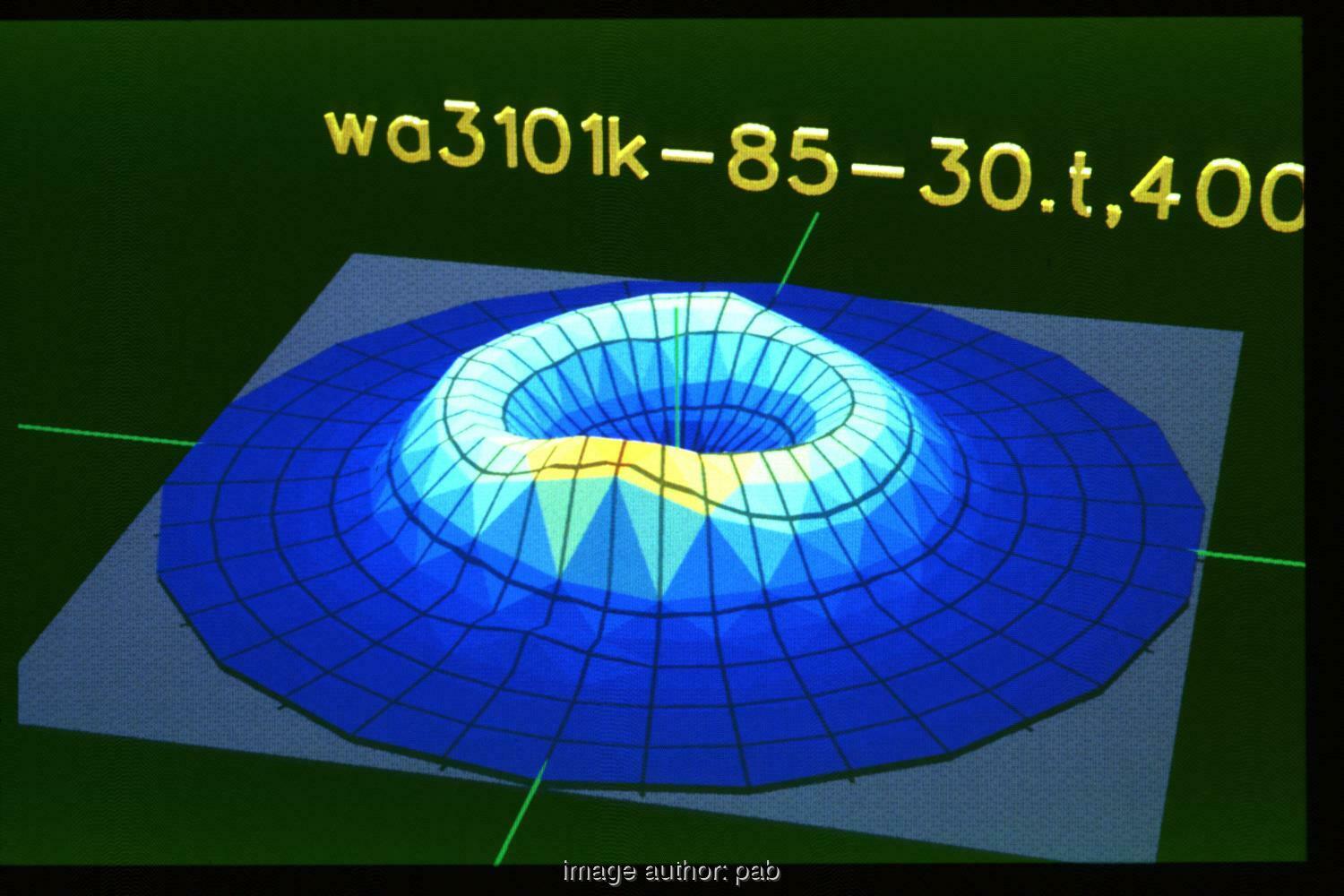}
\caption{\label{rsm-wa3101-p85}{\bf wa3101}, $\phi_{in}=85^o$, increasing forward peak}
\end{center}
\end{figure}

\begin{figure}[h!]
\begin{center}
\includegraphics[height=75mm]{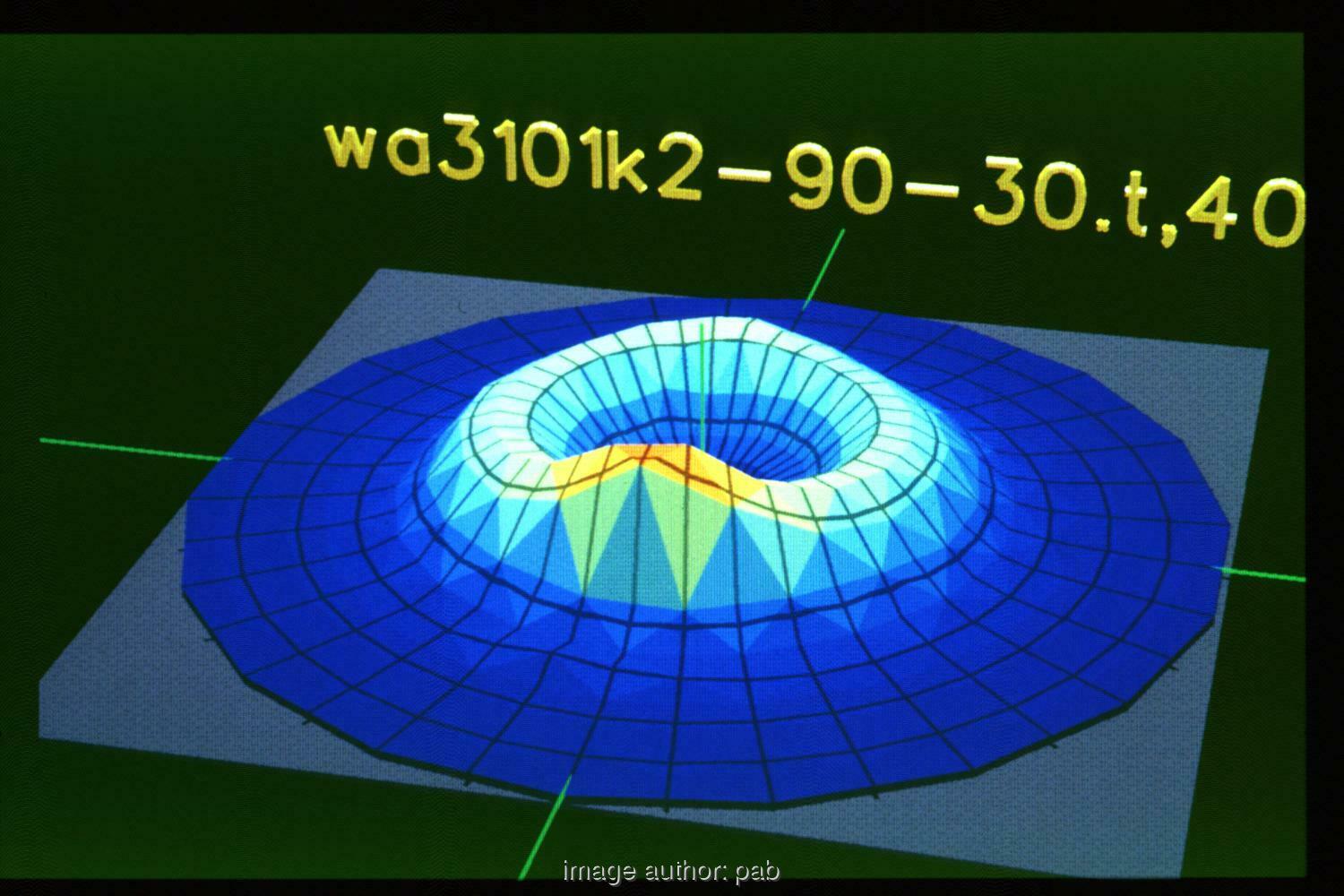}
\caption{\label{rsm-wa3101-p90}{\bf wa3101}, $\phi_{in}=90^o$, maximum forward peak}
\end{center}
\end{figure}

\begin{figure}[h!]
\begin{center}
\includegraphics[height=75mm]{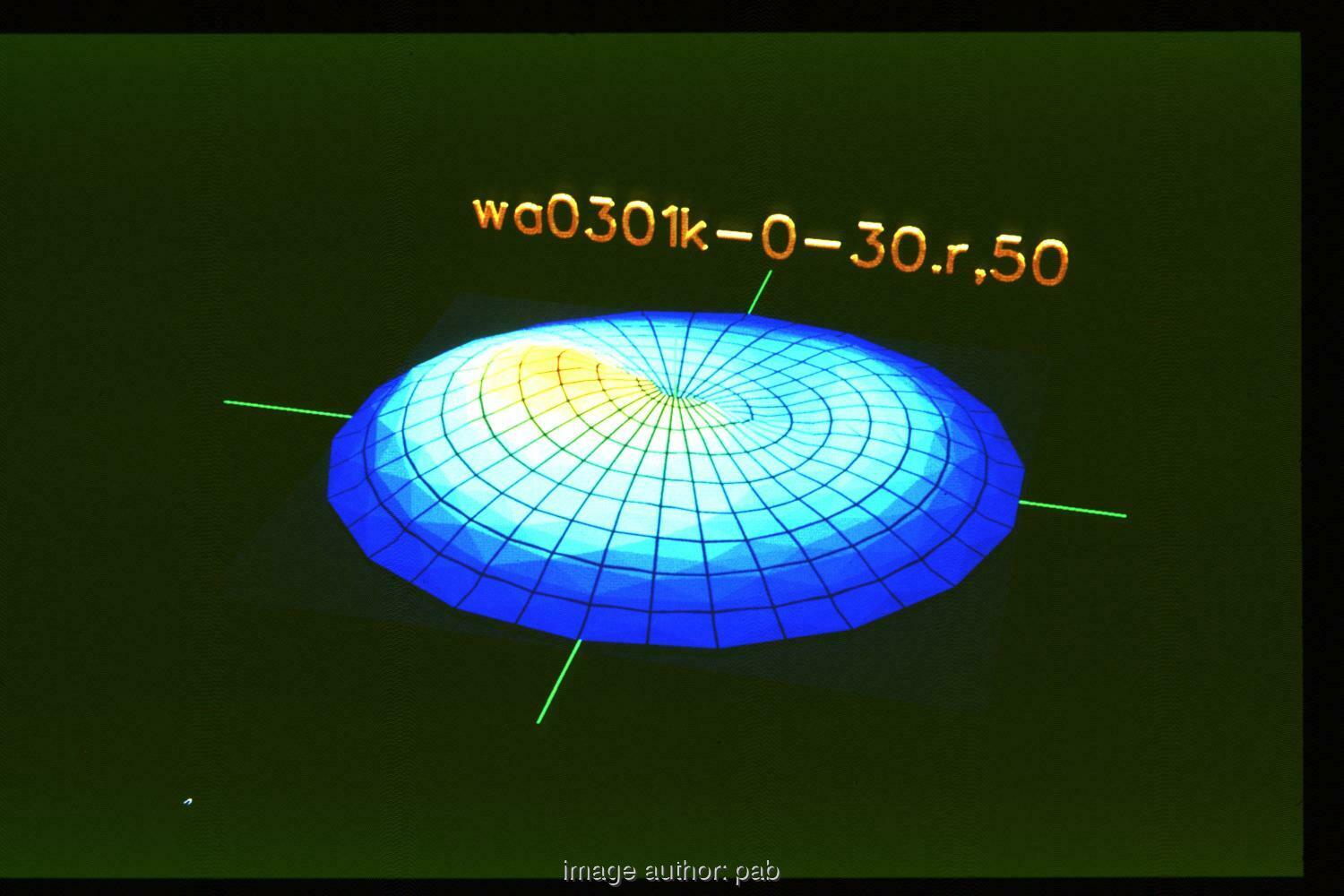}
\caption{\label{rsm-R-wa0301-30}{\bf wa0301}, $\theta_{in}=30^o$, back-scatter due to the melted surface, $Z_{max}=50$}
\end{center}
\end{figure}

\begin{figure}[h!]
\begin{center}
\includegraphics[height=75mm]{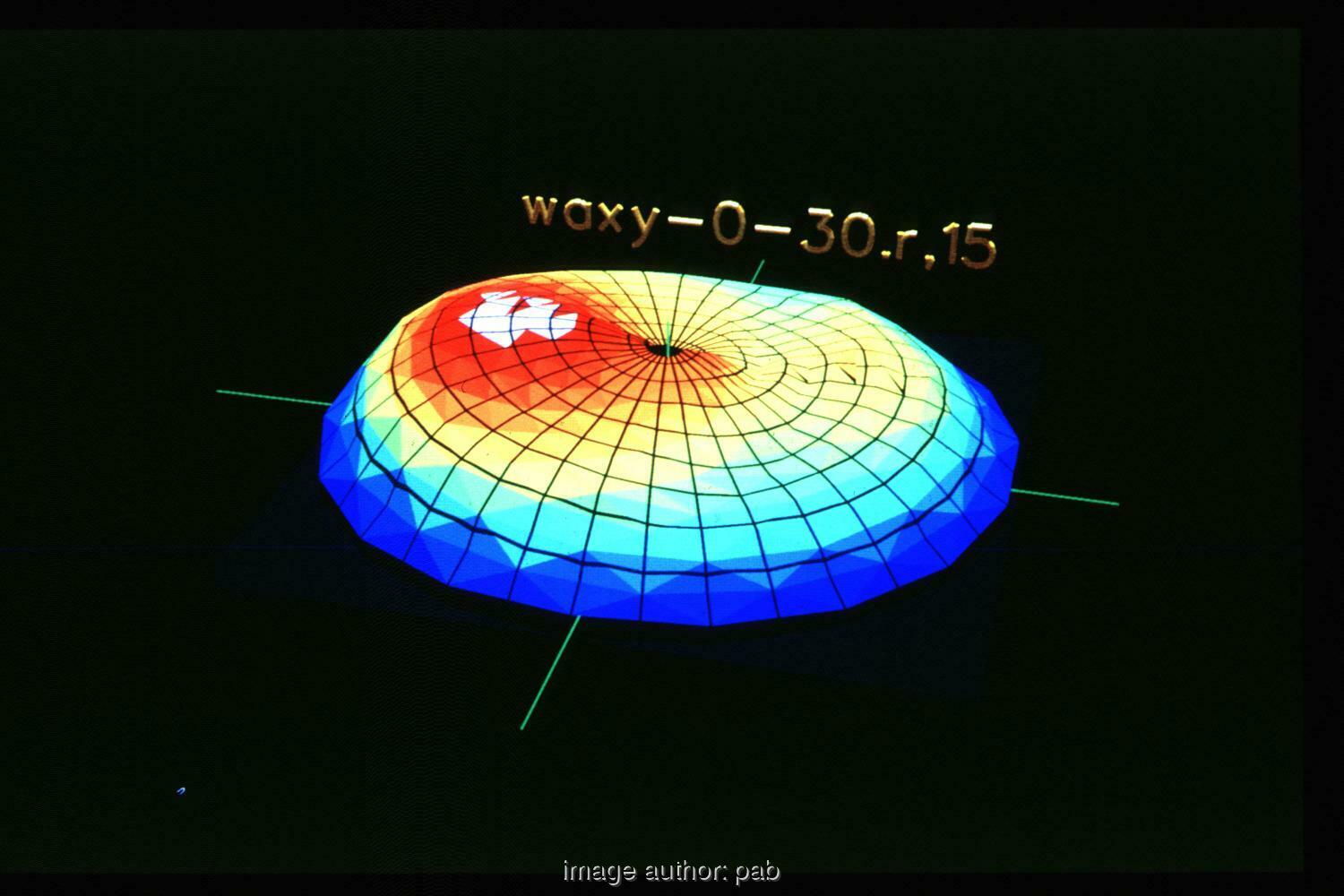}
\caption{\label{rsm-R-waxy-30}{\bf waxy}, $\theta_{in}=30^o$, less back-scatter, $Z_{max}=15$}
\end{center}
\end{figure}

\begin{figure}[h!]
\begin{center}
\includegraphics[height=75mm]{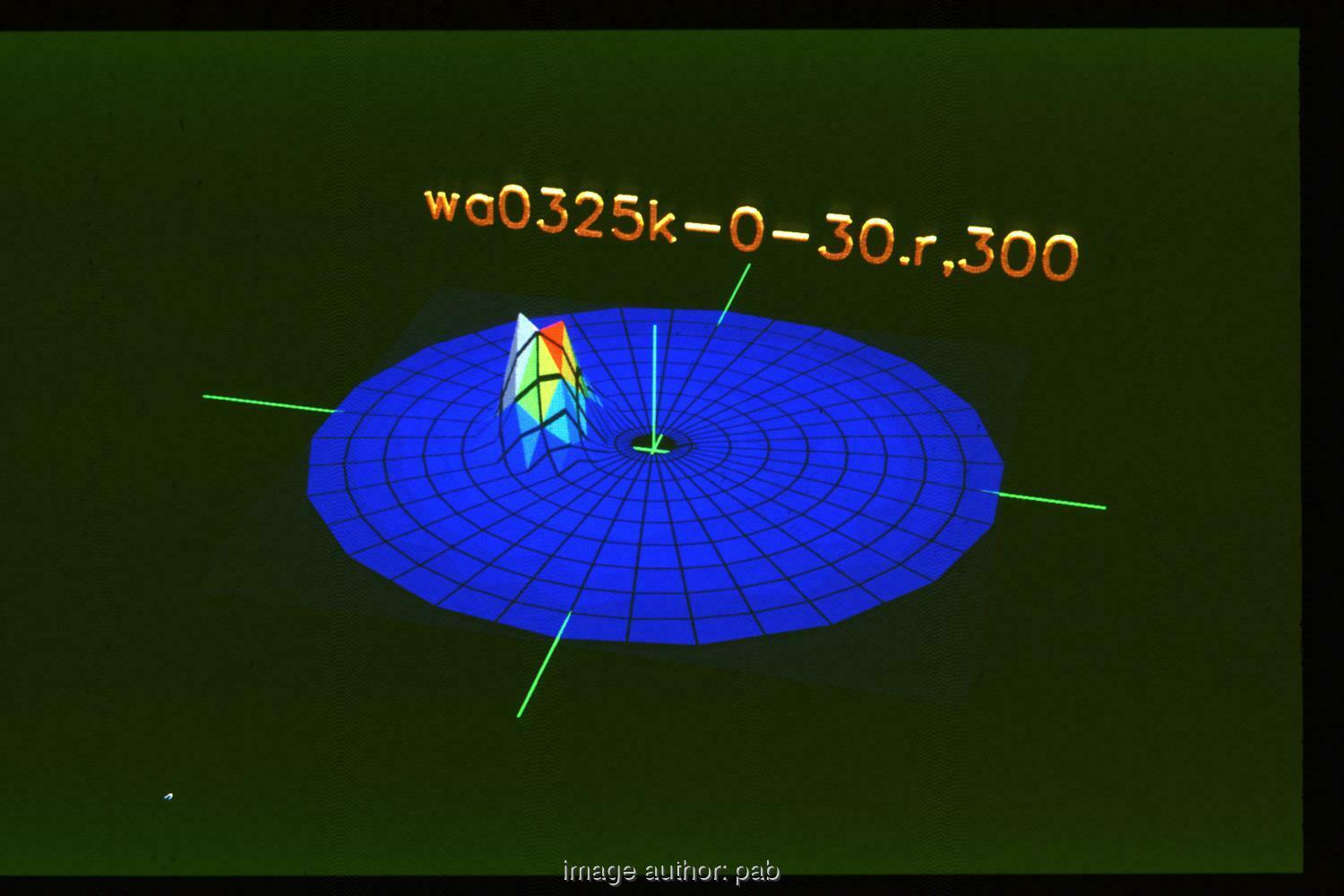}
\caption{\label{rsm-R-wa0325-30}{\bf wa0325}, $\theta_{in}=30^o$, directional reflection off the glass pane, $Z_{max}=300$}
\end{center}
\end{figure}

The following coordinate system ($\theta,\phi$) is used for the visualisations in Fig.\ref{rendering1} - Fig.\ref{rsm-R-wa0325-30}:
$\phi=0^o$ is at the green line at the right hand side (X-axis), and increases clockwise (as seen from above), with radial lines every
$\Delta\phi=10^o$.  $\theta=90^o$ is at the perimeter, $\theta=180^o$ is at the centre, with concentric circles in $\Delta\theta=10^o$
steps. The maximum of the linear Z-scale is given as the last value in the title text (Zmax). The incident direction and the sample name are
included in the title as ''sample-$\phi$-$\theta$''.

Images \ref{rsm-wa3101-30}, \ref{rsm-wa3101-40}, \ref{rsm-wa3101-50} show the angular distribution of a typical TI-honeycomb ''ring'' for
some incident angles. The ''ring'' gets larger and weaker with increasing incident angle $\theta_{in}$. The material is {\bf wa3101}.

The next data sets ( \ref{rsm-wa3101-30} , \ref{rsm-wa0325-30}, \ref{rsm-waxy-30}, \ref{rsm-wa0301-30} ) show a comparison between four
materials ({\bf wa3101 wa0325 waxy wa0301}) at $\theta_{in}=30^o$.

The material {\bf wa3101} is {\em not} rotationally symmetric, as shown in measurements with varying $\phi_{in}$: Fig.~\ref{rsm-wa3101-p80},
\ref{rsm-wa3101-p85}, \ref{rsm-wa3101-p90} . This is caused by its structure being glued together from batches, leading to periodic thicker
walls (see introduction in \ref{meas-intro}). Thicker walls cause broader light scattering orthogonal to the walls. The model of TI
scattering has to be extended, to match {\bf wa3101} fully. These results confirm the importance of 3D visualisation of the measured data to
provide useful information.

As a starting point, the following image shows scattering of a non-diffuse, translucent tracing paper: Transmission (Fig.\ref{rsmountain-paper})
shows a clear ''peak'' of forward scattering at $\theta=150^o, \phi=180^o$.

\clearpage

\subsection{Reflexion data}

\label{rdata}
In these visualisations $\theta=0^o$ is at the centre. Otherwise the coordinates are the same as in \ref{tdata}.

\begin{figure}[h!]
\begin{center}
\includegraphics[height=85mm]{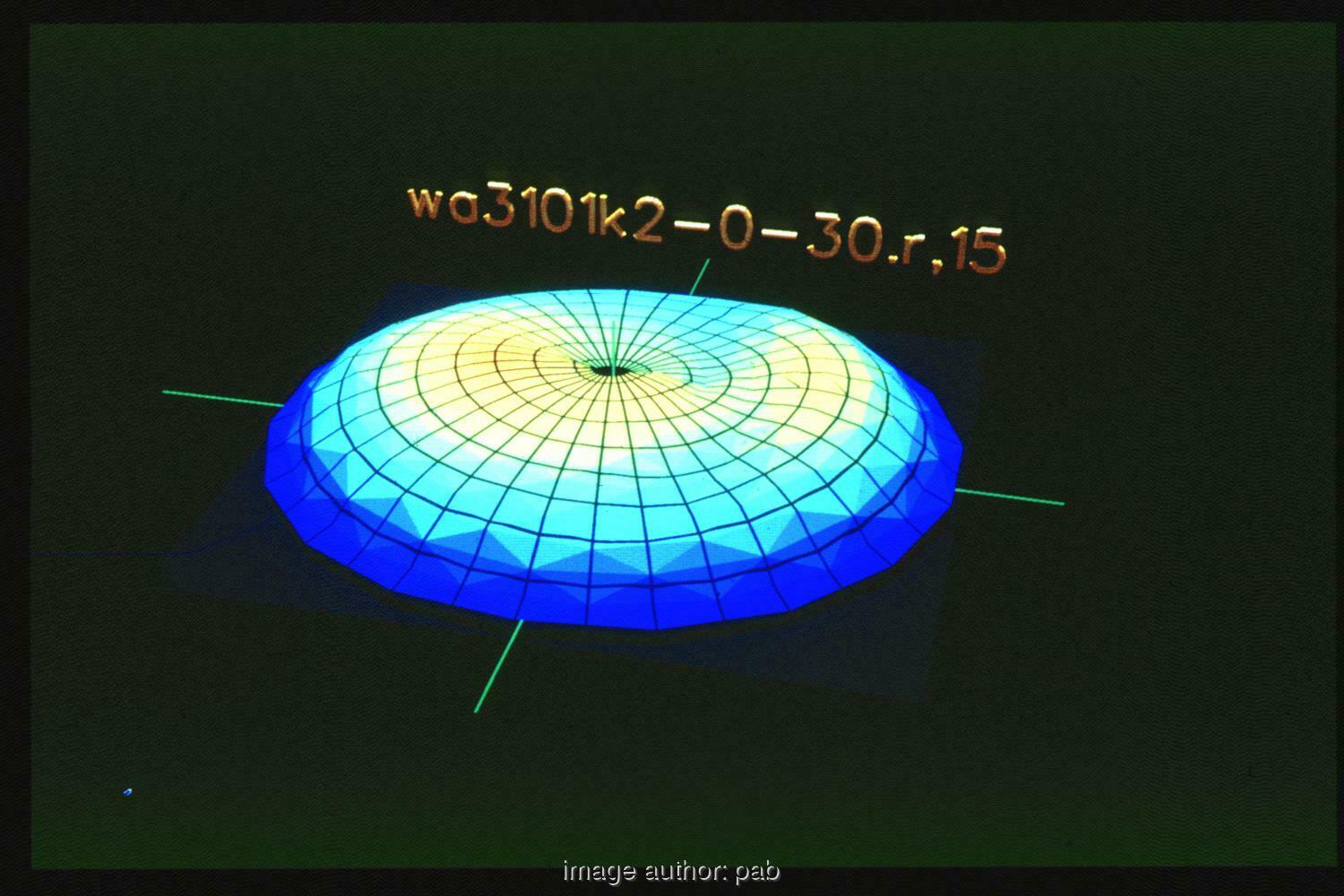}
\caption{\label{rsm-R-wa3101-30}{\bf wa3101}, $\theta_{in}=30^o$, $Z_{max}=15$}
\end{center}
\end{figure}

\begin{figure}[h!]
\begin{center}
\includegraphics[height=85mm]{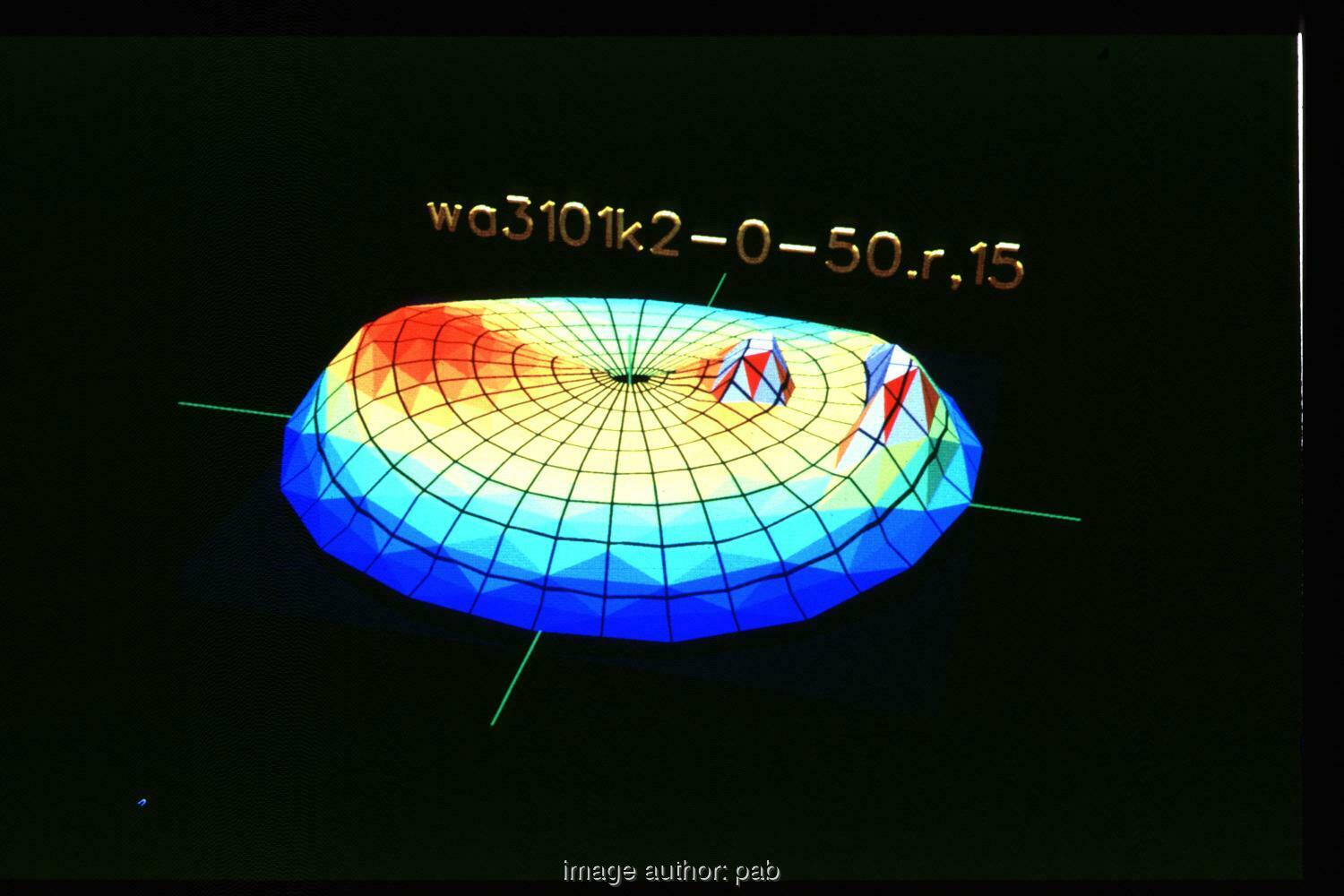}
\caption{\label{rsm-R-wa3101-50}{\bf wa3101}, $\theta_{in}=50^o$, increasing directional reflexion, reflections off second mirror cause
two extra ''peaks'''}
\end{center}
\end{figure}

Fig.~\ref{rsm-R-wa3101-30} and \ref{rsm-R-wa3101-50} show the increase in reflexion with larger $\theta_{in}$ for material {\bf wa3101}.
With the usual law of reflection (incident direction = outgoing direction), the maximum is at $\theta_{out}= 30^o , 50^o$ and
$\phi_{out}=180^o$.

The final three diagrams show a comparison of the reflexion for three materials for $\theta_{in}=30^o, \phi=0^o$ .

\chapter{Error Discussion}

\section{Misalignment of mechanical components}
\label{errors-mech}

The device features five axes, which should all intersect in one point. The misalignment errors are:
\begin{enumerate}
\itemsep0em
\item	parallel offset between axes, two axes have a closest distance $a$
\item	angular offset, the angle $\alpha$ between the axes is not the ideal ($0^o$ or $90^o$)
\end{enumerate}

The get an estimate for $a$, a large wire cross is mounted in the sample mount: The point of intersection between the two wires is at the
sample centre and, a cross-check, should remain fixed in space under rotation of the sample mount.
The centre point should also be invariant with rotation around the vertical axis. The deviation was found to be smaller than 3 mm. For these
two axis, $a<3 mm$ was found to be neglectable in the practice.

The detector axis, $a$ was checked using the CCD camera: A black square piece of paper (1 cm size) was affixed at the sample centre and verified
to remain central on the CCD image. No deviation was found, and $a<1 cm$ was concluded.
\new{Looking back, a bit coarse.}

The estimate for $\alpha$ has not been estimated for orthogonal axis yet. By machining tolerances,  $\alpha < 1^o$ seems a reasonable boundary.

\section{Precision of position}
\label{errors-pos}

The unit conversion (steps per angular degree) of the stepper-motor drives is determined based on the gear ratio between motor-axis and
driven axis. This is subject to 3 errors:
\begin{enumerate}
\itemsep0em

\item	The positional error of the motor axis is specified as $\pm3'$ by the manufacturer of the stepper motors. With a gear ratio of
10:1, the positional error of the driven axis is smaller than $\pm0.3'$.

\item	Non-ideal tooth-pitch of the timing wheel and belt and a non-concentric bore at the gear-wheels. This error depends on the position of the
motor axis and is not possible to measure without an absolute reference.

\item	Backlash in the timing-belt gear, since the tension in the timing-belt is not infinitely large. This can be tested easily by moving
to the same position from two sides. For the vertical axis, this was tested using a mirror as sample and HeNe laser-beam: The same position
was moved to from a starting position by $1^o$ on either side. The largest difference of the reflected laser spot is 11 mm, which, at a distance
between screen and mirror of 7.5 m, gives an angular backlash of $2.5'$ (of the axis, in [arc-minutes]). \label{backlash}

\item	Errors in the ''zero-mark'' detectors. While testing \ref{backlash}, fluctuations where found as $5'$. To check this: The distance
between the small light-barrier and the centre-of-rotation is about 15 cm, so $5'$ correspondents to 0.2 mm, which is a realistic value.

\end{enumerate}

The mechanical error in position is hence estimated to be 5\% of the smallest angular grid that was used ($2^o$).

\section{Precision of Keithley picoammeter}
\label{errors-sig}

\begin{figure}[h!]
\begin{center}
\input{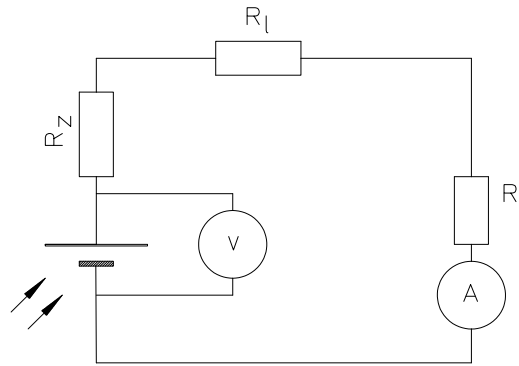}
\caption{\label{keith}Schematic of Keithley-485 picoammeter, Si cell and wiring resistance}
\end{center}
\end{figure}

Fig.~\ref{keith} shows the electrical wiring between the Si solar cell and the Keithley-485 picoammeter.

There are two errors in the signal measurement:
Firstly the difference between the measured value of the short-circuit current and
the ''true'' current, and secondly the digitisation error of the picoammeter itself. For the first error type, the characteristic of a solar cell is
relevant (Fig.~\ref{cell}):

\begin{figure}[h!]
\begin{center}
\input{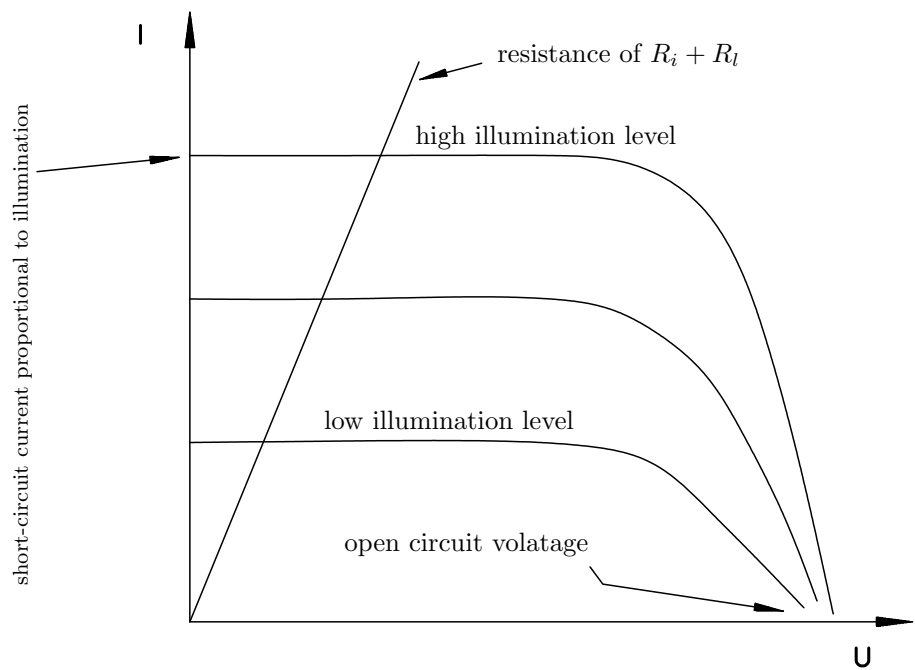}
\caption{\label{cell}Characteristic of a silicon cell, with the incident illumination as parameter}
\end{center}
\end{figure}

The short-circuit current of a solar cell is proportional to the irradiance on its surface. Due to the wiring resistance $R_l$ and the
inner resistance $R_i$ of the cell however, the measured current lies further to right on the cell characteristic (Fig.~\ref{cell}).

\vfill
\pagebreak[10]

Instead of its inner resistance, Keithley specifies an {\em input voltage burden} of $U_{ivb}=200\, uV$, which is the voltage drop at the input.
Thus, the following relation follows for linearity between incident irradiance and measured current:
\begin{displaymath}
	I \; R_l + U_{ivb} < U_l
\end{displaymath}
Where $I$ is measured current, $R_l$ the resistance, $U_{ivb}$ the voltage drop at instrument and $U_l$ the open circuit voltage of the cell.

Resistance of the wiring was measured before the cell was added: $R_l = 2.0 \pm 0.2 \Omega$. $U_l$ and $I$ were measured for to irradiance
levels:

\begin{tabular}{lll}
$I$			&$U_l$			&$I \; R_l + U_{ivb}$\\
$1.0227\pm0.002mA$	&$375.0\pm0.2mV$	&$2.2mV$\\
$107.40\pm0.1uA$	&$190.40\pm0.1mV$	&$0.4mV$\\
\end{tabular}

The voltage drop along the wiring and the inner resistance of the ammeter is less than 1\% of the open-circuit voltage of the cell. Hence the
linearity of the measured current to the incident irradiation was considered to be sufficient.

Measurement errors of the Keithley-485 picoammeter are specified as:

\begin{tabular}{lll}
range		&error		&maximum reading\\
$10uA$		&$\pm0.5\%$	&$19.99\pm0.03uA$\\
$100uA$		&$\pm0.5\%$	&$199.9\pm0.3uA$\\
$1mA$		&$\pm0.5\%$	&$1.999\pm0.003mA$\\
\end{tabular}

Thus, for typical measurements, the errors are:

\begin{tabular}{llll}
value	&range	&error abs	&error relative\\
21uA	&100uA	&0.4uA		&2\%\\
190uA	&100uA	&1.25uA		&0.6\%\\
\end{tabular}

\vfill
\pagebreak[3]

\chapter{Summary}

To research the angular scattering of new facade materials, the described device was designed and built. This included the mechanical
design, writing the control software, data acquisition and visualisation of the resulting data. New ideas included the simulation of transparent
insulation and the visualisation of measurement with the aid of methods borrowed from computer-graphics. Results show some previous unknown
details (the asymmetric scattering of {\bf wa3101} and the very low scattering of massive aerogel blocs).

Data generated by this device can be used to improve models on the radiation transport within transparent insulation material (TI). Absolute
measurements in the future will allow to measure absorption inside the sample. The results allow the simulation of indoor spaces that are
partly or fully illuminated by TI facade material. Two methods are suggested for this and a simple model of TI scattering is presented.

Regarding technical details, there were a number of firsts at the institute: The first use of the HP ME10 mechanical CAD program,
the first UNIX workstation to control an experiment, the first experiment to use VME bus and 5-phase stepper motors.

First results were published in \cite{platzer:90}.

Questions, suggestions, questions regarding the source-code are welcome and should be addressed to {\tt apian@ise.fhg.de}.

\new{Above email is obsolete since 2007, please feel free to contact me at \href{mailto:info@pab-opto.de}{info@pab-opto.de}} .

\vfill

\begin{appendix}
\chapter{3D interpolation of scattered measurement data}
\label{interpol-problem}

Given are data-points of the type $z_i=f(x_i,y_i)$. If the $x_i,y_i$ are not equidistant, points on a regular grid could be constructed by
{\em averaging}. However, this shows the expected results only if the grid is coarser then the original $x_i,y_i$ distribution of points.
If the grid is finer, the resulting averaged data shows artifacts (non existing steps and edges) that are not contained in the original
data.

An interpolation of this data is more difficult than in the 2D case $y_i=f(x_i)$, since a set of data-points is {\em not defining a surface
uniquely}. Even 4 simple points can lead to two different surfaces: Let $f(x,y)$ be sample as:
$f(0,0)=2$, $f(0,10)=1$, $f(10,0)=1$, $f(10,10)=2$, then this allows a saddle, or a ''valley'' between the points. So called ''minimal
surfaces'' are not unique either, since there's no connection between the points defined.

\new{The next work following this diploma work already used Delaunay triangulation, which generates are uniquely defined surface by
adding extra constraints on the triangles. Vertices of its triangles are the measured points, without averaging for visualisation.}

\chapter{Angular encoding of an axis}
\label{ang-encode}

\begin{figure}[h!]
\begin{center}
\fbox{
\input{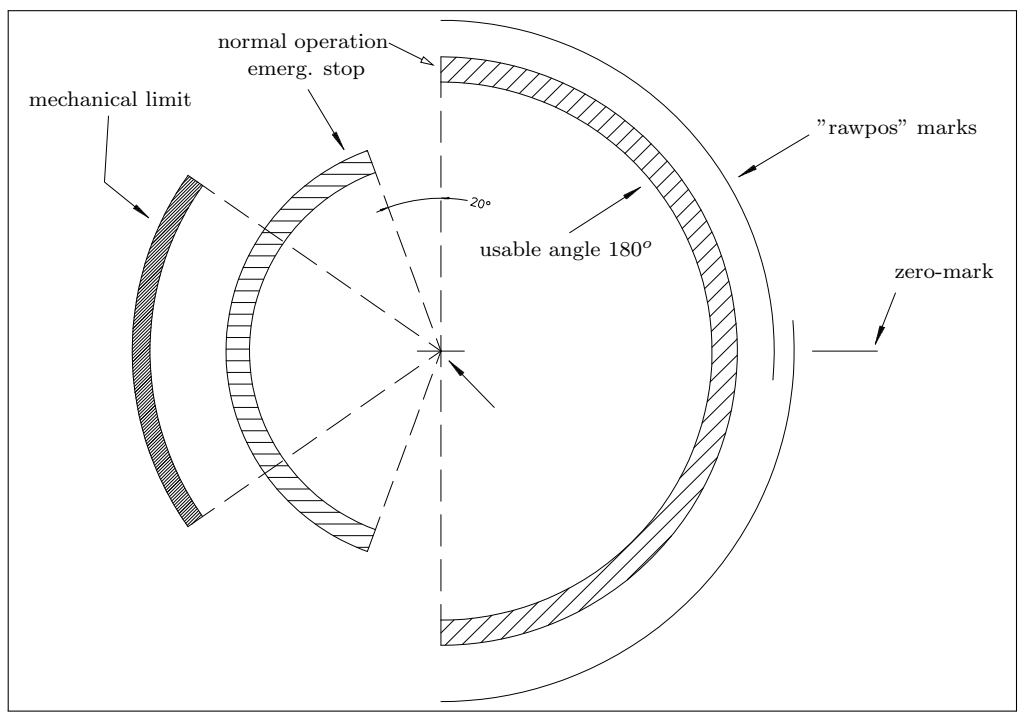}
}
\caption{\label{schwenk}The angular ranges of the detector axis}
\end{center}
\end{figure}

The active scan angular range of the detector axis is defined as $180^o$. A movement to a position outside this area is primarily not
offered by the control program on the UNIX side. Should a malfunction move the axis outside this scan area by more than $20^o$, a hardware
micro-switch interrupts the emergency-stop-loop at the power supply of the drives. The {\em mechanical} limit is given by the steel
''backbone'' of the device, and this limit is protected by the hardware micro-switch circuit. Also shown in the drawing are the location of
the zero-mark and extra ''rawpos'' marks.

\chapter{Motivation of TI model}

The model described in \ref{ti-model} is motivated by two observations:

\begin{enumerate}

\item When only a single cell of TI is illuminated, transmission shows a single small ''ring''. The of the rings for all honeycombs results
in the broad ring.

\item When a HeNe laser beam is pointed inside a rolled tube of aluminium foil (length 10cm, diameter 8mm), the outgoing distribution shows
a ''ring'' whose diameter depends on the incident angle. This tube represents a single cell of the honeycombs.

\end{enumerate}

\bibliographystyle{alpha}
 \bibliography{iselib,lbl,optics,arch,delaunay,sig1}

\chapter{Thanks}

A big ''thank you'' to all who helped:

\begin{itemize}

\item	Professor A. G\"otzberger for assigning this diploma work and for his institute with a good working environment
\item	Dr. V. Wittwer for the trust he put in me and numerous support
\item	P. J\"agle for suggestions and a problem-free institute network
\item	all members of the ISE workshop for the exact mechanical parts and advice
\item	B. Tritsch for fast production of various printed circuit boards (PCB) used in the stepper control
\item	Dr. Papamichael and Dr. Klems (both LBL) for long discussions about their radiometer
\item	my colleagues at the department for the pleasant working environment
\item	the department ''System Technology'' at FhG-ISE for purchasing the ME10 CAD software
\item	my parents for suggestions and morale support

\end{itemize}

\chapter{The ''behind the scene'' special tour}

\setlength{\parindent}{0mm}

\section{Work following diploma thesis 1990}
\label{follow-on}

After completion of my diploma thesis in April 1990, the following thesis covered the topic of BRDF measurements at Fraunhofer ISE,
with my light oversight:
\begin{enumerate}

\item	diploma thesis Thomas Schmidt {\em De-convolution of measured signals}, (near- to far-field conversion, 2D),
	Fraunhofer Institute for Solar Energy and University of Freiburg,
	March 1993
	\begin{enumerate}
	\item in depth analysis of de-convolution of near-field signal, mostly in 2D
	\end{enumerate}

\item	PhD thesis Peter Apian-Bennewitz, {\em Modelling of new window materials for daylight simulation programs} \cite{apian:95},
	Fraunhofer Institute for Solar Energy and University of Freiburg,
	August 1995 (in German, English version coming soon)

\item	diploma thesis Jochen von der Hardt, {\em Enhancing a device for measuring micro-structured materials},
	Fraunhofer Institute for Solar Energy and University of Freiburg,
	1995-1996 (in German)
	\begin{enumerate}
	\item replaced the linear detector rail (coverage of quarter sphere) with a commercial circular rail with radius 1m (half-sphere)
	\item upgraded the data-acquisition by replacing the HP multi-meter with a much faster precision I/U converter
	\item and most importantly: wrote a reliable program for Voronoi-cells on the sphere, which integrated scattered data correctly
	\end{enumerate}

\item	diploma thesis Matthias Braun, {\em De-convolution of measured signals by regularisation},
	Fraunhofer Institute for Solar Energy and University of Freiburg,
	August 1998 (in German)
	\begin{enumerate}
	\item tested large-scale off-axis parabolic mirror machined from solid block of brass
	\item extended the work on de-convolution by Thomas Schmidt to 3D
	\end{enumerate}

\end{enumerate}

\clearpage
\vfill
\pagebreak[10]

\section{TI sample photos of 1989}
\label{ti-photos}

These had not been included in the diploma work. Inexcusable, no scale was included in the pictures when taking them. Nevertheless, images
may be of historical interest.

\begin{figure}[h!]
\begin{center}
\includegraphics[height=10cm]{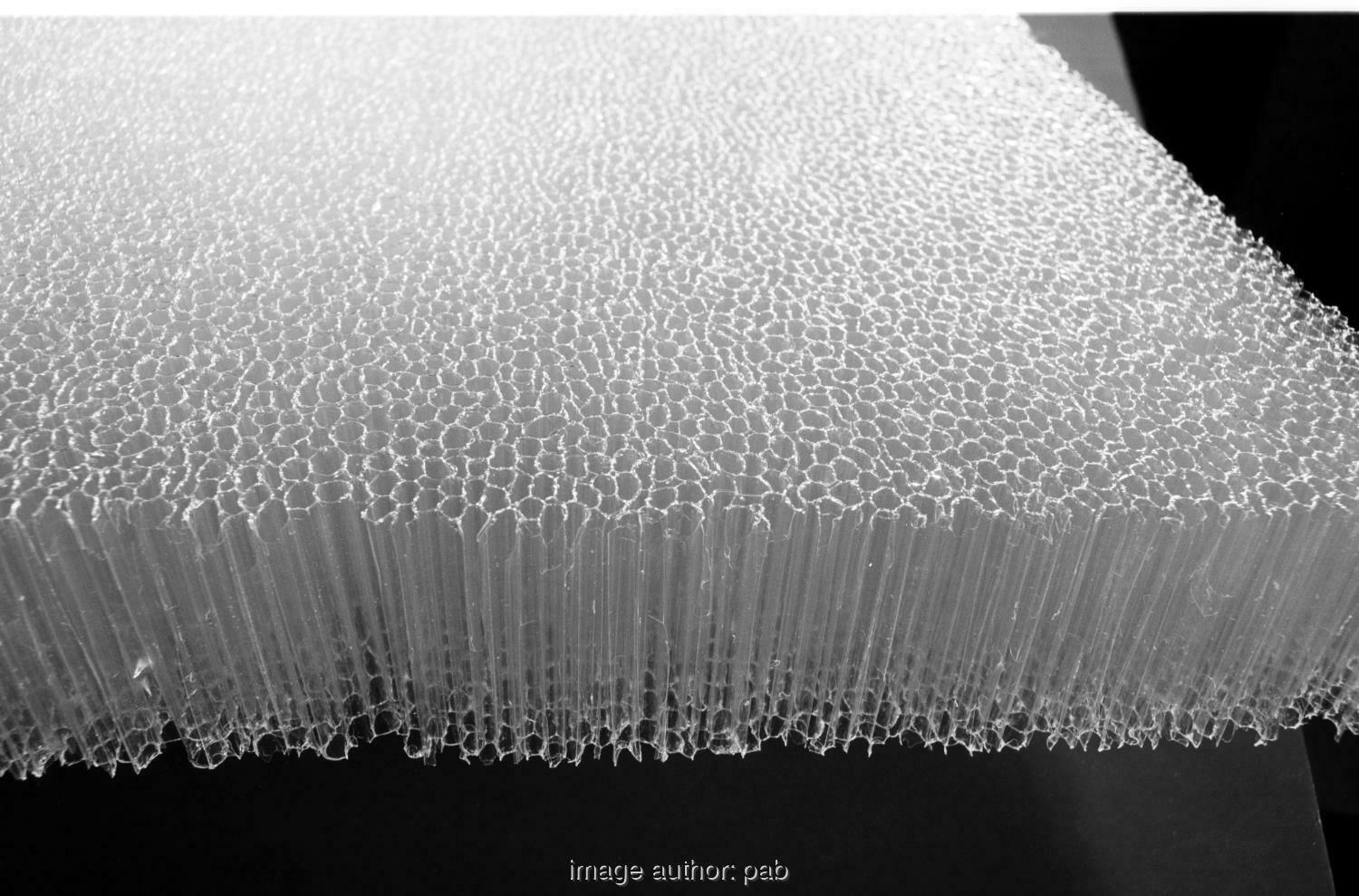}
\caption{\label{photo-waxy}TI material, ISE number {\bf waxy}}
\end{center}
\end{figure}

\begin{figure}[h!]
\begin{center}
\includegraphics[height=10cm]{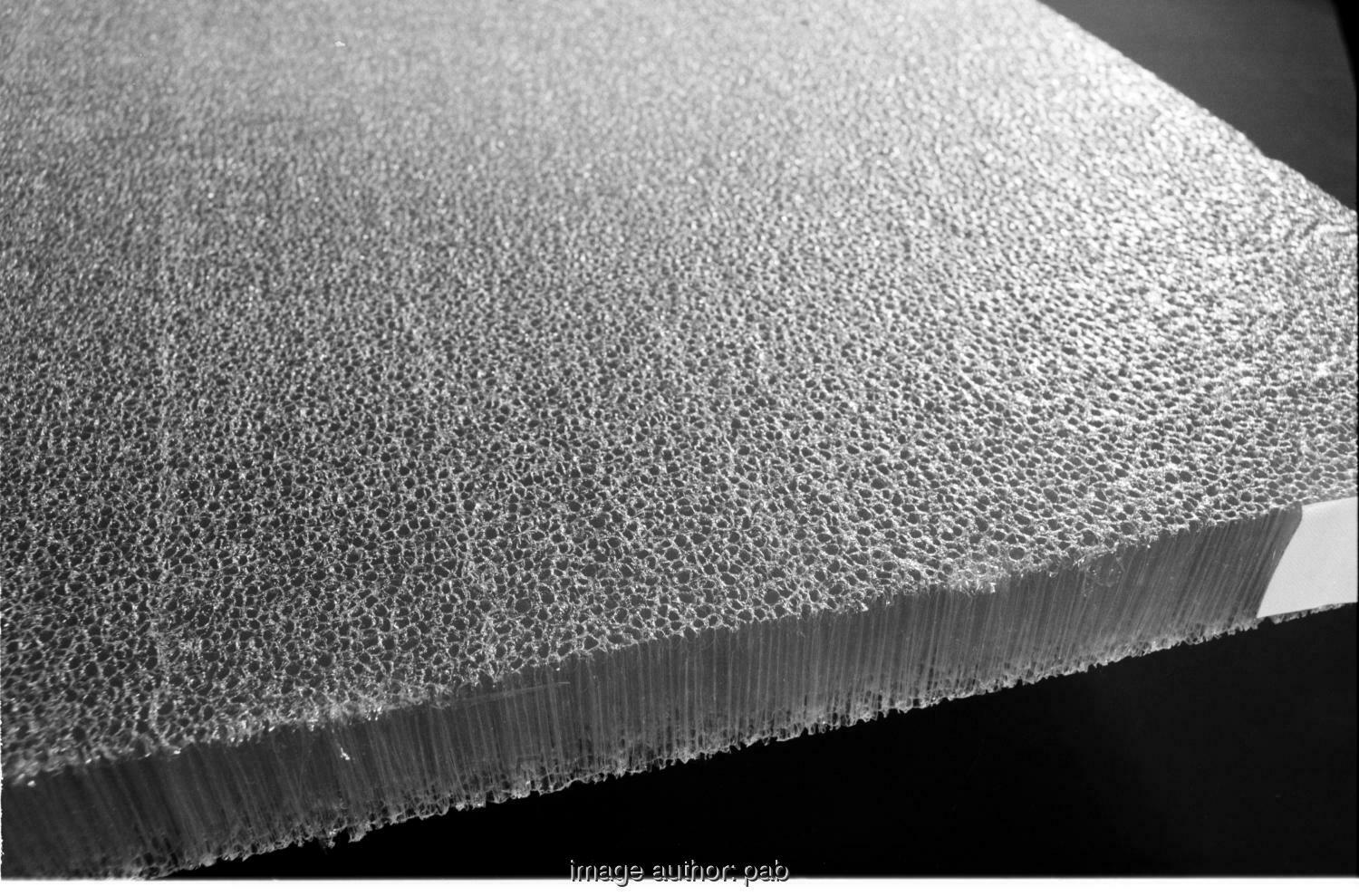}
\caption{\label{photo-wa0301}TI material, ISE number {\bf wa0301}}
\end{center}
\end{figure}

\clearpage

\begin{figure}[h!]
\begin{center}
\includegraphics[height=10cm]{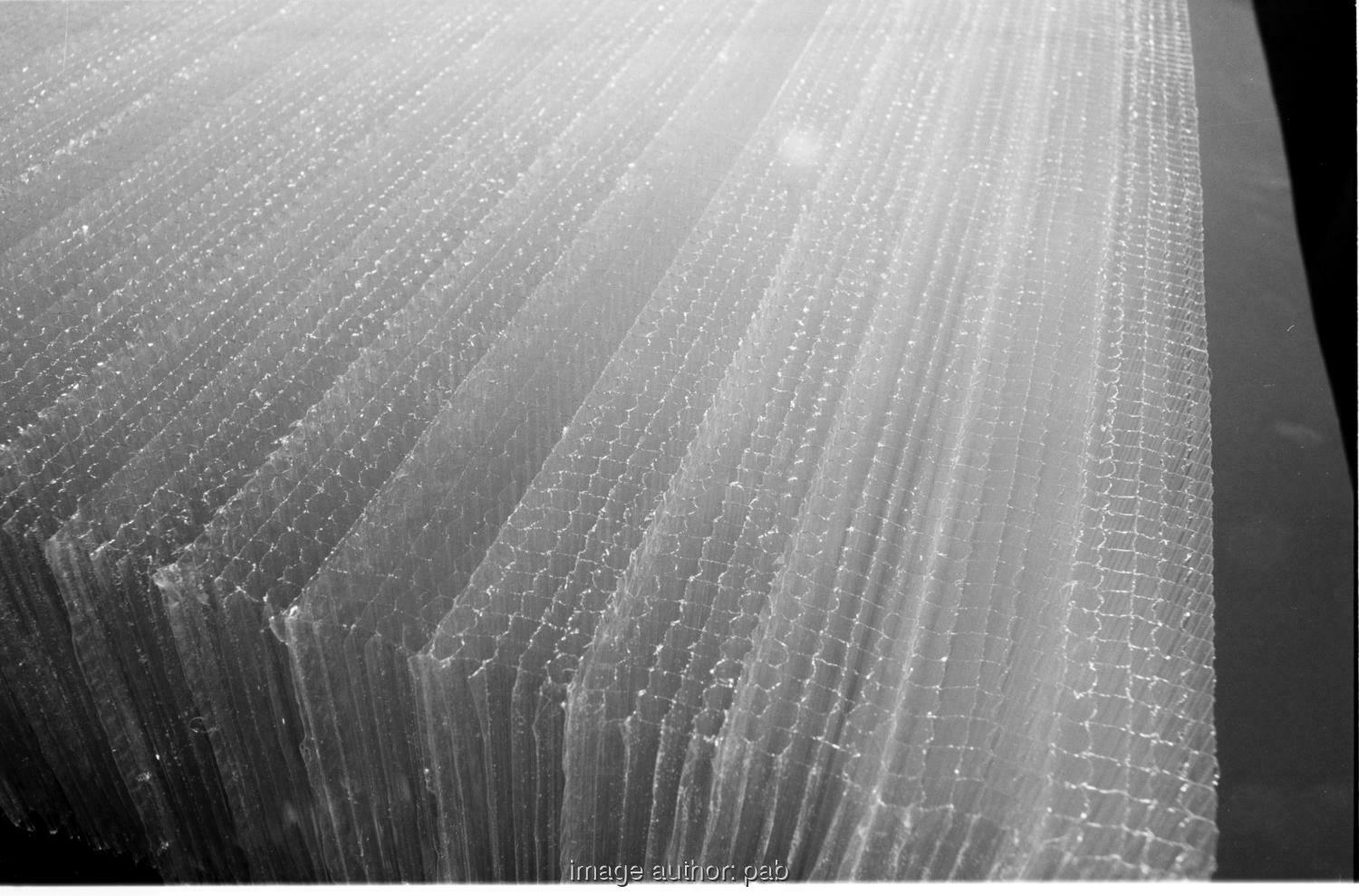}
\caption{\label{photo-wa3101}TI material, ISE number {\bf wa3101}}
\end{center}
\end{figure}

\begin{figure}[h!]
\begin{center}
\includegraphics[height=10cm]{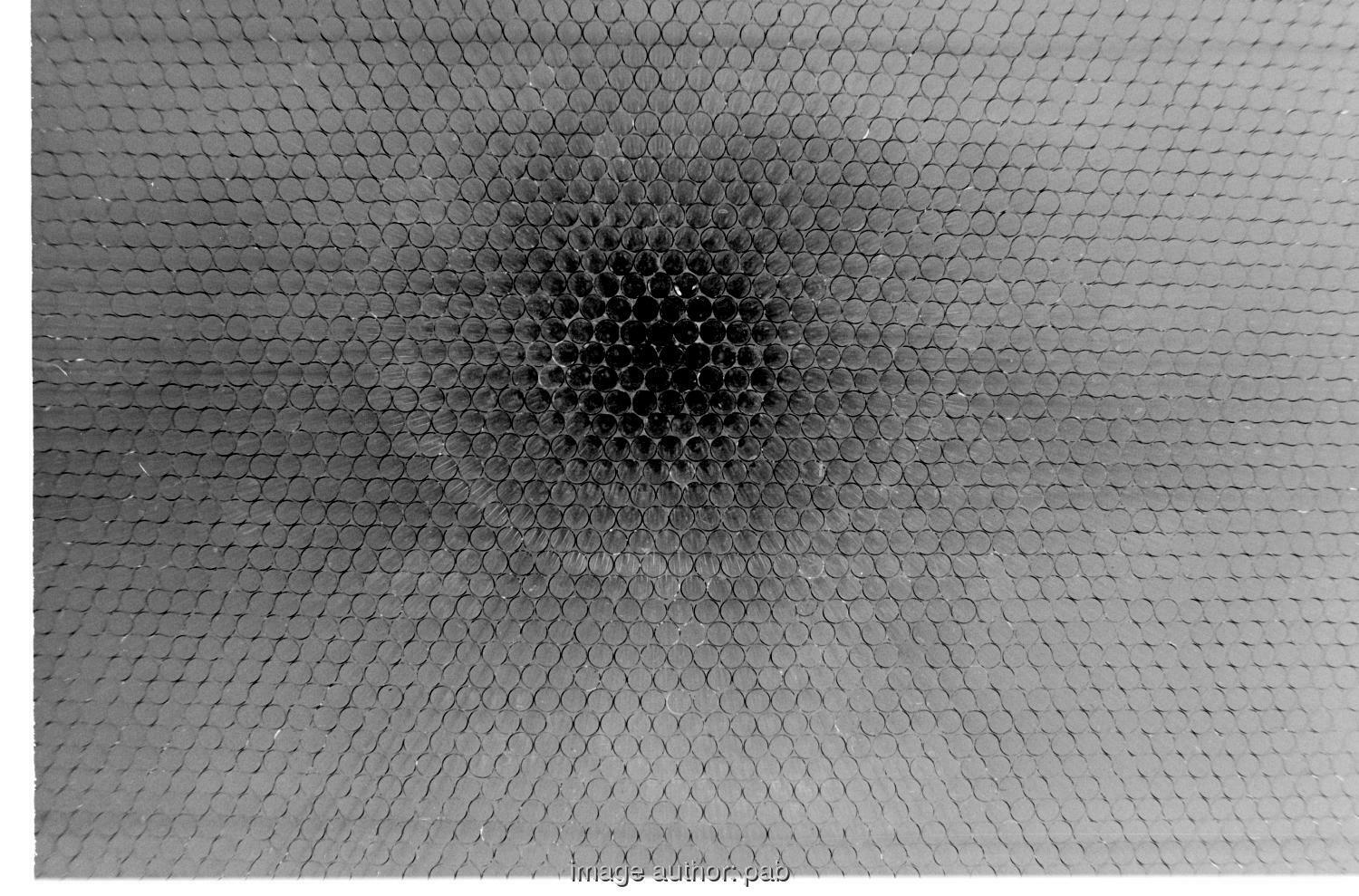}
\caption{\label{photo-wa0325}TI material, ISE number {\bf wa0325}}
\end{center}
\end{figure}

\clearpage
\vfill
\pagebreak[10]

\section{The history of {\tt mountain} versions}
\label{mountain-hist}

Custom programs to visualise 3D scattering data (BRDF, scattered data-points of the form $f_i(\theta_i,\phi_i)$ ), generated by
gonio-photometers, have been written and re-written a number of times.  This is short list of programs written at FhG-ISE, pab-opto or by
users of the PG2 gonio-photometer:

\begin{enumerate}
	\item 1989: version ''zero'', the version used for renderings in this thesis, building a 3D scene from data-points and rendering an
		image in software
	\item 1991: to test 3D accelerated graphic workstations, a minimal version was written by me that used Hewlett-Packards ''Starbase''
		library on a HP TVRX workstation,\\
		and a second version using Silicon-Graphics (SGI) ''IRIS-GL'' library on a SGI GTX,\\
		both machines had been on loan from manufacturers, before purchasing of a 3D-accelerated graphic workstation at FhG-ISE, July 1991
	\item 1992-2001 FhG-ISE production version, using ''IRIS-GL'', on the SGI VGX320 workstation at FhG-ISE, later ported to an SGI ONYX
	\item 2001+ rewrite from scratch by me at pab-opto, using Open-GL, with optional feature to read PG2 data from its  database
		directly. Highly featured, but no GUI. This became the standard tool in batch processing data of the PG2 gonio-photometer.
	\item ca 2008 rewrite from scratch by Christian Reetz at and for Fraunhofer-ISE, with Open-GL and GUI
		(Graphical User Interface using the QT library)
	\item 2015 implementation by \href{http://rgl.epfl.ch/}{EPFL RGL (Realistic Graphics Lab)} as a browser-plugin,
		for their research on BSDF
\end{enumerate}

It would be nice to compare images of each version with the same data-set, but most of the old programs are not available by now. Either the required
libraries or hardware don't exist anymore.

\clearpage
\vfill
\pagebreak[10]

\section[Converting the legacy version of 1990]{Technical aspects of converting a version of 1990 to 2021}
\label{conversion-technical}

The original text of this thesis had been written as ASCII text, imported and formatted with {\em Word} as requested by the department. Bad
decision. Images had been photographed off a monitor, printed extra and glued in the master copy manually, since that bloody software
wouldn't handle the file size with the images included. Typing scientific work with a program sold for office use is a waste of time. The
master copy had then been copied on a colour copier.
If the original formatting had been in \LaTeX, already well available then, things would have be a little easier, e.g. with 
cross-references and citations.

Drawings and images are converted with as little loss of technical quality as possible:
Drawings had been generate using Hewlett-Packards ME10 CAD package and a plot package called DISSPLA, both generating output files in HP-GL
format for HP pen-plotters, and these can be converted to modern line-drawing formats rather easily.

Illustrations have been re-rendered with the {\tt rayshade} program from original source files.
However, most files of the early {\tt mountain} data visualisation, rendered with Rayshade also, are presumed lost. Only few data files of
the very first measurements have survived, likely most of them had been thrown out after the device was changed to a circular detector rail
in the early nineties, together with my backups on HP QIC tape. These first visualisations are re-scanned from the still existing 35mm slides.

In the case of my own photographs of my first gonio-photometer at FhG-ISE, the original film negatives have been preserved intact and have
been scanned.

The honour roll:

HP-GL files are converted to {\tt xfig} by \href{https://www.gnu.org/software/hp2xx/}{hp2xx}.
However, some HP-GL graphics could only be read completely after conversion with CERN's
\href{http://service-hpglview.web.cern.ch/service-hpglview/hpglviewer.html}{hpglview}, thanks for making it available.

\href{https://www.xfig.org/}{xfig}, a program stable to the passage of time, allows editing the text language and export to modern SVG
or PDFTEX formats. With special thanks to Brian Smith for nursing {\tt xfig} for a long time, advice in need and excellent steaks.

3D illustrations had been done with \href{https://graphics.stanford.edu/~cek/rayshade/rayshade.html}{Rayshade} on HP400 workstations running UNIX
(HP-UX) in 1989. The scene files had been archived and images were re-generated using the same {\tt Rayshade} program, compiled and run on
Linux now. The aspect ratio was now set to match the 3:2 ratio of the 36x24mm film used in prints, the original ratio matched the 1280x1024
pixels of the then high-end monitors.

\vfill
\pagebreak[10]

\section[Assembling and testing the sample mount]{Assembling and testing the sample mount as homework assignment}
\label{homework}

The lower support frame of the sample mount and the stepper drives were machined and tested to a large extend at my flat, while all other parts
were built in the FhG-ISE workshop.
Some more images are found on my web page \url{http://pab-opto.de/pers/phd/}.

\begin{figure}[h!]
\begin{center}
\includegraphics[height=10cm]{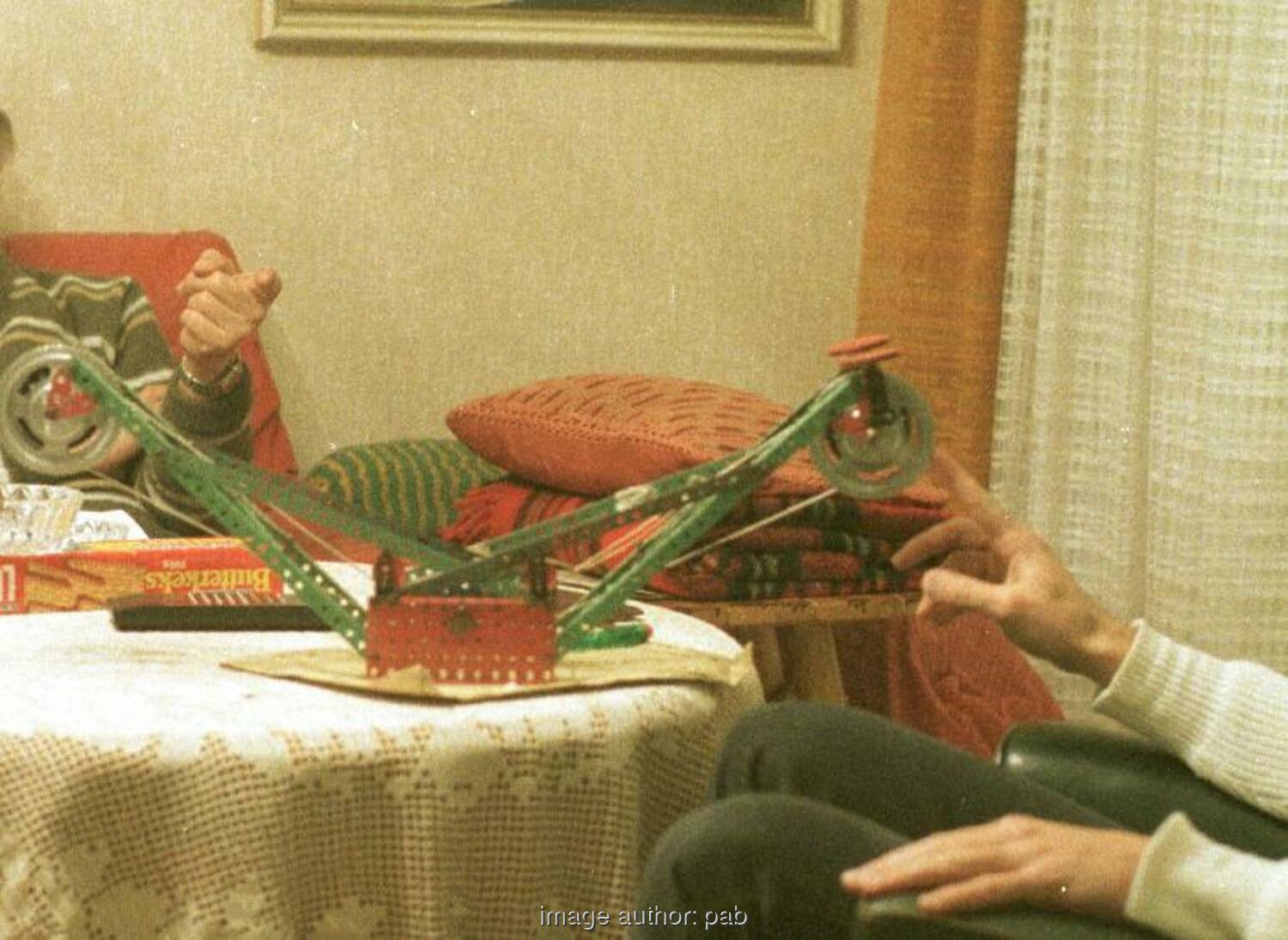}
\caption{\label{sample-0} Presenting the initial layout of the sample support within the family, using M\"arklin metal construction kit, Autumn 1988}
\end{center}
\end{figure}

\begin{figure}[h!]
\begin{center}
\includegraphics[height=9cm]{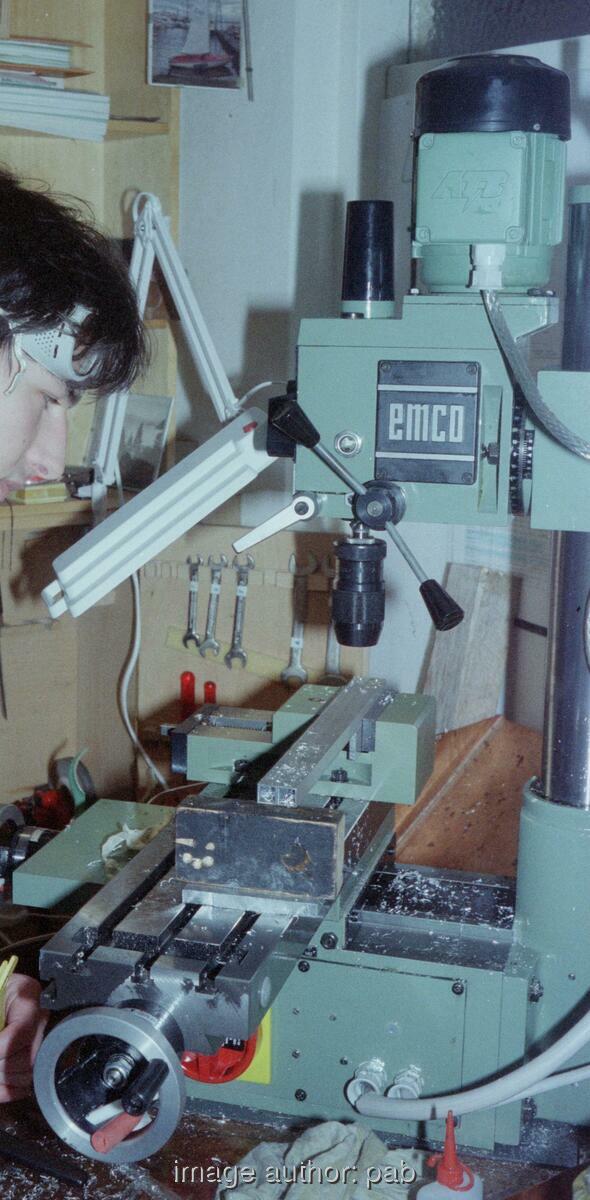}
~~
\includegraphics[height=9cm]{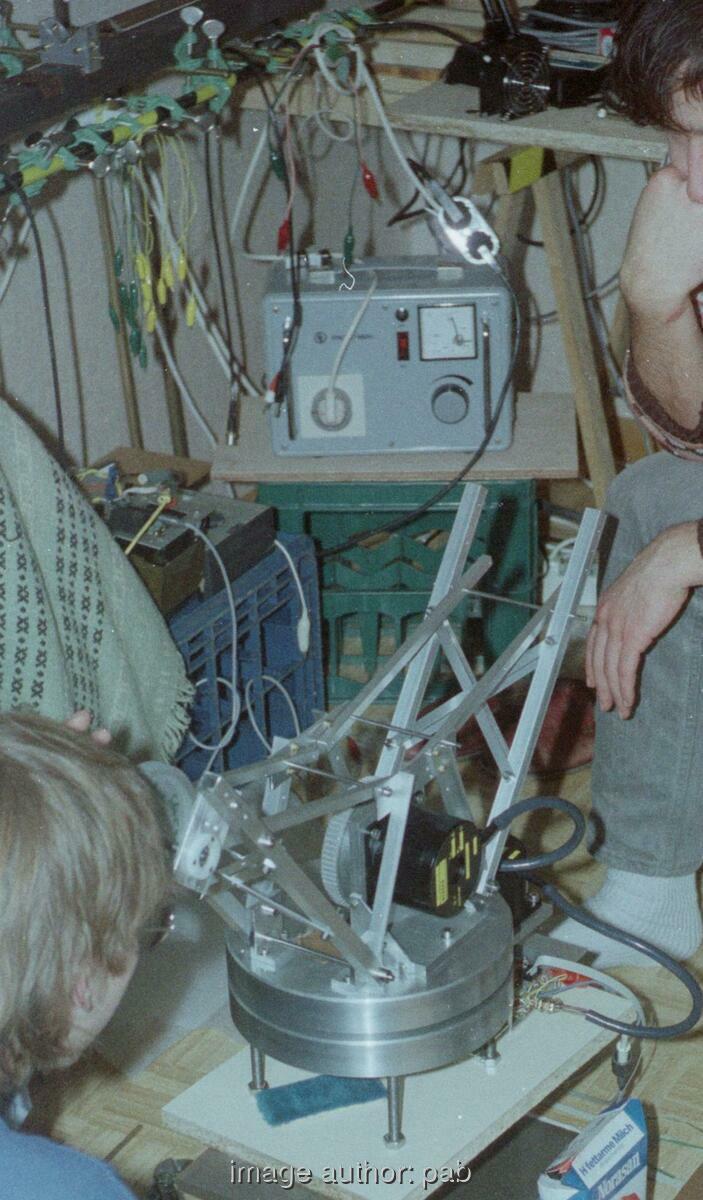}
\caption{\label{sample-1}Machining the struts of the sample support and discussing stepper-drivers}
\end{center}
\end{figure}

\begin{figure}[h!]
\begin{center}
\includegraphics[height=24cm]{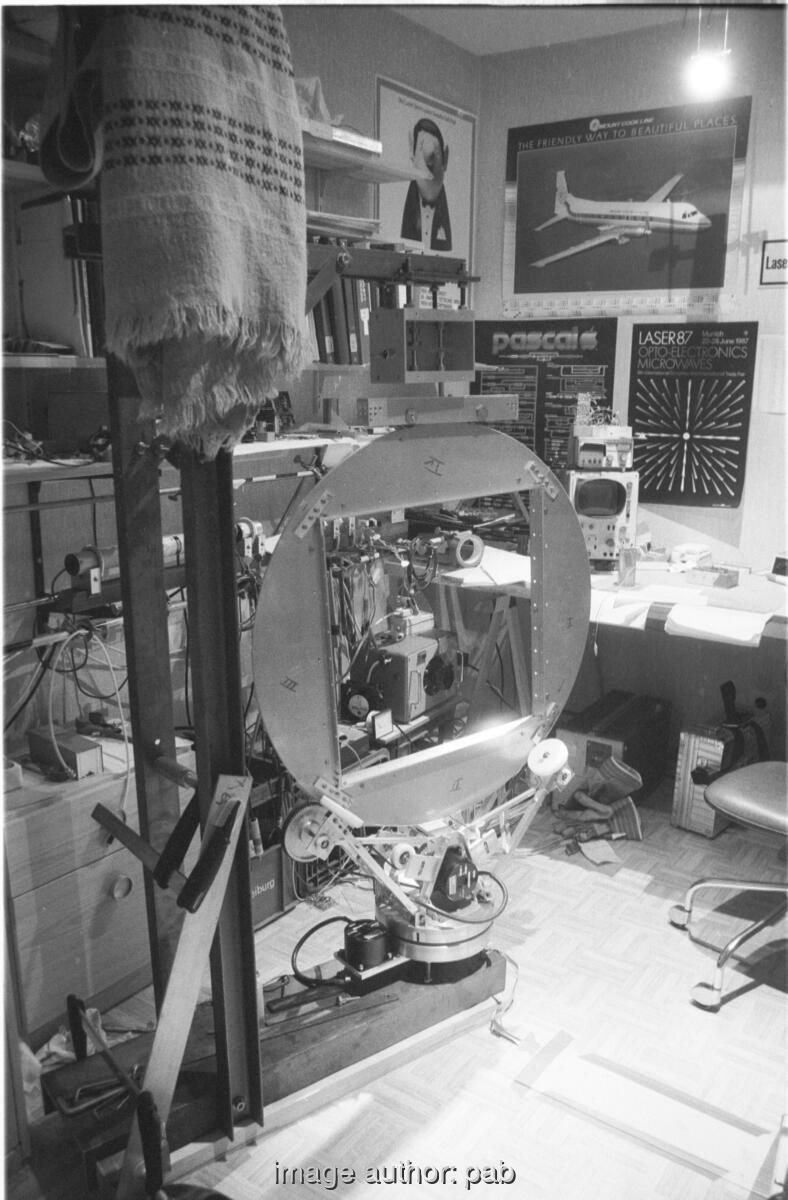}
\caption{\label{sample-2}Testing the sample mount in Spring 1989, not an FhG-ISE location, obviously.}
\end{center}
\end{figure}

\clearpage
\vfill
\pagebreak[10]

\section{Putting it together}
\label{assembling-it}

At first, ceiling, walls or floor were not black. But that was to change soon thereafter (Fig.\ref{overall-photo}) to a working habitat inside
black rooms for the next 30 years:

\begin{figure}[h!]
\begin{center}
\includegraphics[height=10cm]{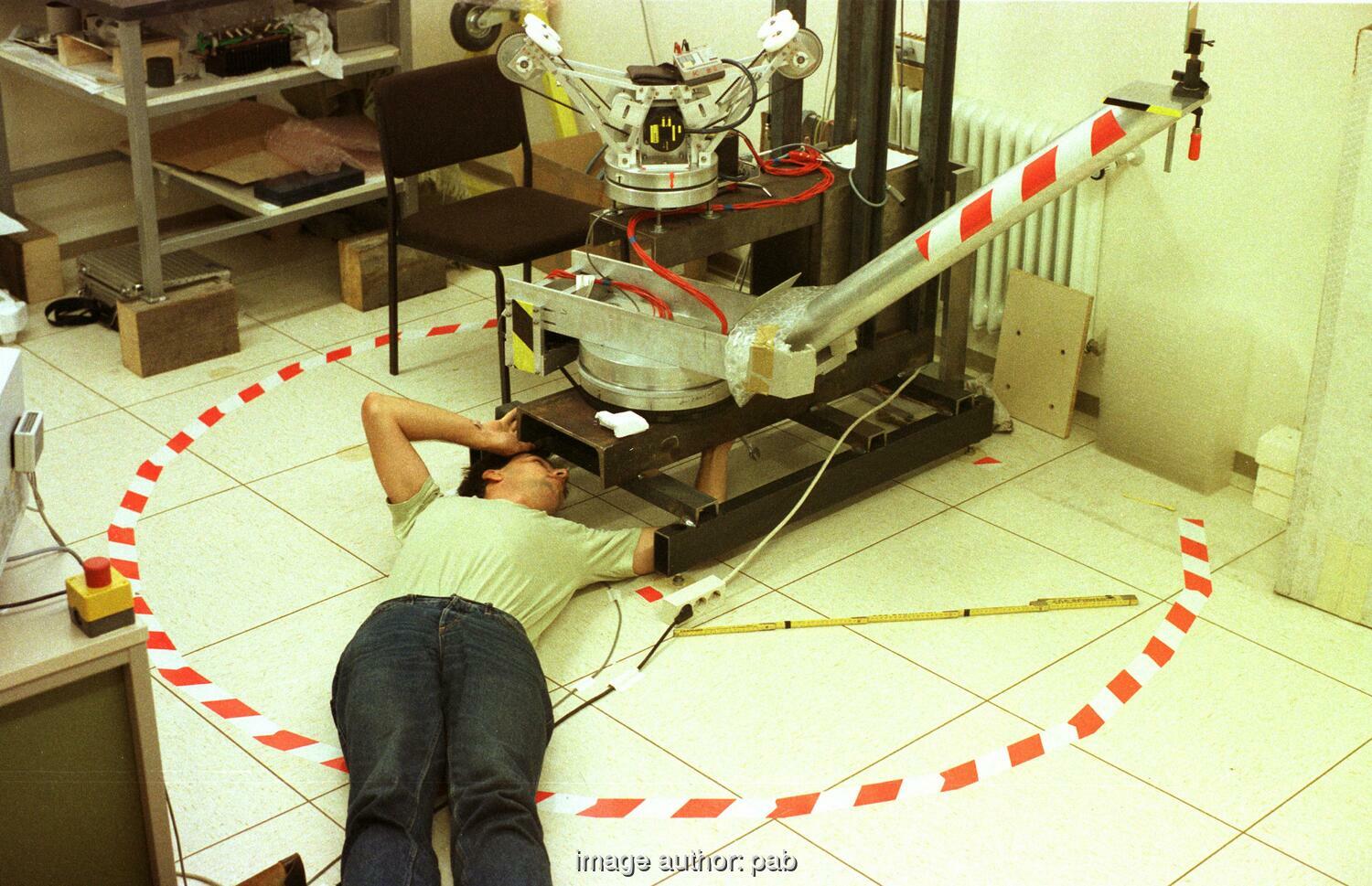}
\caption{\label{pg1-1} Assembling it all 1st time, FhG-ISE, summer 1989. The small mirror clamped at the top of the arm serves to check positional
precision with the HeNe laser.}
\end{center}
\end{figure}

\begin{figure}[h!]
\begin{center}
\includegraphics[height=10cm]{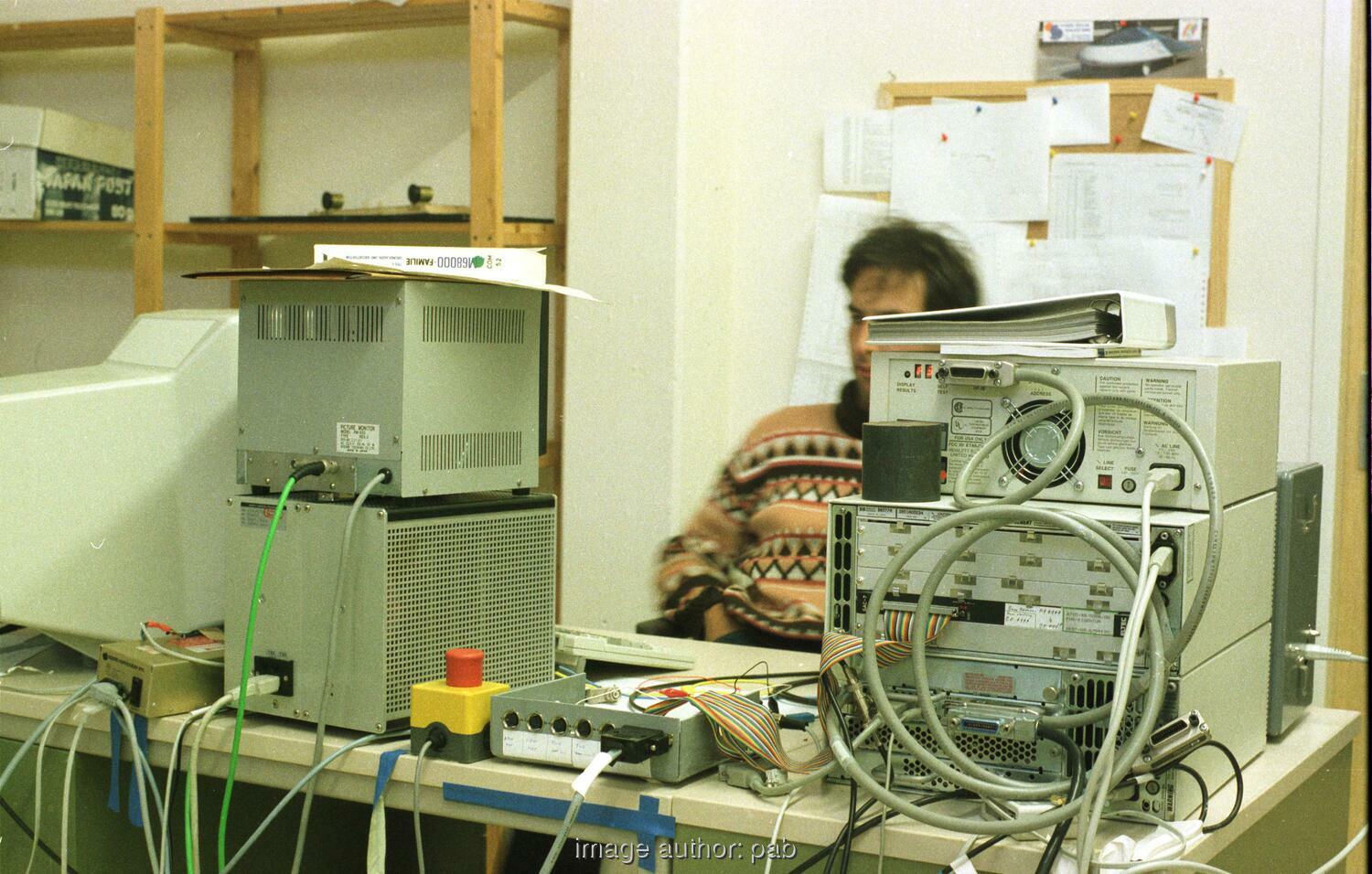}
\caption{\label{pg1-2} The UNIX workstation, HP360 and VME box, PCB with opto-couplers, monitor to view CCD image, lab power supplies and
designer/operator}
\end{center}
\end{figure}

\begin{figure}[h!]
\begin{center}
\includegraphics[height=11cm]{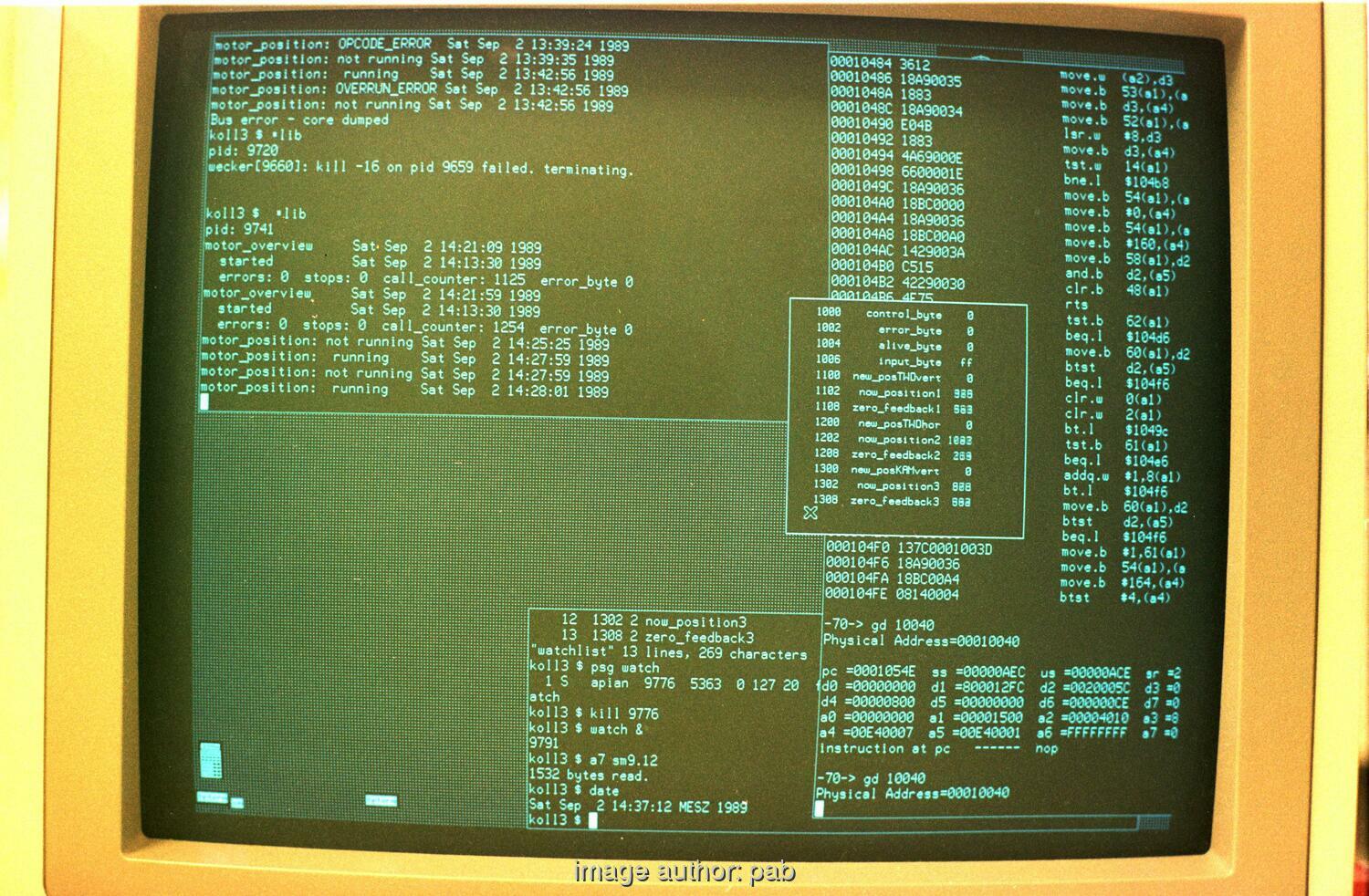}
\caption{\label{pg1-3} Workstation X11-windows on HP-UX with control program, assembler code and monitor, 2nd September 1989, Saturday}
\end{center}
\end{figure}

\begin{figure}[h!]
\begin{center}
\includegraphics[height=11cm]{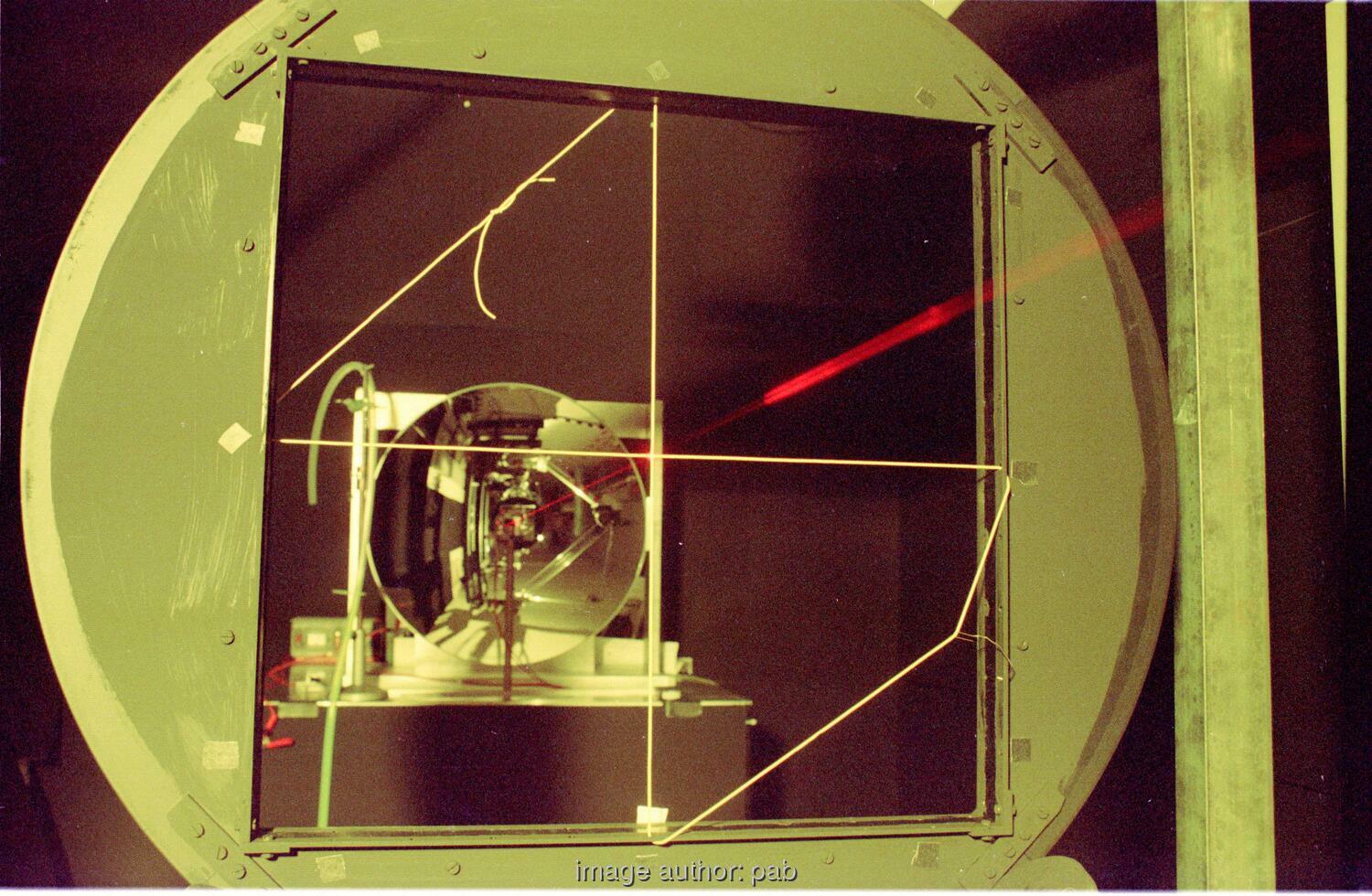}
\caption{\label{pg1-4} Adjusting sample centre and mirrors using a HeNe laser}
\end{center}
\end{figure}

\begin{figure}[h!]
\begin{center}
\includegraphics[height=11cm]{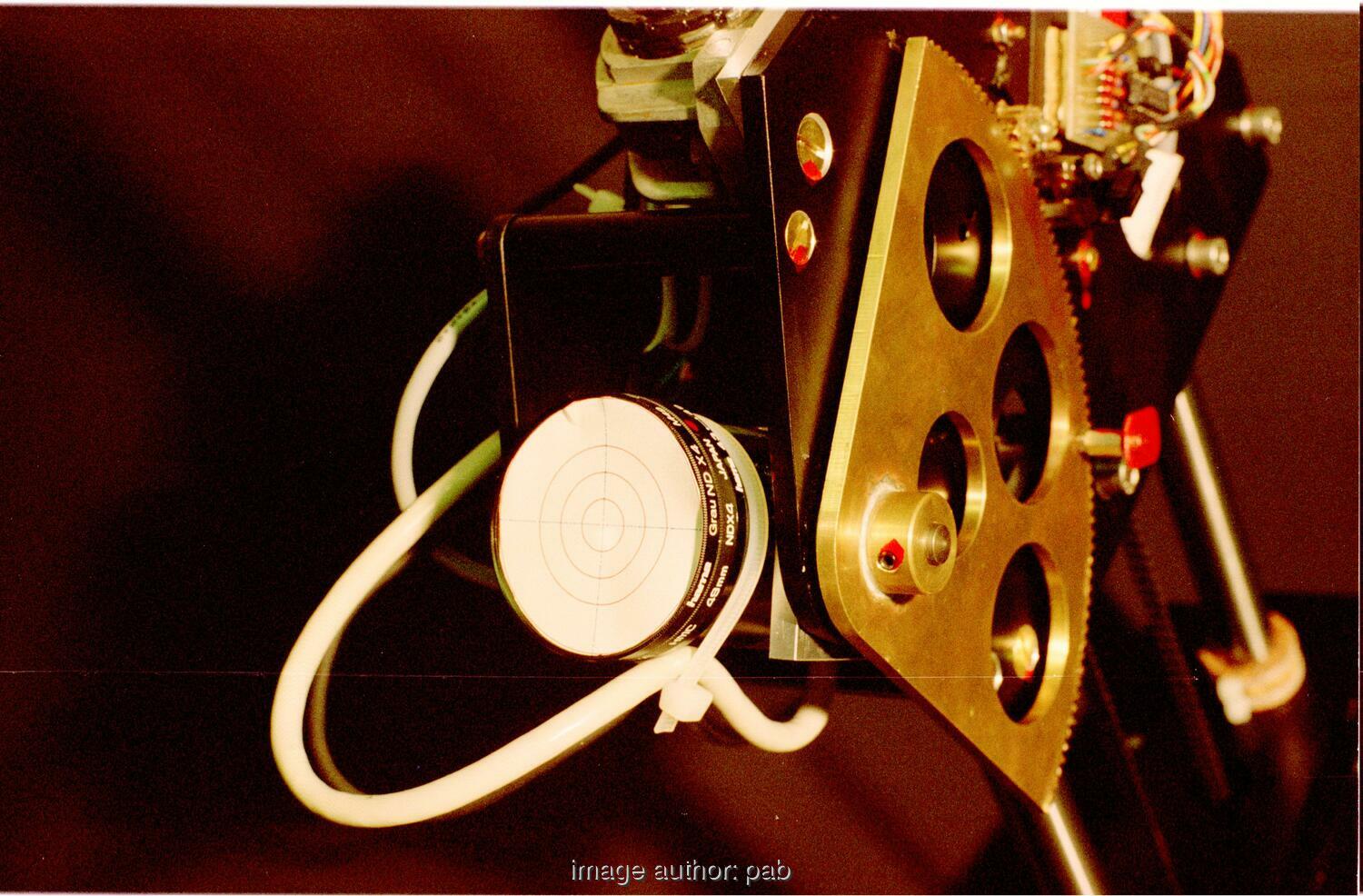}
\caption{\label{pg1-5} Camera module, tiltable, with lens, grey filters and target for adjustment}
\end{center}
\end{figure}

\begin{figure}[h!]
\begin{center}
\includegraphics[height=11cm]{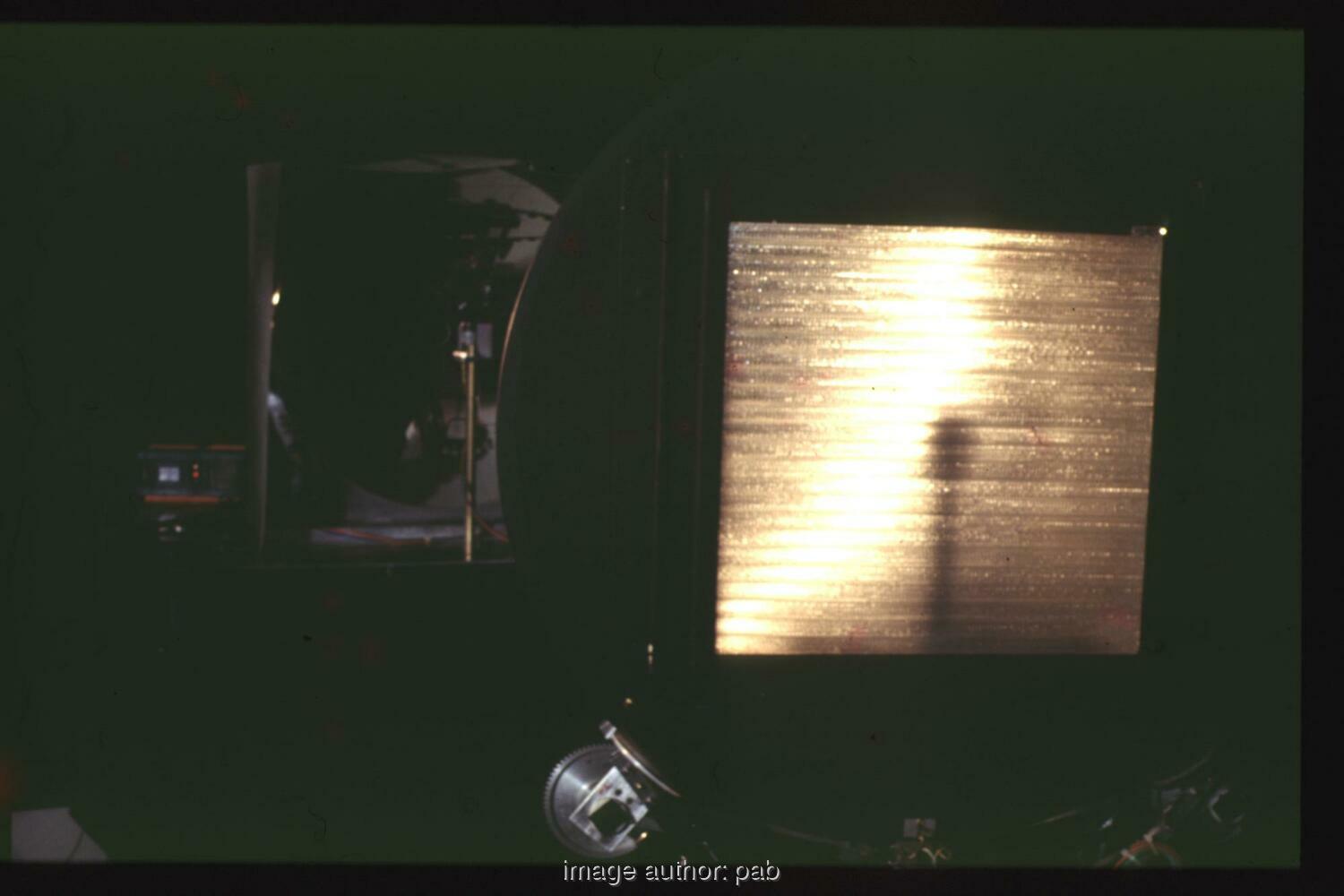}
\caption{\label{honeycomb-real1}
	Photo of mounted {\bf wa3101}, with typical ''ring'', lit by first lamp-design with parabolic mirror (in background left).
	The shadow of the lamp post was avoided in next design step, which had the lamp suspended on steel-wires.}
\end{center}
\end{figure}

\begin{figure}[h!]
\begin{center}
\includegraphics[height=24cm]{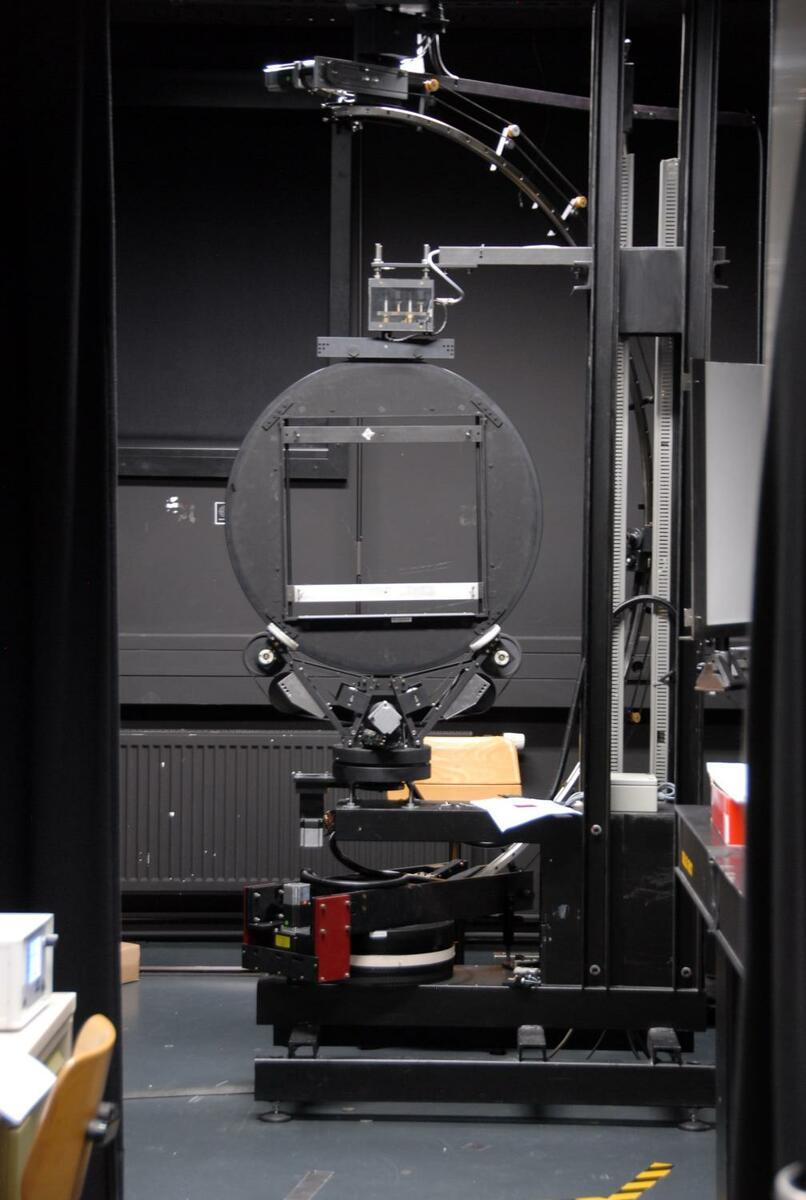}
\caption{\label{workshop2010}
	One of the last photos of my first FhG-ISE gonio-photometer, taken by Greg Ward during the institute-tour at the
	{\em Radiance Workshop 2010} in Freiburg.
	The device was dismantled, by its designer, in 2013 and replaced by a {\em pab PG2 gonio-photometer}.}
\end{center}
\end{figure}

\begin{figure}[h!]
\begin{center}
\includegraphics[height=11cm]{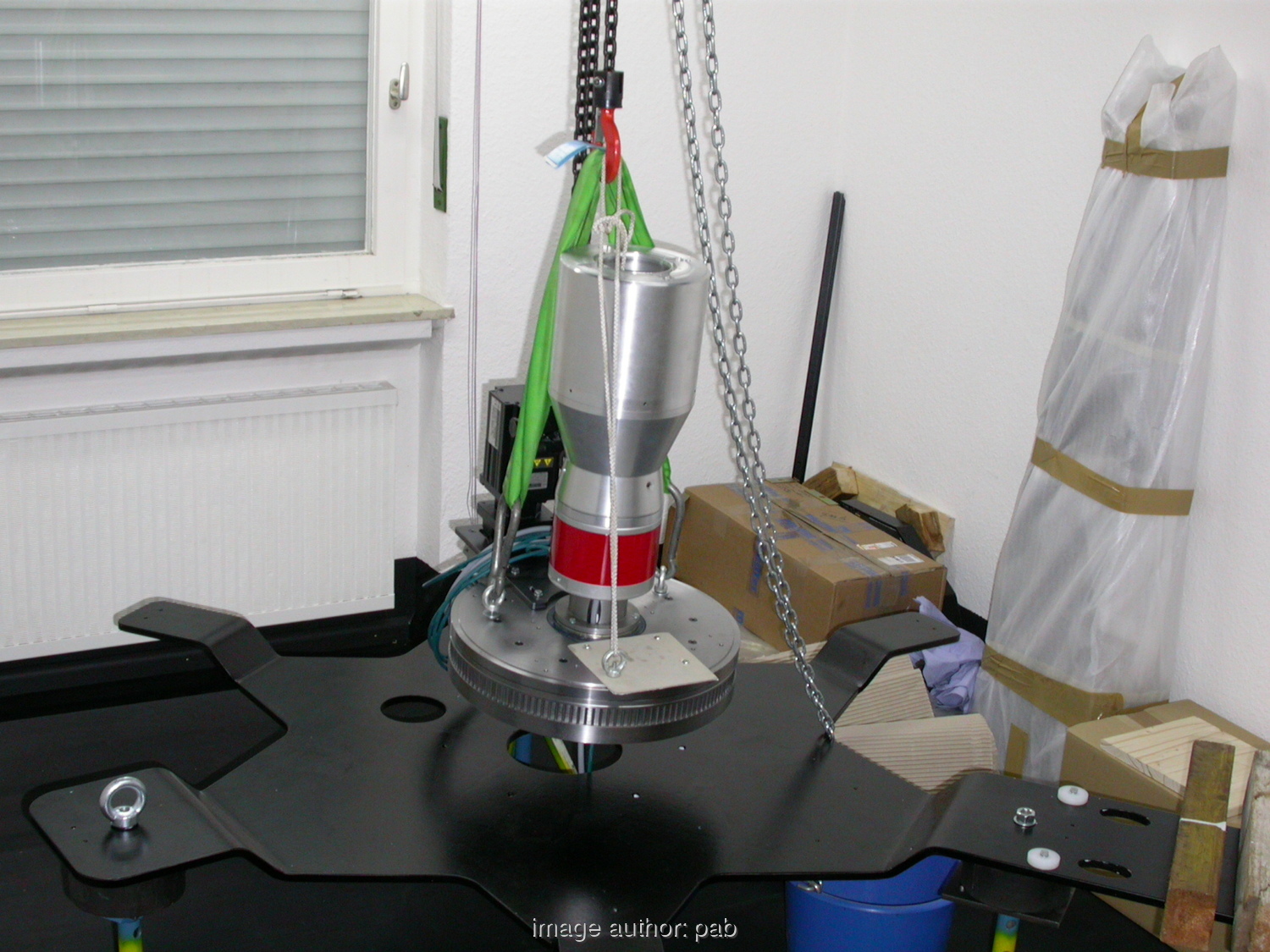}
\caption{\label{pg2-zero} The grand-child of my first gonio-photometer: The assembly of the prototype of my PG2 gonio-photometer,
15th Juli 2004 at pab-opto}
\end{center}
\end{figure}

\begin{figure}[h!]
\begin{center}
\includegraphics[height=11cm]{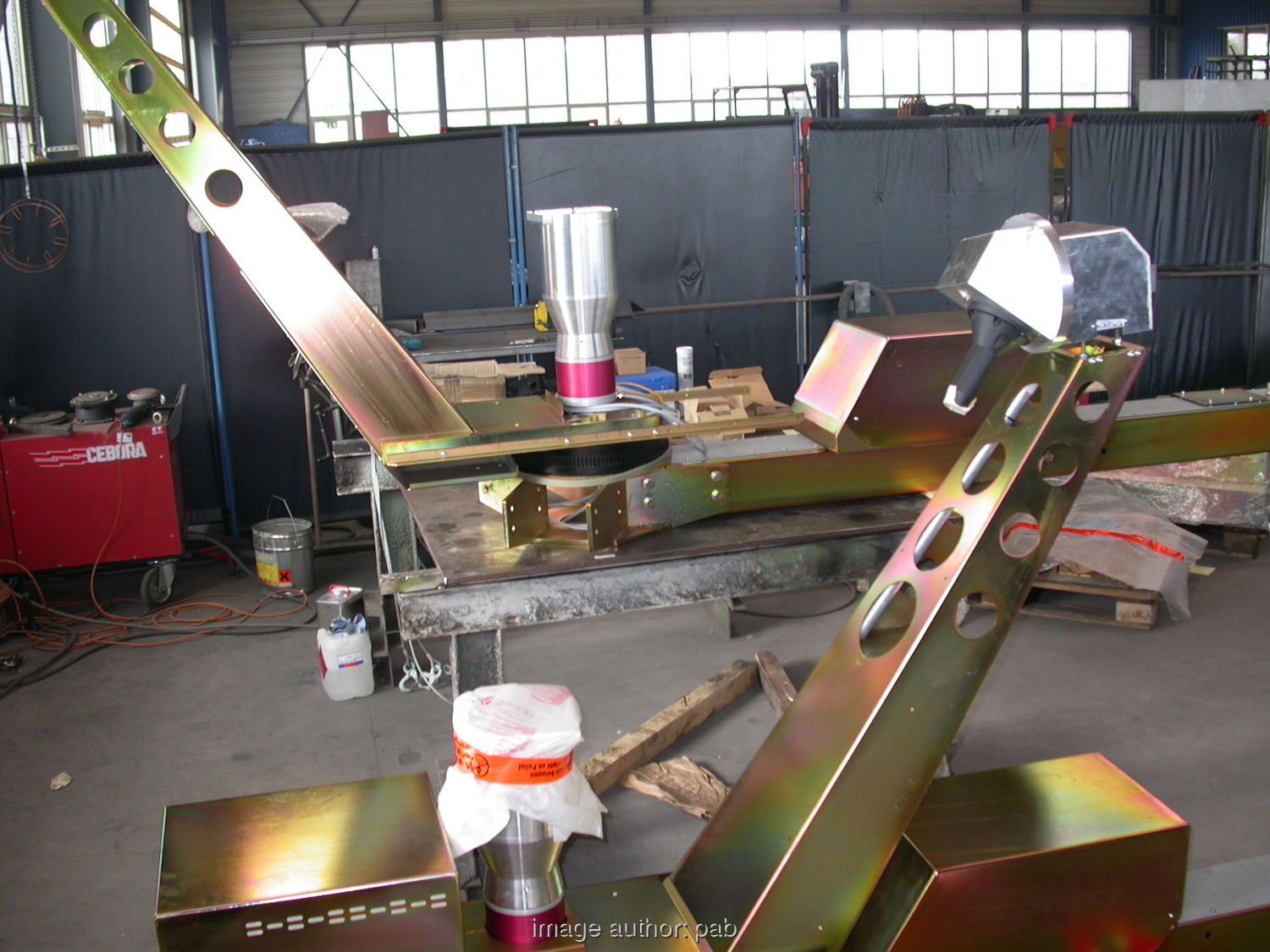}
\caption{\label{pg2-witec} Assembling two \href{http://pab.eu}{PG2 gonio-photometer}, 1st March 2008, \href{https://www.witec-ag.com/}{Witec workshop}}
\end{center}
\end{figure}

\end{appendix}

\vspace*{5cm}
Institutional address as of 2021:
~\\

Institute of Physics\\
Albert-Ludwigs-University Freiburg\\
Hermann-Herder-Str. 3\\
79104 Freiburg im Breisgau\\
Germany\\
\url{https://www.physik.uni-freiburg.de/}
~\\

Fraunhofer Institute for Solar Energy Systems, FhG-ISE\\
Heidenhofstraße 2\\
79110 Freiburg im Breisgau\\
Germany\\
\url{http://www.ise.fraunhofer.de/}

\vspace*{5cm}

Author's affiliation as of 2021:

\href{http://www.pab.eu/}{\includegraphics[height=2cm]{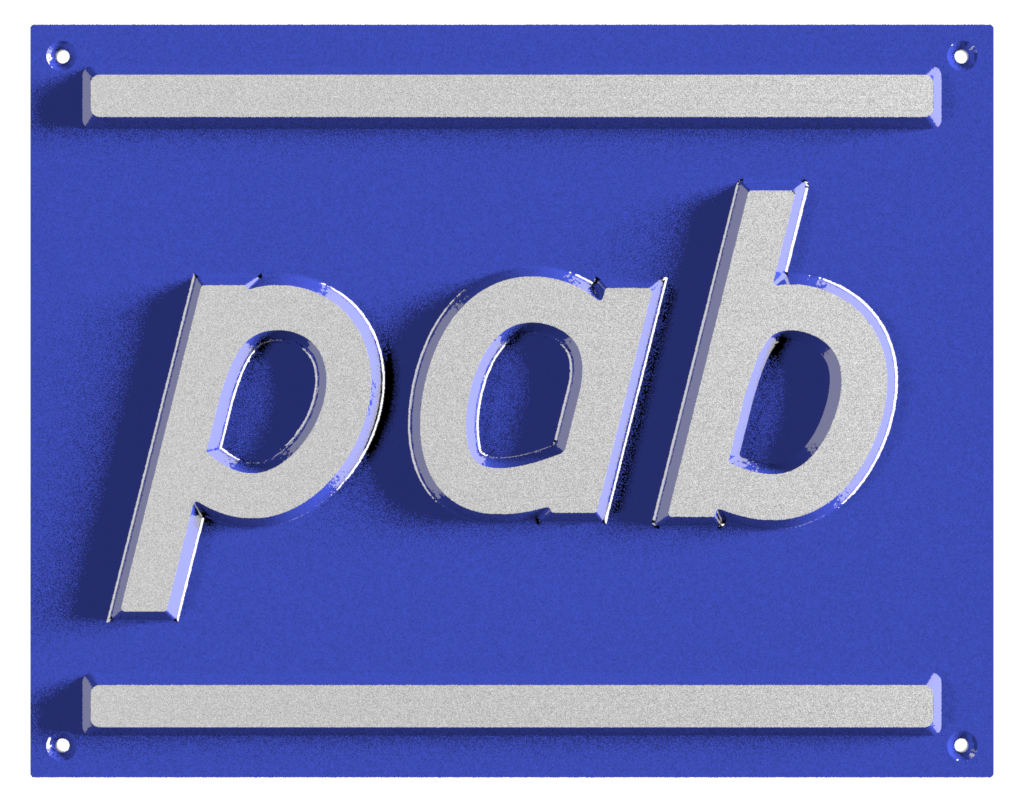}}

last page. Thank you, have a nice day.

\end{document}